\documentclass[numberedappendix,twocolappendix,iop,apj]{emulateapj}
\usepackage{psfig,amsfonts,amsmath,graphicx,natbib,nicefrac,url,hyperref,longtable}

\usepackage{booktabs}
\hypersetup{colorlinks=true, urlcolor=blue, citecolor=blue}

\newcommand{\Swift}{\textit{Swift}}

\newcommand{\EK}{\ensuremath{E_{\rm K}}}
\newcommand{\EKiso}{\ensuremath{E_{\rm K,iso}}}
\newcommand{\Egamma}{\ensuremath{E_{\gamma}}}
\newcommand{\Egammaiso}{\ensuremath{E_{\gamma,\rm iso}}}	     
\newcommand{\epse}{\ensuremath{\epsilon_{\rm e}}}
\newcommand{\epsb}{\ensuremath{\epsilon_{\rm B}}}
\newcommand{\dens}{\ensuremath{n_{0}}}
\newcommand{\Astar}{\ensuremath{A_{*}}}

\newcommand{\tjet}{\ensuremath{t_{\rm jet}}}
\newcommand{\thetajet}{\ensuremath{\theta_{\rm jet}}}
\newcommand{\tnr}{\ensuremath{t_{\rm NR}}}
\newcommand{\AV}{\ensuremath{A_{\rm V}}}

\newcommand{\pcmsq}{\ensuremath{{\rm cm}^{-2}}}
\newcommand{\pcc}{\ensuremath{{\rm cm}^{-3}}}

\newcommand{\nua}{\ensuremath{\nu_{\rm a}}}
\newcommand{\nusa}{\ensuremath{\nu_{\rm sa}}}
\newcommand{\nuac}{\ensuremath{\nu_{\rm ac}}}
\newcommand{\numax}{\ensuremath{\nu_{\rm m}}}
\newcommand{\nuc}{\ensuremath{\nu_{\rm c}}}

\newcommand{\nuopt}{\ensuremath{\nu_{\rm opt}}}
\newcommand{\nux}{\ensuremath{\nu_{\rm X}}}

\newcommand{\lbt}{LBT14}

\shorttitle{GRB~140311A}
\shortauthors{Laskar et al.}

\def\nrao{1}
\def\ucb{2}
\def\cfa{3}
\def\ou{4}
\def\northwestern{5}
\def\einsteinfellow{6}
\def\ariz{7}
\def\ash{8}

\begin{document} 

\title{A VLA Study of High-redshift GRBs I --\\
       Multi-wavelength Observations and Modeling of GRB\,140311A}
\author{Tanmoy Laskar\altaffilmark{\nrao,\ucb},
        Edo Berger\altaffilmark{\cfa},
        Ryan Chornock\altaffilmark{\ou},
        Raffaella Margutti\altaffilmark{\northwestern},        
        Wen-fai Fong\altaffilmark{\einsteinfellow,\ariz},
        \\and Ashley Zauderer\altaffilmark{\ash}
}
\altaffiltext{\nrao}{National Radio Astronomy Observatory,
520 Edgemont Road, Charlottesville, VA 22903, USA} 
\altaffiltext{\ucb}{Department of Astronomy, University of California, 501 Campbell Hall, 
Berkeley, CA 94720-3411, USA} 
\altaffiltext{\cfa}{Department of Astronomy, Harvard University, 60 Garden Street, Cambridge, MA 
02138, USA}
\altaffiltext{\ou}{Astrophysical Institute, Department of Physics and 
Astronomy, 251B Clippinger Lab, Ohio University, Athens, OH 45701, USA}
\altaffiltext{\northwestern}{Center for Interdisciplinary Exploration and Research in Astrophysics 
(CIERA) and Department of Physics and Astrophysics,Northwestern University, Evanston, IL 60208, USA}
\altaffiltext{\einsteinfellow}{Einstein Fellow}
\altaffiltext{\ariz}{Steward Observatory, University of Arizona, 933 N. Cherry Ave, Tucson, AZ 
85721, USA}
\altaffiltext{\ash}{Center for Cosmology and Particle Physics, New York University, 4 Washington 
Place, New York, NY 10003, USA}

\begin{abstract}
We present the first results from a recently concluded study of GRBs at $z\gtrsim5$ with the Karl 
G. Jansky Very Large Array (VLA). Spanning $1$ to $85.5$\,GHz and 7 epochs from 1.5 to 82.3\,d, our 
observations of GRB~140311A are the most detailed joint radio and millimeter 
observations of a GRB afterglow at $z\gtrsim5$ to date. In conjunction with optical/near-IR and 
X-ray data, the observations can be understood in the framework of radiation from a single blast 
wave shock with energy 
$\EKiso\approx8.5\times10^{53}$\,erg expanding into a constant density environment with density, 
$\dens\approx8$\,\pcc. The X-ray and radio observations require a jet break at 
$\tjet\approx0.6$\,d, yielding an opening angle of $\thetajet\approx4^{\circ}$ and a 
beaming-corrected blast wave kinetic energy of $\EK\approx2.2\times10^{50}$\,erg. The results from 
our radio follow-up and multi-wavelength modeling lend credence to the hypothesis that 
detected high-redshift GRBs may be more tightly beamed than events at lower redshift. We do not find 
compelling evidence for reverse shock emission, which may be related to fast cooling driven by
the moderately high circumburst density.
\end{abstract}

\section{Introduction}
Long-duration $\gamma$-ray bursts (GRBs) have been firmly established to originate from the
catastrophic death of massive stars \citep[e.g.][]{wb06}. The large luminosities of GRB afterglows 
makes these energetic events premier probes of the high-redshift Universe, ranging from the parsec 
scale environments of the progenitors to the properties of the intergalactic medium 
\citep{tkk+06,ioc07,tfl+09,wbg+12,cbf+13,cbf+14}.
Additionally, high-redshift GRBs have been speculated to possibly differ from lower redshift events 
in their energy scale, durations, and circumburst media \citep{fwh01,byh03,hfw+03,si11,tsm11}.
Due to time dilation, high-redshift GRBs also afford an opportunity to capture rapidly-evolving 
reverse shock emission, and thereby a means of probing the Lorentz factor and composition of the 
relativistic ejecta powering the afterglow \citep{pir05,mes06,lbz+13,lbt+14,lab+16,pcc+14,alb+17}.

Detailed multi-wavelength observations of GRB afterglows spanning the X-rays to the radio 
bands yield a measurement of the explosion properties and circumburst environments
\citep[e.g.][]{spn98}. 
Whereas the rapid response of \Swift\ has yielded prompt X-ray afterglow localization and rich 
X-ray light curves, and ground-based facilities have improved the detection and 
characterization of optical light curves 
\citep{nkg+06,lzz07,lrz+08,ebp+09,mzb+13,zbm+13}, 
detailed observations of afterglows in the radio and millimeter have resulted in a low detection 
rate of about $30\%$, with sensitivity being the primary challenge \citep{cf12,duplm+12}.

Due to these limitations, the multi-wavelength properties of GRBs at $z\gtrsim5$ remain
poorly characterized \citep{tac+05,hnr+06,kka+06,gkf+09,sdvc+09,tfl+09,clf+11}.
We carried out a comprehensive analysis of three GRBs at $z\gtrsim6$ with radio detections, 
and demonstrated that these events exhibit explosion energies typical of 
the lower redshift population but tend to exhibit narrower jet opening angles
\citep[][henceforth LBT14]{lbt+14}.
However, the radio observations for two out of the three events only yielded an upper bound on the 
synchrotron self-absorption frequency, resulting in 
an order of magnitude or greater uncertainty in the circumburst density and 
degeneracies between the various physical parameters. 

With the upgrade of the Karl G.~Jansky Very Large Array (VLA) providing an order of magnitude 
improvement in sensitivity and continuous frequency coverage from 1 to 40\,GHz, detailed 
observations of GRB afterglows in the cm band are now feasible. Taking advantage of this 
opportunity, we targeted all events with secure spectroscopic redshifts of $z\gtrsim5$ at multiple 
VLA frequencies, supported by mm-band data from the Combined Array for Millimeter Astronomy 
(CARMA). Our sample consists of four events: GRBs 130606A, 140304A, 140311A, and 140515A. In this 
series of papers, we present the results of our observations and characterize the multi-wavelength 
afterglows of high-redshift GRBs, focusing in particular on whether the inferred explosion 
properties and circumburst environments are suggestive of evolution in the nature of the 
progenitors. Here, we present our observations and analysis of GRB~140311A at $z=4.954$. We employ 
standard cosmological parameters of $\Omega_{m}=0.31$, $\Omega_{\Lambda}=0.69$, and 
$H_0=68$\,km\,s$^{-1}$\,Mpc$^{-1}$ \citep{aaa+16}; all magnitudes are in the AB system, all 
uncertainties are at $1\,\sigma$, and all times are in the observer frame, unless otherwise 
specified.

\section{GRB Properties and Observations}
\label{text:GRB_Properties_and_Observations}
GRB~140311A was discovered by the \Swift\ \citep{gcg+04} Burst Alert Telescope 
\citep[BAT,][]{bbc+05} on 2014 March 11 at 21:05:16\,UT \citep{gcn15944}. The burst duration  
is $T_{90} = 71.4\pm9.5$\,s, with a fluence of $F_{\gamma} = (2.3 \pm 0.3) 
\times10^{-6}$\,erg\,\pcmsq\ \citep[15--150\,keV, $90\%$ confidence;][]{gcn15962}. 
The optical afterglow, discovered by the 1\,m Nanshan telescope at Xinjiang Observatory 
\citep{gcn15947}, was subsequently observed with several other telescopes
\citep{gcn15952,gcn15953,gcn15954,gcn15956}. Spectroscopic observations 11.3\,hr after the burst 
with the Gemini-South 8\,m telescope provided a redshift  of $z = 4.95$ \citep{gcn15961}, which was 
confirmed by the Nordic Optical Telescope \citep{gcn15964}.

At this redshift, the \Swift/BAT $\gamma$-ray fluence corresponds to an isotropic energy release of 
$\Egammaiso = (1.0\pm0.1)\times10^{53}$\,erg (89--890\,keV, rest frame). In the absence of 
observations by a wide-band $\gamma$-ray satellite and the consequent lack of information about the 
$\gamma$-ray spectrum, we adopt a K-correction to the rest-frame $1$--$10^4$\,keV band of a factor 
of $2.7\pm0.9$ to determine $\Egammaiso \approx (2.7\pm0.9)\times10^{53}$\,erg, where the 
uncertainty is dominated by the uncertainty in the K-correction (\lbt). 

\begin{deluxetable}{lc}
 \tabletypesize{\footnotesize}
 \tablecolumns{2}
 \tablecaption{XRT Spectral Analysis for GRB 140311A\label{tab:xrtspect}}
 \tablehead{   
   \colhead{Parameter} &
   \colhead{Value} 
   }
 \startdata 
 $T_{\rm start}$ (s)                                    & $9.5\times10^{3}$ \\
 $T_{\rm end}$ (s)                                      & $6.3\times10^{5}$ \\
 $N_{\rm H, gal}$ ($10^{20}$ \pcmsq)                    & 2.81  \\
 $N_{\rm H, int}$ ($10^{22}$ \pcmsq)                    & $< 3.9$\tablenotemark{\dag}\\
 Photon index, $\Gamma$                                 & $1.66\pm0.08$\\
 Flux (0.3--10\,keV, observed; erg\,\pcmsq\,s$^{-1}$)   & $2.0\times10^{-12}$ \\
 Flux (0.3--10\,keV, unabsorbed; erg\,\pcmsq\,s$^{-1}$) & $2.1\times10^{-12}$ \\
 Counts to flux (observed; erg\,\pcmsq\,ct$^{-1}$)      & $4.5\times10^{-11}$ \\
 Counts to flux (unabsorbed; erg\,\pcmsq\,ct$^{-1}$)    & $4.4\times10^{-11}$ \\
 C statistic (dof) & 193 (214)
 \enddata
 \tablecomments{${}^\dag3\sigma$ upper limit.}
\end{deluxetable}

\subsection{X-ray: \Swift/XRT}\label{text:data_analysis:XRT}                            
\Swift\ X-ray Telescope \citep[XRT,][]{bhn+05} observations of GRB~140311A were delayed due to an 
Earth limb constraint and began 0.11~d after the BAT trigger. The X-ray afterglow was 
localized to RA = 13h\,57m\,13.25s, Dec = +00$^\circ$\,38\arcmin\,30.8\arcsec\ (J2000), with an
uncertainty radius of 1.5\arcsec\ (90\% 
containment)\footnote{\url{http://www.swift.ac.uk/xrt_positions/00591390/}}. XRT continued 
observing the afterglow for 4.7\,d in photon counting mode. We extract XRT PC-mode spectra using 
the on-line tool on the \Swift\ website 
\citep{ebp+07,ebp+09}\footnote{\url{http://www.swift.ac.uk/xrt_spectra/00591390/}}.
We downloaded the event and response files generated by the on-line tool in these time bins, and 
fit them using the HEASOFT (v6.19) software package 
and corresponding calibration files. We used Xspec to fit all 
available PC-mode data, assuming a photoelectrically absorbed power law model (\texttt{tbabs 
$\times$ ztbabs $\times$ pow}) and fixing the galactic absorption column to $N_{\rm H, Gal} = 
2.81\times10^{20}\,\pcmsq$ \citep{wsb+13}. We find no evidence for excess absorption in the host 
galaxy, and therefore freeze the intrinsic absorption column to $N_{\rm H, int} = 0$ in the fit.
The parameters of our best fit spectral model are listed in Table \ref{tab:xrtspect}.
In the following analysis, we take the 0.3 -- 10\,keV count rate light curve from the 
\Swift\ website and compute the 1\,keV flux density using our spectral model.
We combine the uncertainty in flux calibration based on our spectral analysis (6\%) 
in quadrature with the statistical uncertainty from the on-line light curve. 

\begin{deluxetable}{ccc}
\tabletypesize{\scriptsize}
\tablecaption{\Swift\ UVOT Observations of GRB~140311A \label{tab:data:UVOT}}
\tablehead{
\colhead{$\Delta t$} & \colhead{Filter} & \colhead{$3\sigma$ Flux Upper Limit} \\ 
\colhead{(d)} & \colhead{} & \colhead{(mJy)} }

\startdata
$1.16\times10^{-1}$ &	white &	$3.6\times10^{-3}$ \\
$4.52\times10^{-1}$ &	white &	$4.6\times10^{-3}$ \\
$4.43\times10^{-1}$ &	b &	$1.6\times10^{-2}$ \\
\ldots & \ldots & \ldots
\enddata
\tablecomments{This is a sample of the full table available on-line.}
\end{deluxetable}

\begin{deluxetable*}{ccccccccc}
\tabletypesize{\scriptsize}
\tablecaption{Optical Observations of GRB~140311A \label{tab:data:GCN}}
\tablehead{
\colhead{$\Delta t$} & \colhead{Observatory} & \colhead{Instrument/} &
\colhead{Filter} & \colhead{Frequency} & \colhead{Flux density} & 
\colhead{Uncertainty\tablenotemark{\dag}} & \colhead{Detection?} & \colhead{GCN}\\ 
\colhead{(d)} & & Telescope & & \colhead{(Hz)} & \colhead{(mJy)} & \colhead{(mJy)} & 
\colhead{1=Yes} }
\startdata
$1.16\times10^{-4}$ & Blagoveschensk & MASTER & \textit{CR} 
& $4.56\times10^{14}$ & $8.28$ & $2.76$ & 0 & GCN 15946 \\
$3.13\times10^{-4}$ & Tunka & MASTER & \textit{CR}
 & $4.56\times10^{14}$ & $1.02$ & $3.41\times10^{-1}$ & 0 & GCN 15946 \\
$1.03\times10^{-3}$ & Blagoveschensk & MASTER & \textit{CR} 
& $4.56\times10^{14}$ & $8.28$ & $2.76$ & 0 & GCN 15946 \\
$1.47\times10^{-3}$ & Tunka & MASTER & \textit{CR} 
& $4.56\times10^{14}$ & $1.02$ & $3.41\times10^{-1}$ & 0 & GCN 15946 \\
$3.76\times10^{-3}$ & Gingin & Zadko & \textit{R} 
& $4.56\times10^{14}$ & $3.61\times10^{-2}$ & $1.15\times10^{-2}$ & 1 & GCN 15952 \\
\ldots & \ldots & \ldots & \ldots & \ldots & \ldots & \ldots & \ldots & \ldots
\enddata
\tablecomments{\textit{CR} indicates clear filter calibrated to $R$-band. ${}^\dag$An uncertainty 
of 0.2~AB mag is assumed where not provided. All upper limits are $3\sigma$. The data have not been 
corrected for Galactic extinction. This is a sample of the full table available 
on-line.}\end{deluxetable*}

\subsection{Optical}
\label{text:data_analysis:optical}
The \Swift\ UV/Optical Telescope \citep[UVOT;][]{rkm+05}
observed GRB~140311A beginning 0.11\,d after the burst 
\citep{gcn15973}. We analyzed the UVOT data using HEASOFT (v. 6.19) and corresponding calibration 
files and list our derived upper limits in Table \ref{tab:data:UVOT}. 

We further analyze $i^{\prime}$- and $r^{\prime}$-band acquisition images of the field taken at 
Gemini-North and Gemini-South, respectively\footnote{\url{https://archive.gemini.edu}}. 
We downloaded the images and performed photometry in a 2.5\arcsec\ aperture calibrated to SDSS. We 
present the results and a compilation of all other optical observations reported in GCN 
circulars in Table \ref{tab:data:GCN}.

\begin{figure}
 \includegraphics[width=\columnwidth]{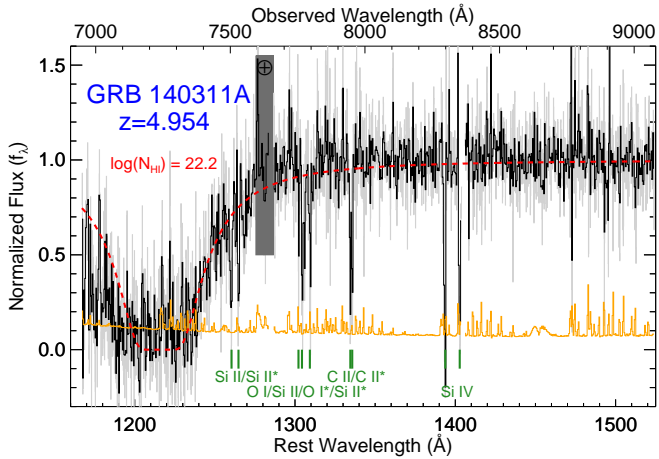}
\caption{Gemini spectrum of the afterglow of GRB 140311A.  The black spectrum has been binned for 
display purposes only, and the original spectrum is shown as gray in the background.  Gaps in the 
data reflect the GMOS-N CCD chip gaps and the region most adversely affected by telluric absorption 
is marked with the dark gray box.
The formal 1$\sigma$ uncertainty for the binned spectrum is shown in orange.
Absorption lines from the host galaxy are identified in green and a fit to a host DLA model is 
shown 
as the red dashed line.}
\label{specfig}
\end{figure}

\subsection{Optical Spectroscopy}
We obtained a single epoch of optical spectroscopy of the afterglow using the Gemini Multi-Object 
Spectrograph (GMOS; \citealt{hja+04}) on the 8~m Gemini-North telescope through proposal 
GS-2014A-Q-36 (PI: Berger).  A dithered pair of 900~s exposures were taken at a midpoint of 13:11 UT 
on 2014 March 12 (0.67\,d after the burst) using a 1$\arcsec$ slit, the R831 grating, and an OG515 
order-blocking filter.  Our setup covered the wavelength range 6945--9070\,\AA\ with a resolution 
of $\sim$2.9\,\AA\ (full width at half maximum).  We applied standard data analysis tasks using a 
combination of IRAF\footnote{IRAF is distributed by the National Optical Astronomy Observatories, 
which are operated by the Association of Universities for Research in Astronomy, Inc., under 
cooperative agreement with the National Science Foundation.} and custom IDL scripts to provide a 
flux calibration and telluric correction based on observations of archival standard stars. We 
normalized the continuum by fitting a power law at long wavelengths ($\lambda>8100$\,\AA), excluding 
strong absorption features. The final combined spectrum has a median signal-to-noise ratio per 
resolution element of $\sim$12 in the continuum (Figure~\ref{specfig}).

The spectrum exhibits a broad damped Ly$\alpha$ absorber (DLA) centered near 7200~\AA, as well 
as several narrow absorption lines redward of Ly$\alpha$. We fit individual Gaussian profiles to 
each absorption line except in cases of blends, where double Gaussian profiles are used. We report 
the line identifications and equivalent widths (EW) in Table~\ref{tab:lines}, along with inferred 
column densities using atomic data collected by \citet{pcdb07}.  Since most of the lines are 
saturated, our derived column densities represent only lower limits.  
We take a weighted average of the narrower lines and derive a mean redshift $z$=4.9540.  The 
presence of excited fine-structure transitions from \ion{C}{2}*, \ion{O}{1}*, and \ion{Si}{2}* in 
this system mark this as the redshift of the GRB. We fit a DLA model to the continuum in the range 
7200--8200\,\AA, excluding absorption lines and a region of strong telluric absorption. We fix the 
redshift of the absorber to match the narrow lines and find $\log(N_{\mathrm{H 
I}}/\mathrm{cm}^{-2})\approx22.2$ for the host DLA, consistent with the upper limit on $N_{\rm 
H,int}$ derived from the X-ray afterglow (Table \ref{tab:xrtspect}).
                                                                                            
\begin{deluxetable*}{cccccc}
\tabletypesize{\scriptsize}
\tablecaption{Absorption Lines in GRB 140311A Spectrum}
\tablehead{
\colhead{$\lambda_{\mathrm{obs}}$ (\AA)\tablenotemark{a}} &
\colhead{Line ID} &
\colhead{$\lambda_{\mathrm{rest}}$ (\AA)} &
\colhead{Redshift} &
\colhead{EW (\AA)\tablenotemark{b}} &
\colhead{log($N_{\mathrm{X}}$/cm$^{-2}$)\tablenotemark{c}}
}
\startdata
7503.55$\pm$0.48 &  SiII & 1260.42 & 4.9532$\pm$0.0004 & 0.85$\pm$0.21 & 13.78$\pm$0.11\\
7530.28$\pm$0.78 & SiII* & 1264.74 & 4.9540$\pm$0.0006 & 0.69$\pm$0.26 & 13.73$\pm$0.17\\
7753.71$\pm$0.29 &    OI & 1302.17 & 4.9545$\pm$0.0002 & 0.34$\pm$0.14 & 14.67$\pm$0.18\\
7765.95$\pm$0.28 &  SiII & 1304.37 & 4.9538$\pm$0.0002 & 0.35$\pm$0.08 & 14.39$\pm$0.10\\
7769.02$\pm$0.24 &   OI* & 1304.86 & 4.9539$\pm$0.0002 & 0.41$\pm$0.09 & 14.74$\pm$0.09\\
7775.69$\pm$0.31 & unidentified & \nodata & \nodata & 2.40$\pm$0.74\tablenotemark{d} & \nodata \\
7795.77$\pm$0.20 & SiII* & 1309.28 & 4.9543$\pm$0.0002 & 0.32$\pm$0.11 & 14.16$\pm$0.15\\
7945.71$\pm$0.25 &   CII & 1334.53 & 4.9539$\pm$0.0002 & 0.69$\pm$0.12 & 14.54$\pm$0.08\\
7953.23$\pm$0.28 &  CII* & 1335.71 & 4.9543$\pm$0.0002 & 0.57$\pm$0.12 & 14.50$\pm$0.09\\
8296.77$\pm$0.35 &  SiIV & 1393.76 & 4.9528$\pm$0.0003 & 1.46$\pm$0.16 & 14.21$\pm$0.05\\
8351.96$\pm$0.37 &  SiIV & 1402.77 & 4.9539$\pm$0.0003 & 1.41$\pm$0.15 & 14.49$\pm$0.05\\
\enddata
\tablenotetext{a}{Vacuum wavelengths}
\tablenotetext{b}{Rest frame}
\tablenotetext{c}{Lower limit due to optically-thin assumption}
\tablenotetext{d}{$EW$ for unidentified line is in observer frame}
\label{tab:lines}
\end{deluxetable*}

\subsection{Millimeter: CARMA}                                                   
\label{text:data_analysis:millimeter}
We observed GRB\,140311A with the Combined Array for Research in Millimeter Astronomy (CARMA) 
beginning on 2014 March 13.32 UT (1.66\,d after the burst; PI: Zauderer) in continuum wideband mode 
with 8 GHz bandwidth (16 windows, 487.5 MHz each) at a mean frequency of 85.5 GHz. Following an 
initial detection, we obtained two additional epochs. All observations utilized J1337-139 as phase 
calibrator, 3C279 as bandpass calibrator and Mars as flux calibrator. We derived a linelength 
calibration to account for thermal changes in the delays through the optical fibers connecting the 
CARMA antennas to the correlator using MIRIAD \citep{stw95}, and performed the  rest of the data 
analysis using the Common Astronomy Software Applications (CASA; \citealt{mws+07}). We summarize 
our mm-band observations in Table \ref{tab:data:radio}.

\subsection{Centimeter: VLA}\label{text:data_analysis:radio}
We observed the afterglow using the Karl G. Jansky Very Large Array (VLA) starting 2.48\,d after 
the 
burst through program 14A-344 (PI: Berger). We detected and tracked the flux density of the 
afterglow from 1.2\,GHz to 37\,GHz over six epochs until 82.3\,d after the burst, when it faded 
beyond detection at all frequencies. We used 3C286 as the flux and bandpass calibrator and 
J1354-0206 as gain calibrator. We carried out data reduction using CASA, and list the results of 
our VLA observations in Table \ref{tab:data:radio}.

\begin{deluxetable}{clcccc}
\tabletypesize{\scriptsize}
\tablecaption{GRB\,140311A: Log of radio observations \label{tab:data:radio}}
\tablehead{
\colhead{$\Delta t$} & \colhead{Facility} & \colhead{Frequency} & \colhead{Flux density} & 
\colhead{Uncertainty} & \colhead{Det.?}\\
\colhead{(d)} & & \colhead{(GHz)} & \colhead{($\mu$Jy)} & \colhead{($\mu$Jy)}
}
\startdata
1.66  &  CARMA      &  85.5  &  699   &  184    &  1\\
2.48  &  VLA        &  4.9   &  40.1  &  13.4   &  0\\
2.48  &  VLA        &  7.0   &  33.6  &  11.2   &  0\\
2.63  &  VLA        &  19.2  &  243.0 &  23.0   &  1\\
2.63  &  VLA        &  24.5  &  259.0 &  29.0   &  1\\
\ldots & \ldots & \ldots & \ldots & \ldots & \ldots
\enddata
\tablecomments{The last column indicates a detection (1) or non-detection (0). This is a sample of 
the full table available on-line.}
\end{deluxetable}
                                              
\section{Basic Considerations}
\label{text:basic_considerations}
We now interpret the X-ray, optical, and radio observations in the standard synchrotron framework 
\citep{spn98,gs02}, in which the observed spectra are characterized by power law segments 
connected at characteristic break frequencies: the self-absorption frequency (\nua), the 
characteristic synchrotron frequency (\numax), and the cooling frequency (\nuc). The electrons 
responsible for the observed radiation are assumed to form a power law distribution in energy with 
index $p$. The parameters of the shock creating the radiation are the total isotropic-equivalent 
kinetic energy ($\EKiso$), the circumburst density ($\dens$ in the case of a constant density 
environment, or $A_*$ in the case of a wind environment), the fraction of shock energy imparted to 
relativistic electrons (\epse) and the fraction imparted to magnetic fields (\epsb).

\begin{figure}
 \includegraphics[width=\columnwidth]{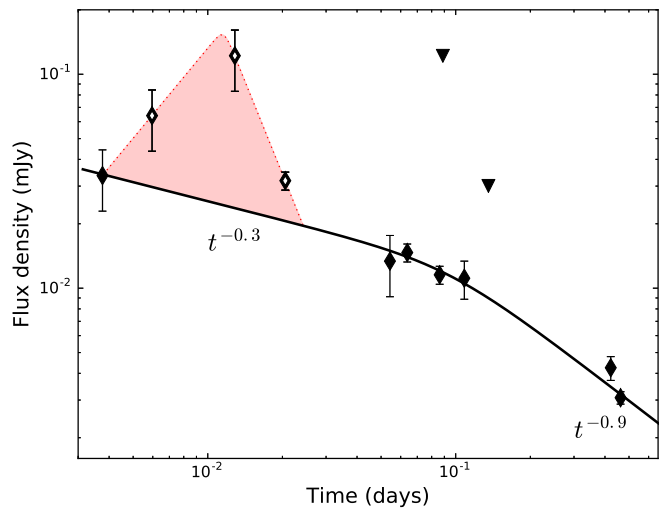}
 \caption{Optical $R$-band light curve of GRB~140311A (black points), together with a 
broken power law model (solid). The smoothness of the break has been fixed to $y=5.0$. The light 
curve exhibits a flare between $\approx3.8\times10^{-3}$ and $\approx2\times10^{-2}$\,d (red 
shaded region), which we exclude from our multi-wavelength modeling.}
\label{fig:optlc}
\end{figure}

\subsection{Optical and X-rays}  
\label{text:basic_optx}
The $R$-band light curve\footnote{We employ the convention, 
$F_{\nu} \propto t^{\alpha}\nu^{\beta}$ throughout.}  exhibits a rapid rise, with $\alpha_{\rm 
R,f1}=1.0\pm0.1$ from 
$3.8\times10^{-3}$ to $1.3\times10^{-2}$\,d followed by a steep decline
with $\alpha_{\rm R,f2} = -2.8\pm0.7$ to $2.0\times10^{-2}$\,d (Figure \ref{fig:optlc}). 
Optical flares with rapid rise and decline have previously 
been observed in GRB afterglows, and cannot be explained under the standard 
synchrotron framework \citep[e.g.][]{llt+12}. Such flares may be related to continued activity of 
the central engine \citep{gngc09,nggc10}, and we therefore do not include the flare 
in our analysis.

The underlying $R$-band light curve can be fit with a single power law from 
$3.8\times10^{-3}$\,d to $1.1\times10^{-1}$\,d, with $\alpha_{\rm R,1} = 0.34\pm0.06$; however, 
this fit over-predicts the RATIR $r^{\prime}$ band data at $\approx0.42$\,d by a factor of 
$\approx1.5$, suggesting a temporal break before $0.42$\,d. A broken power law fit yields a break 
time of $\approx0.1$\,d, and a post-break decay rate of $\alpha_{\rm R,2}\approx-0.9$. In the next 
section, we show that this break is consistent with the passage of \numax\ through the optical 
band. The shallow pre-break decline is only possible under one spectral ordering: 
$\nuc<\nuopt<\numax$, where the light curve declines as $t^{-1/4}$. The location of $\nuc<\nuopt$ is 
typically only expected at early times or when the circumburst density is large. We show below that 
a high-density environment is also demanded by the cm-band data, and is the likely cause for 
a sustained fast cooling evolution ($\nuc < \numax$). Since \nuc\ decreases with time in the ISM 
environment and increases in the wind environment, a low value of \nuc\ at later times is more 
naturally explained in the ISM case.

\begin{figure}
 \includegraphics[width=\columnwidth]{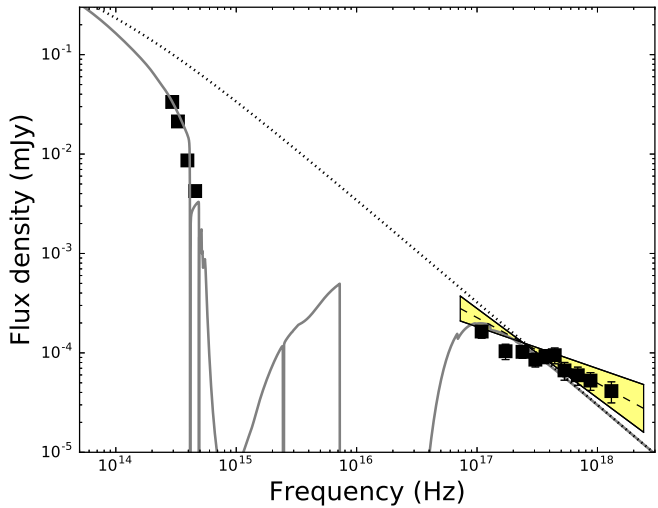}
 \caption{Optical to X-ray spectral energy distribution of the afterglow of GRB~140311A at 
0.42~d (black points) together with the best-fit ISM model (grey, solid). The dotted line is the 
best fit afterglow model, corrected for extinction and IGM absorption (Section \ref{text:model}). 
The optical data are from RATIR \citep{gcn15978}, while the X-ray data have been interpolated to 
0.42~d using a broken power law fit to the \Swift\ XRT light curve (Figure \ref{fig:XRT-lc}). The 
optical spectrum is steep ($\beta \approx-4$), partly due to IGM absorption, but also due to dust 
extinction in the host galaxy (Section \ref{text:basic_considerations}). 
The yellow shaded region indicates the 
$3\sigma$ error bar for the X-ray spectral index (Table \ref{tab:xrtspect}). }
\label{fig:betaoptx}
\end{figure}

The optical spectral index between the RATIR $Y$-band and $r^{\prime}$-band observations at 
0.42\,d is extremely steep, $\beta_{\rm opt} = -4.6\pm0.1$ (Figure \ref{fig:betaoptx}). 
The flux density at  $i^{\prime}$- and $r^{\prime}$-band is 
expected to be suppressed due to IGM absorption, given $z=4.954$. 
However, the spectral index between $Y$- and $z^{\prime}$-band is also steep ($\beta_{\rm Yz} = 
-4.0\pm1.0$), while the spectral index between $Y$-band and the X-rays is much 
shallower ($\beta_{\rm opt,X}=-0.83\pm0.04$). This suggests significant dust extinction  
along the line of sight through the host galaxy. 

\begin{figure}
 \includegraphics[width=\columnwidth]{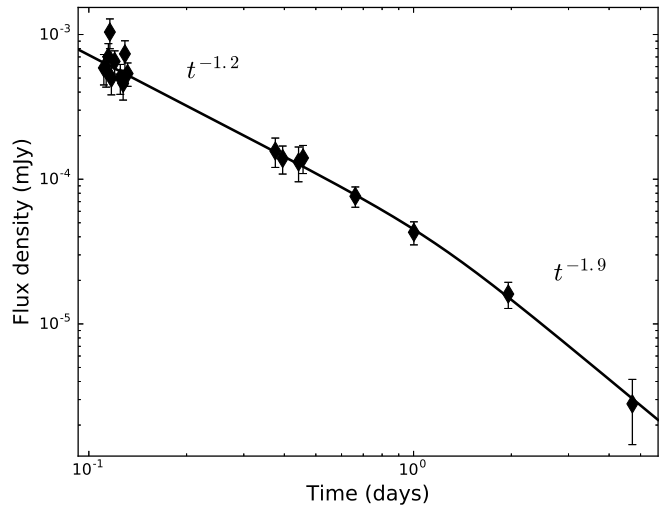}
 \caption{\Swift\ XRT light curve of GRB~140311A at 1\,keV (black points), together with a broken 
power law model (solid; Section \ref{text:basic_considerations}). The smoothness of the break has 
been fixed to $y=5.0$.}
\label{fig:XRT-lc}
\end{figure}

The X-ray light curve is well fit with a broken power law model, with an initial decline 
rate of $\alpha_{\rm X,1} = -1.16\pm 0.08$ steepening to $\alpha_{\rm X,2} = -1.9\pm0.4$ at $t_{\rm 
b} = 1.2\pm0.7$\,d (Figure \ref{fig:XRT-lc}). For $\nuc,\numax< \nux$, the pre-break decline rate 
indicates $p=2.2\pm0.1$ in both the ISM and wind environments. 
The shallow X-ray spectral index ($\beta_{\rm X} = -0.66\pm0.08$) is in tension with this 
interpretation, since the latter requires $\beta_{\rm X} \approx-1.1$. It is possible that 
Klein-Nishina suppression of inverse Compton cooling or the contribution of another component, such 
as inverse Compton radiation, causes the observed flattening of the slope \citep{se01,lem15}. We 
note that the late time decline rate of $\alpha_{\rm X,2}\approx-2$ is consistent with 
post-jet break evolution with $p\approx2$, since $\alpha\approx-p$ at $t>\tjet$. 
In the fast cooling scenario, the X-ray spectral index and light curve 
both  depend on the location of $\numax$ relative to the X-ray band. Unfortunately, no X-ray 
observations are available before the break in the optical light curve at $\lesssim 0.42$\,d. 
However, $\numax$ can be constrained through radio observations, and we return to this point in 
Section \ref{text:basic_radio}.

To summarize, the optical light curve exhibits a shallow decay until $\approx 0.1$\,d, suggesting 
the afterglow SED is in the fast cooling regime with $\nuc<\nuopt<\numax$ until $\approx0.1$\,d. 
The X-ray light curve exhibits a steepening at $\approx 1$\,d indicative of a jet break. The 
post-break decline indicates that $p\approx2$, consistent with the pre-break X-ray light curve for 
$\nuc,\numax<\nuopt,\nux$.

\begin{deluxetable}{ccc}
 \tabletypesize{\footnotesize}
 \tablecolumns{3}
 \tablecaption{Radio spectral fits for GRB 140311A\label{tab:radio_bplfits}}
 \tablehead{   
    \colhead{$\Delta T$} &
    \colhead{$\nu_{\rm{break}}$ (GHz)} &
    \colhead{$F_{\rm{break}}$ ($\mu$Jy)} \\
    }
 \startdata 
 2.5 & $16.7\pm2.8$ & $208\pm27$ \\
 4.5 & $9.6\pm0.7$  & $206\pm11$ \\
 9.5 & $9.9\pm1.1$  & $122\pm10$ \\
 18.5 & $5.7\pm1.4$  & $54\pm7$ 
 \enddata
\end{deluxetable}

\begin{figure*}
\begin{tabular}{ccc}
 \centering
 \includegraphics[width=0.31\textwidth]{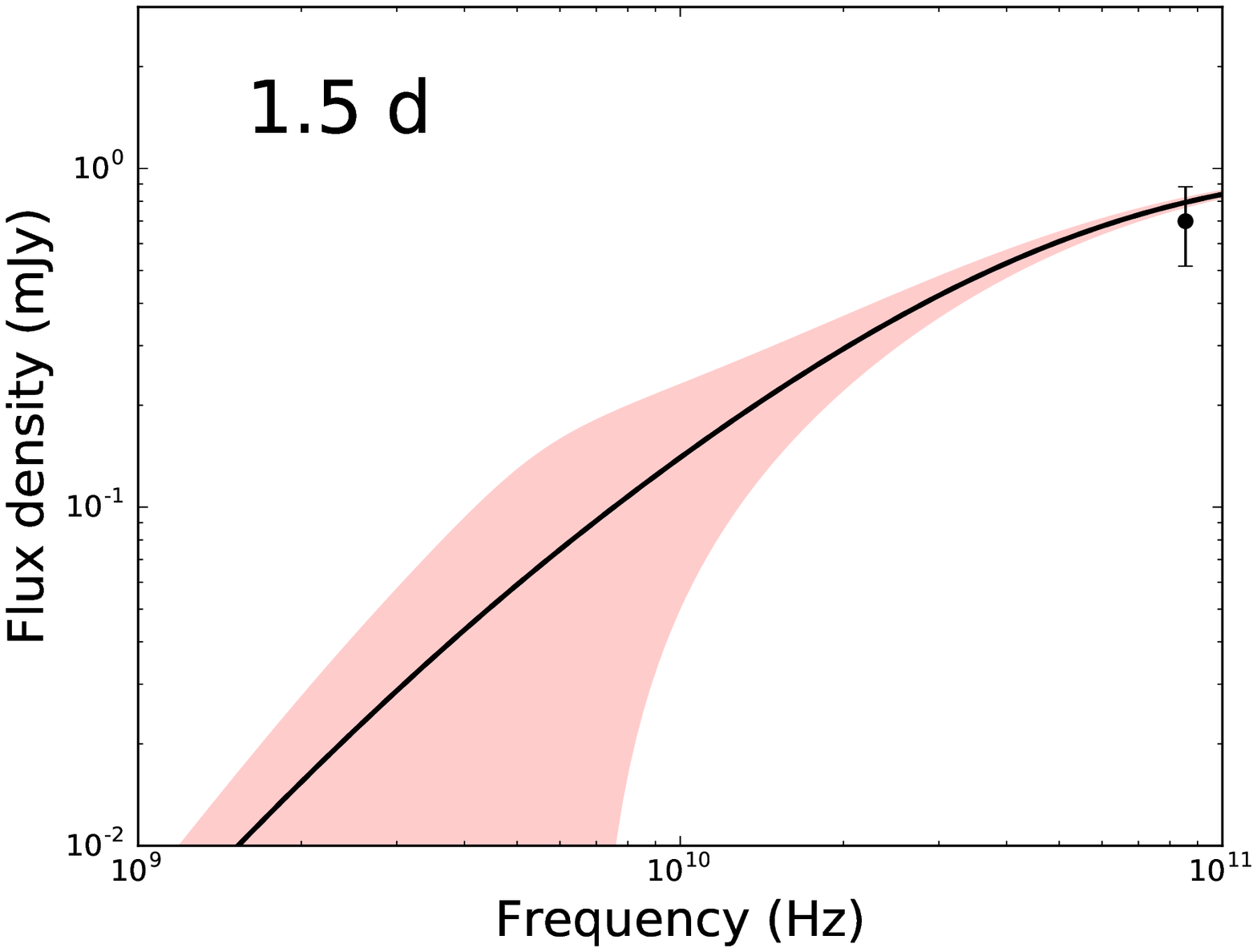} &
 \includegraphics[width=0.31\textwidth]{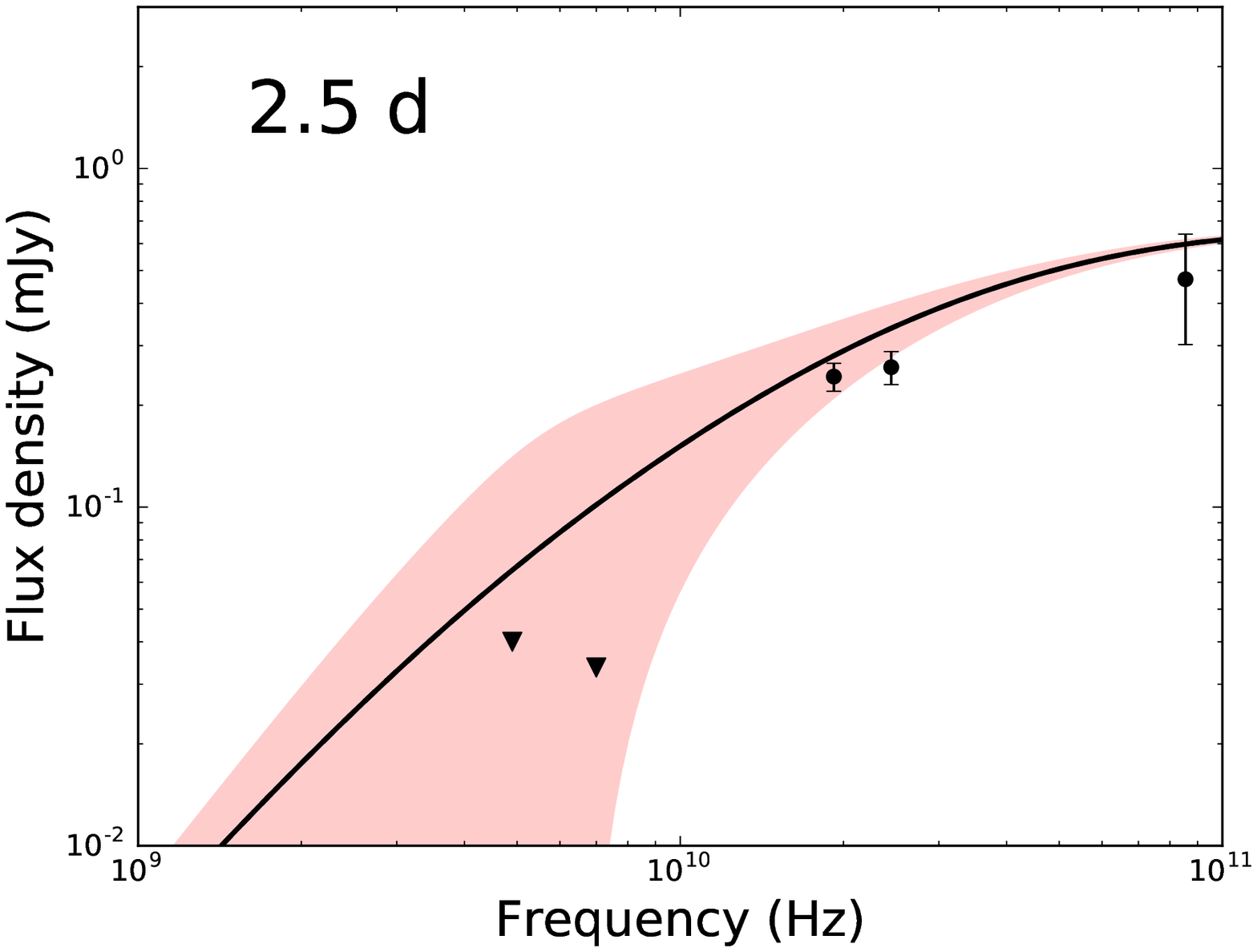} &
 \includegraphics[width=0.31\textwidth]{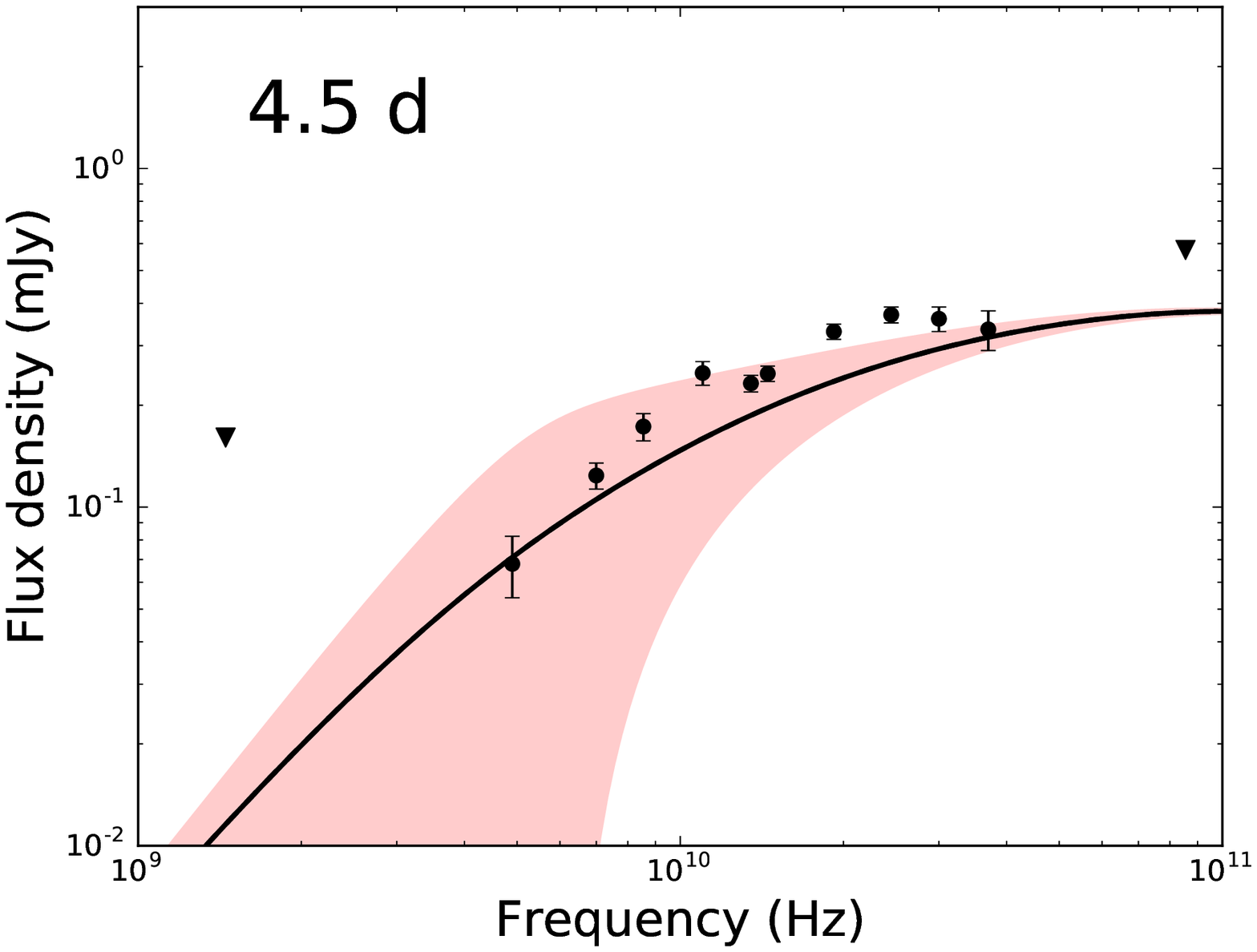} \\
 \includegraphics[width=0.31\textwidth]{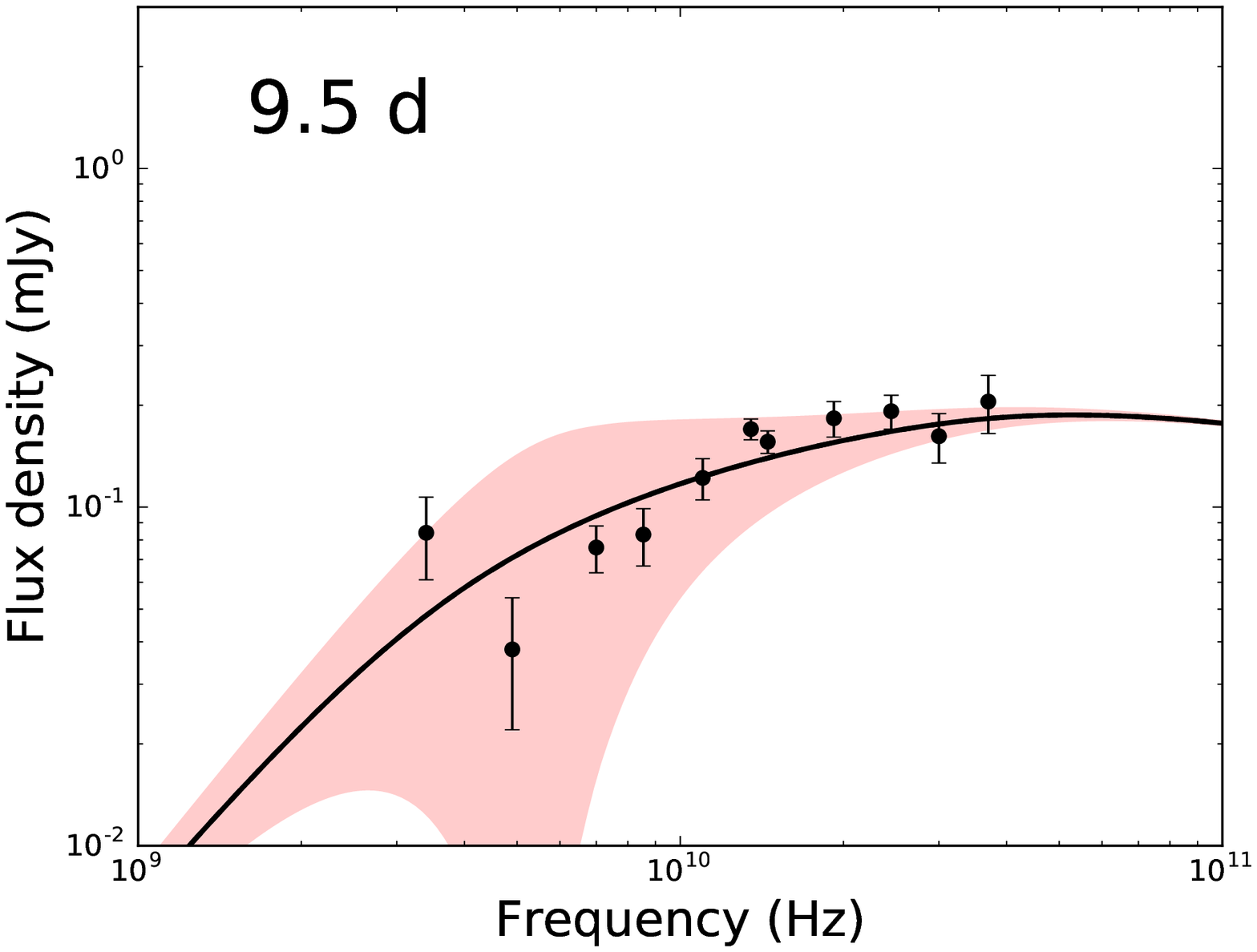} &
 \includegraphics[width=0.31\textwidth]{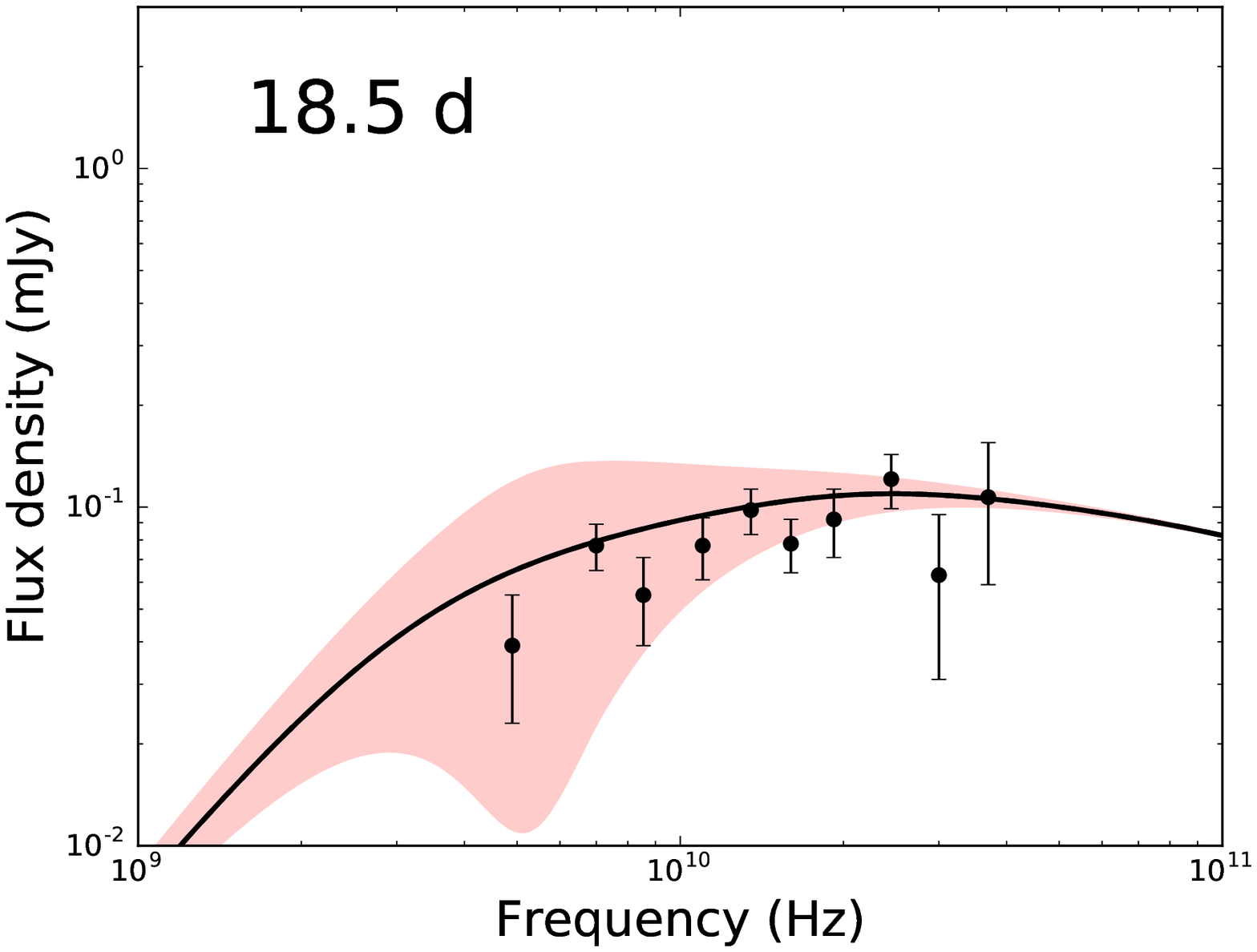} &
 \includegraphics[width=0.31\textwidth]{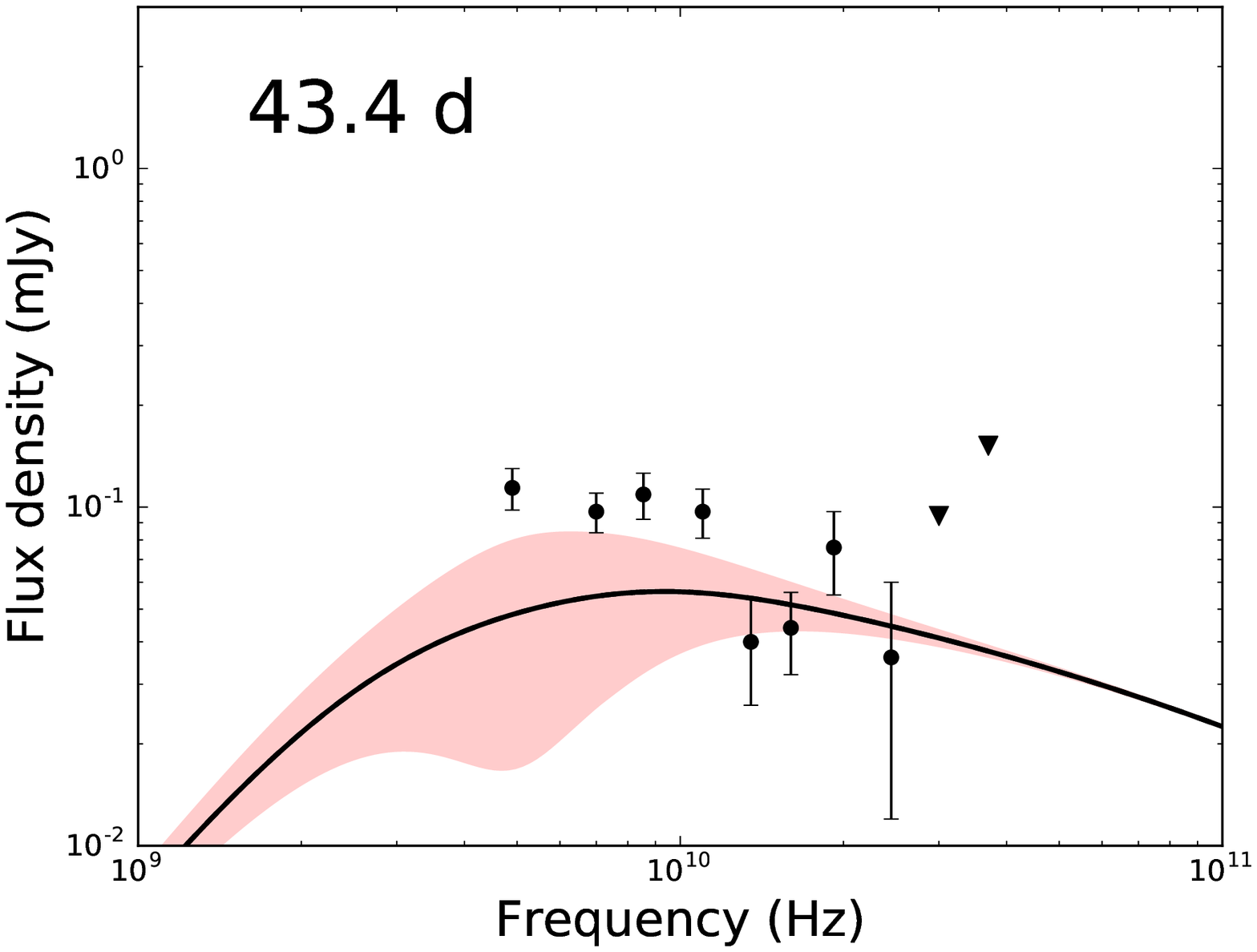} \\
 \includegraphics[width=0.31\textwidth]{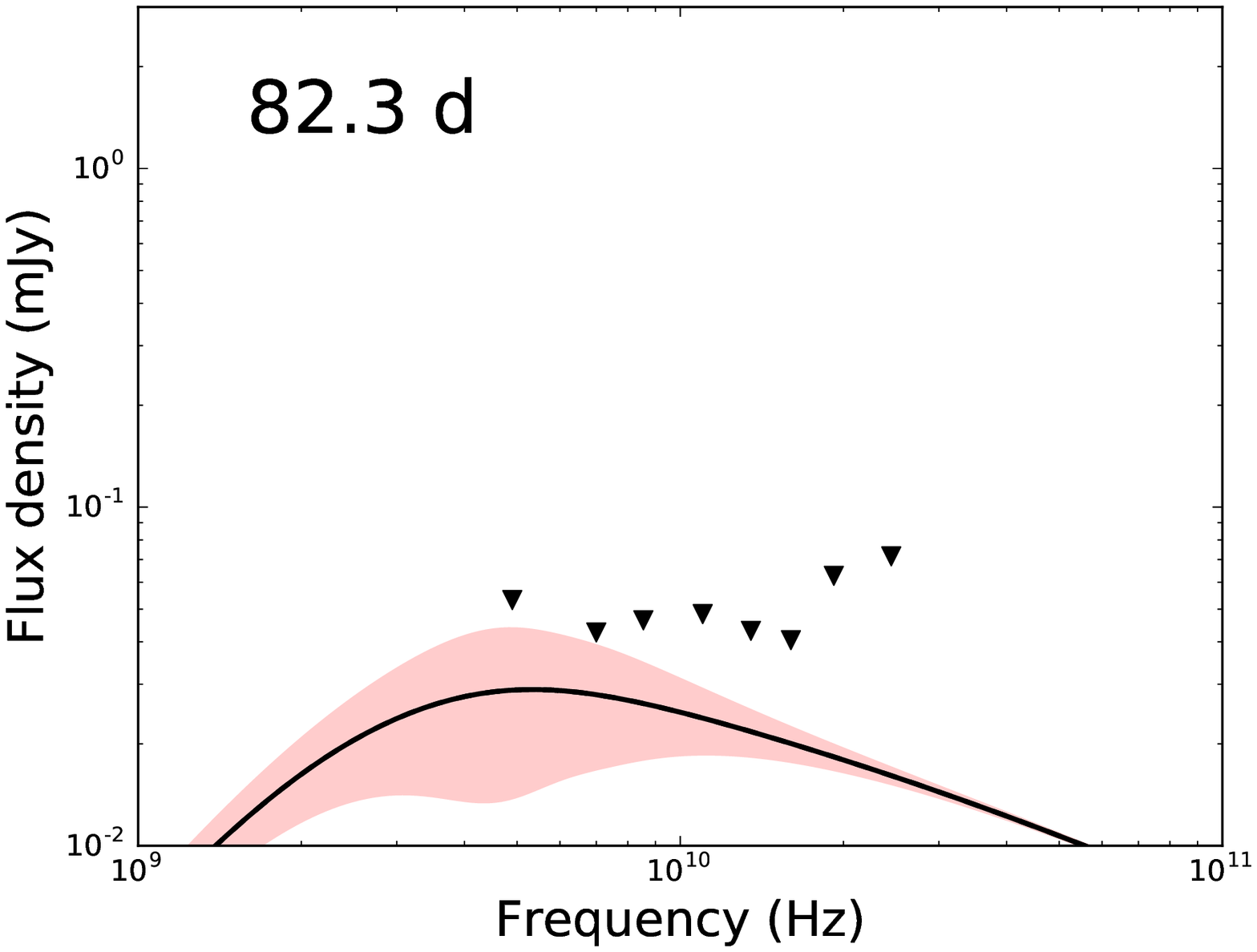} & 
\end{tabular}
\caption{Radio spectral energy distributions of the afterglow of GRB~140311A at multiple epochs 
starting at 1.5~d, together with the same ISM model in Figure \ref{fig:betaoptx}. The red 
shaded regions represent the expected variability due to interstellar scintillation.
}
\label{fig:modelsed_ISM1}
\end{figure*}

\subsection{Radio}
\label{text:basic_radio}
Synchrotron self-absorption is expected to result in a steep spectral index ($\beta=11/8$ to $5/2$) 
at low frequencies. A measurement of the self-absorption frequency yields a strong constraint on the 
circumburst density. In our first joint VLA and CARMA observation of the afterglow at 
$\approx2.5$\,d, we find a steep slope, $\beta_1=2.0\pm0.3$, from 7 GHz to 24.5\,GHz, flattening to 
$\beta_2=0.4\pm0.1$ at $24.5$--$85.5$\,GHz (Figure \ref{fig:modelsed_ISM1}).
The radio SED at 4.5\,d also exhibits a clear transition from a self-absorbed to an optically thin 
slope at $\approx10$\,GHz. To track the evolution of this break frequency, which we identify as 
\nua, we fit the radio SEDs between 2.5 and 18.5\,d with a broken power law model described by 
\begin{equation}
\label{eqn:bpl}
F_{\nu} = F_{\rm b}
\left(\frac{(\nu/\nu_{\rm b})^{-y\beta_1} + (\nu/\nu_{\rm b})^{-y\beta_2}}{2}\right)^{-1/y},
\end{equation}
with $\beta_1=2$, $\beta_2=1/3$, and smoothness, 
$y=5$\footnote{For the purposes of the fit at 2.5\,d, we treat the 4.9 and 7.0\,GHz upper limits
as $3\sigma$ detections, with flux densities equal to three times the map RMS.}.
We present the results in Table \ref{tab:radio_bplfits}. 
Fitting the evolution of the break frequency and break flux density as  power laws 
with time, we find
$\alpha_{\nu} = -0.3\pm0.1$ and $\alpha_{\rm F} = -0.59\pm0.05$, which is more
consistent with post-jet break evolution ($\alpha_{\nu} = -0.2$, $\alpha_{\rm F} = -0.4$) than
a spherical outflow in an ISM ($\alpha_{\nu} = 0$, $\alpha_{\rm F} = 0.5$) or wind ($\alpha_{\nu} 
= -0.6$, $\alpha_{\rm F} = -0.2$) environment. 

The observed value of \nua\ and $F_{\nu,\rm a}$ at 4.5\,d results in a moderately high density. To 
show this, we define\footnote{This expression is relevant for fast cooling and an ISM environment.}
the quantity
\begin{equation}
    \xi_a \equiv \nu_{\rm a, GHz}^2/F_{\nu,\rm a, mJy} = 46.8\epsb\dens d_{\rm L,28}^2 t_d^{-1},
\end{equation}
where $d_{\rm L,28}=14.5$ is the luminosity distance divided by $10^{28}$\,cm, and $t_d$ is the 
observer time in days. Computing $\xi_a\approx4.5\times10^2$ at 4.5\,d where the radio SED is 
particularly well constrained, we obtain $n\approx 2\epsilon_{\rm B,0.1}^{-1}$\,cm$^{-3}$, 
comparable to the mean density of the Milky Way ISM.

The cm-band SED at 43.4\,d is inverted, with $\beta = -0.7\pm0.2$. This spectral inversion between 
18.5 and 43.4\,d is only possible if $\numax$ crossed the radio band between these two epochs. 
Taking $\numax\approx5$\,GHz at 43.4\,d and a jet break time of $\approx 1$\,d from the X-ray light 
curve (Section \ref{text:basic_optx}), we find $\numax$ passes through $r^{\prime}$-band at 
$\approx0.1$\,d. We therefore confirm the steepening in the $r^{\prime}$-band light curve between 
$\approx0.1$ and $0.42$\,d as arising from the passage of $\numax$ through the optical band.

To summarize, the radio SEDs allow us to locate both \nua\ and \numax. The inferred location of 
\numax\ is consistent with the optical and X-ray light curves, while the observed value of \nua\ is 
consistent with post-jet break evolution for the duration of the radio observations. We focus 
in the rest of the paper on the ISM model, and present a wind model in Appendix \ref{appendix:wind}.

\begin{deluxetable}{lcc}
 \tabletypesize{\footnotesize}
 \tablecolumns{2}
 \tablecaption{Parameters for best-fit ISM model}
 \tablehead{   
           \colhead{Parameter} &
           \colhead{Best-fit}  &
           \colhead{MCMC}
   }
 \startdata    
   $p$                  & 2.06 & $2.08^{+0.02}_{-0.01}$  \\[2pt]
   \epse                & 0.63 & $0.60\pm0.10$           \\[2pt]
   \epsb                & 0.34 & $0.22^{+0.23}_{-0.14}$  \\[2pt]
   \dens                & 8.09 & $11.1^{+9.1}_{-3.7}$    \\[2pt]
   $E_{\rm K, iso, 52}$ & 8.46 & $8.7^{+2.5}_{-1.5}$     \\[2pt]
   \tjet\ (d)           & 0.56 & $0.57\pm0.05$           \\[2pt]
   \thetajet\ (deg)     & 3.92 & $4.1\pm0.3$             \\[2pt]
   \AV\ (mag)           & 0.34 & $0.34\pm0.02$           \\[2pt]
   \nuac\  (Hz)         & $8.9\times10^{8}{}^{\dag}$ & \ldots \\[2pt]
   \nusa\  (Hz)         & $4.1\times10^{11}$    & \ldots \\[2pt]
   \nuc\   (Hz)         & $4.3\times10^{11}$    & \ldots \\[2pt]
   \numax\ (Hz)         & $9.0\times10^{14}$    & \ldots \\[2pt]
   $F_{\nu, \rm peak}$ (mJy) & 11.3             & \ldots \\[2pt]
   \hline\\[-4pt]
   $E_{\gamma}$ (erg)   & \multicolumn{2}{c}{$(6.9^{+2.6}_{-2.4})\times 10^{50}$} \\[2pt]
   $E_{\rm K}$ (erg)    & \multicolumn{2}{c}{$(2.2^{+0.4}_{-0.3})\times10^{50}$}  \\[2pt]
   $E_{\rm tot}$ (erg)  & \multicolumn{2}{c}{$\approx9\times10^{50}$}  \\[2pt]
   $\eta_{\rm rad}$     & \multicolumn{2}{c}{$\approx76\%$}
 \enddata
 \tablecomments{All break frequencies are listed at 0.1\,d. 
 ${}^\dag$This break frequency is not directly constrained by the data.}
\label{tab:params}
\end{deluxetable}

\begin{figure*} 
 \begin{tabular}{cc}
  \includegraphics[width=0.47\textwidth]{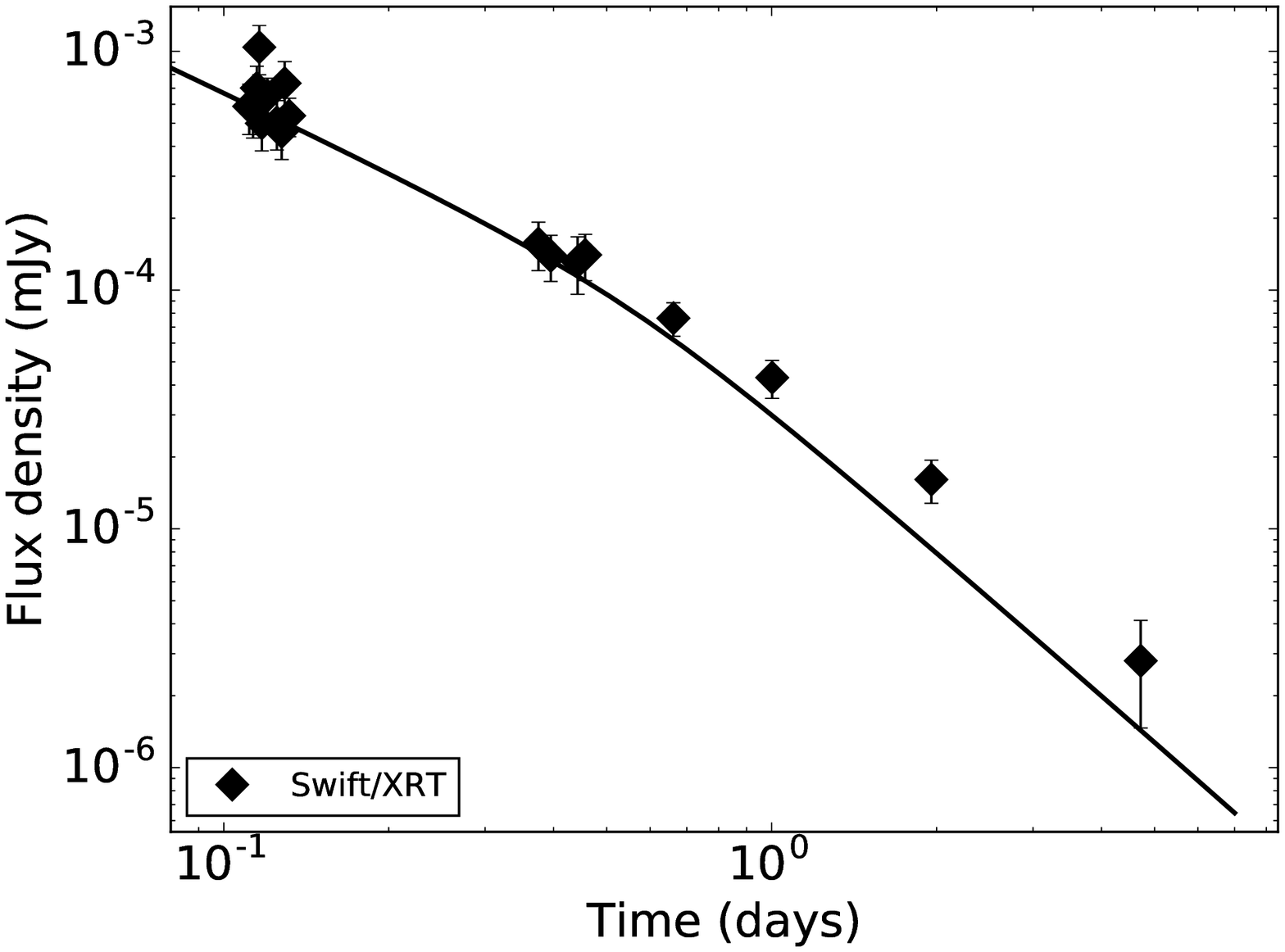} &
  \includegraphics[width=0.47\textwidth]{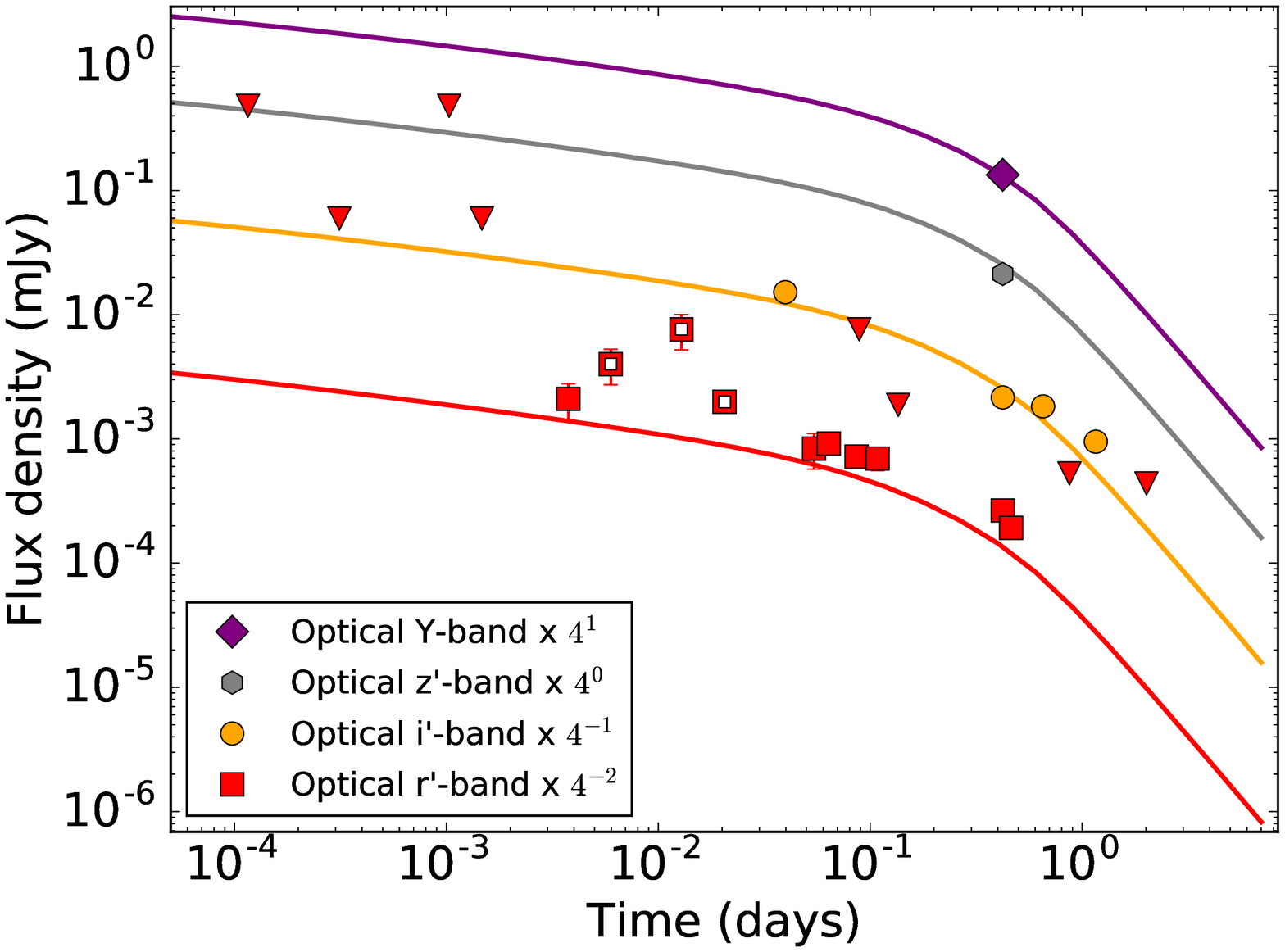} \\
  \includegraphics[width=0.47\textwidth]{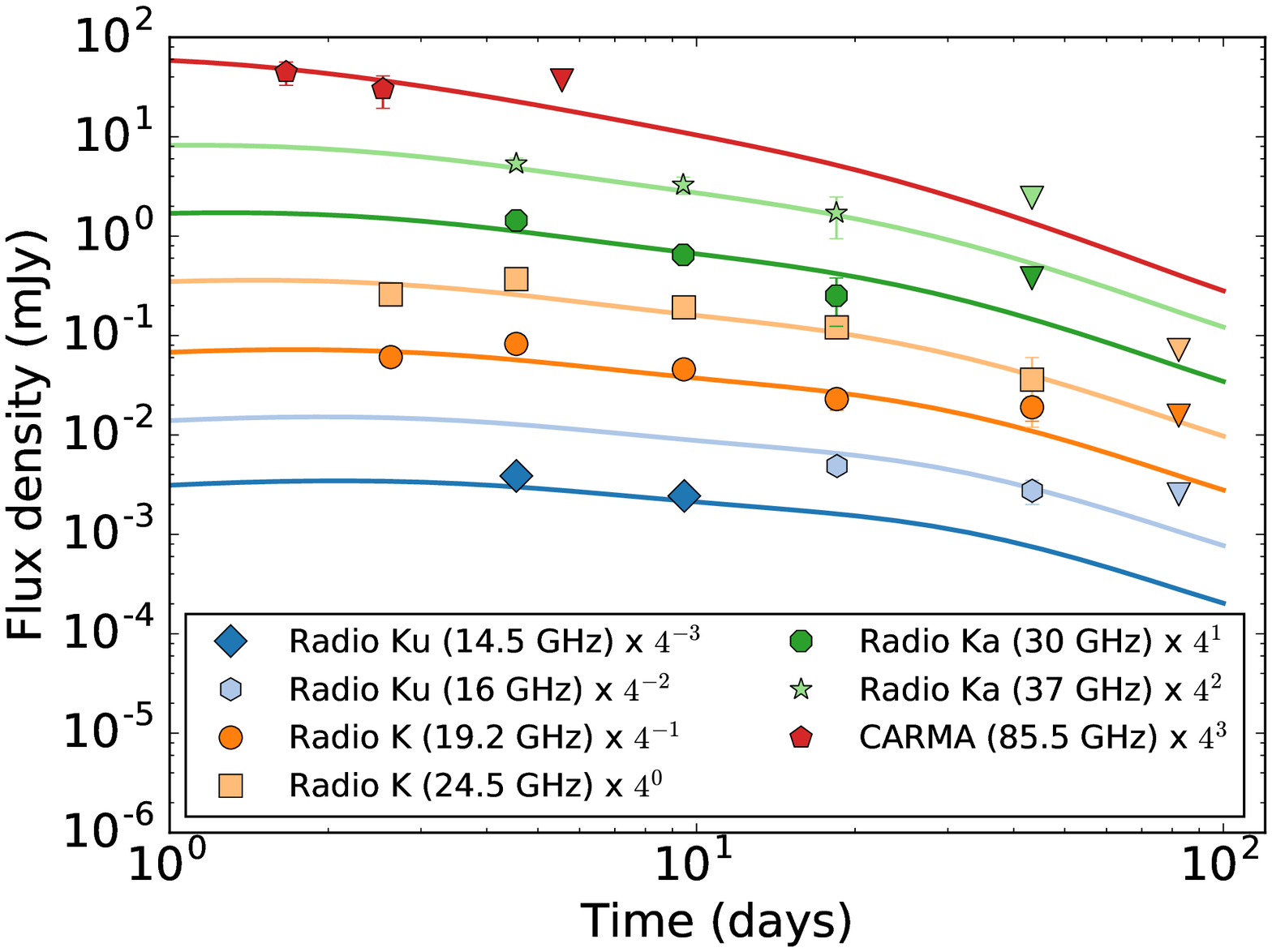} &
  \includegraphics[width=0.47\textwidth]{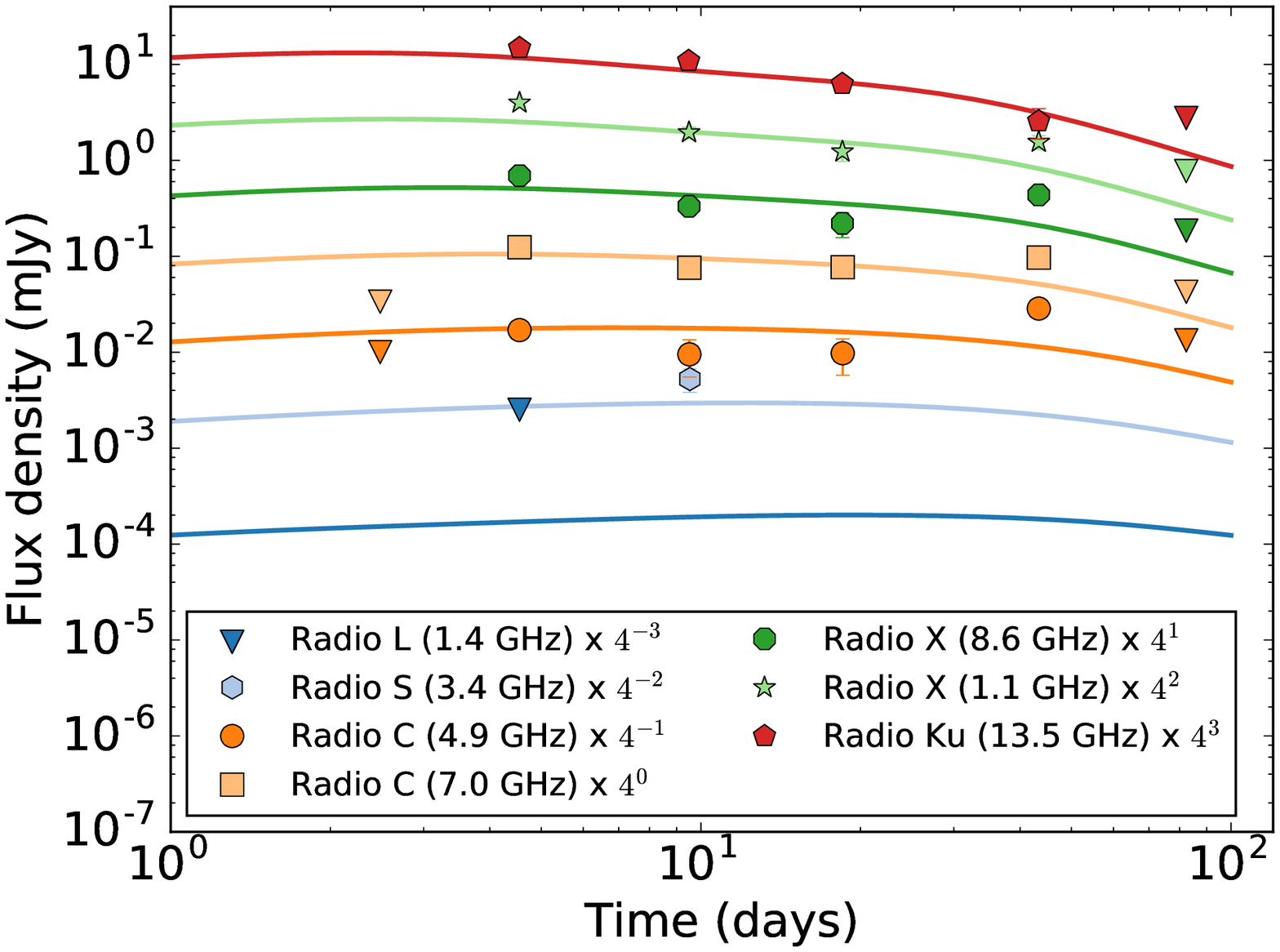} \\
 \end{tabular}
 \caption{X-ray (top left), optical (top right) and radio (bottom) light curves of the 
afterglow of GRB 140311A, together with an ISM model (Section \ref{text:model}). The data points 
with open symbols are not included in the fit.}
\label{fig:modellc_ISM}
\end{figure*}

\section{Multi-wavelength modeling}
\label{text:model}
Following the considerations outlined in Section \ref{text:basic_considerations}, we now perform a 
Markov Chain Monte Carlo analysis to determine the physical parameters of the afterglow. We fit 
all available photometry (with the exception of the optical flare at 
$1.3\times10^{-2}$--$2.0\times10^{-2}$\,d) with a forward shock model using the prescription of 
\cite{gs02} with $p$, \epse, \epsb, \dens, \EKiso, \tjet\, and the extinction in the host galaxy 
(\AV) as free parameters. The details of our modeling procedure are described in \lbt\ and 
\cite{lbm+15}.

\begin{figure*}
\begin{tabular}{ccc}
 \centering
 \includegraphics[width=0.31\textwidth]{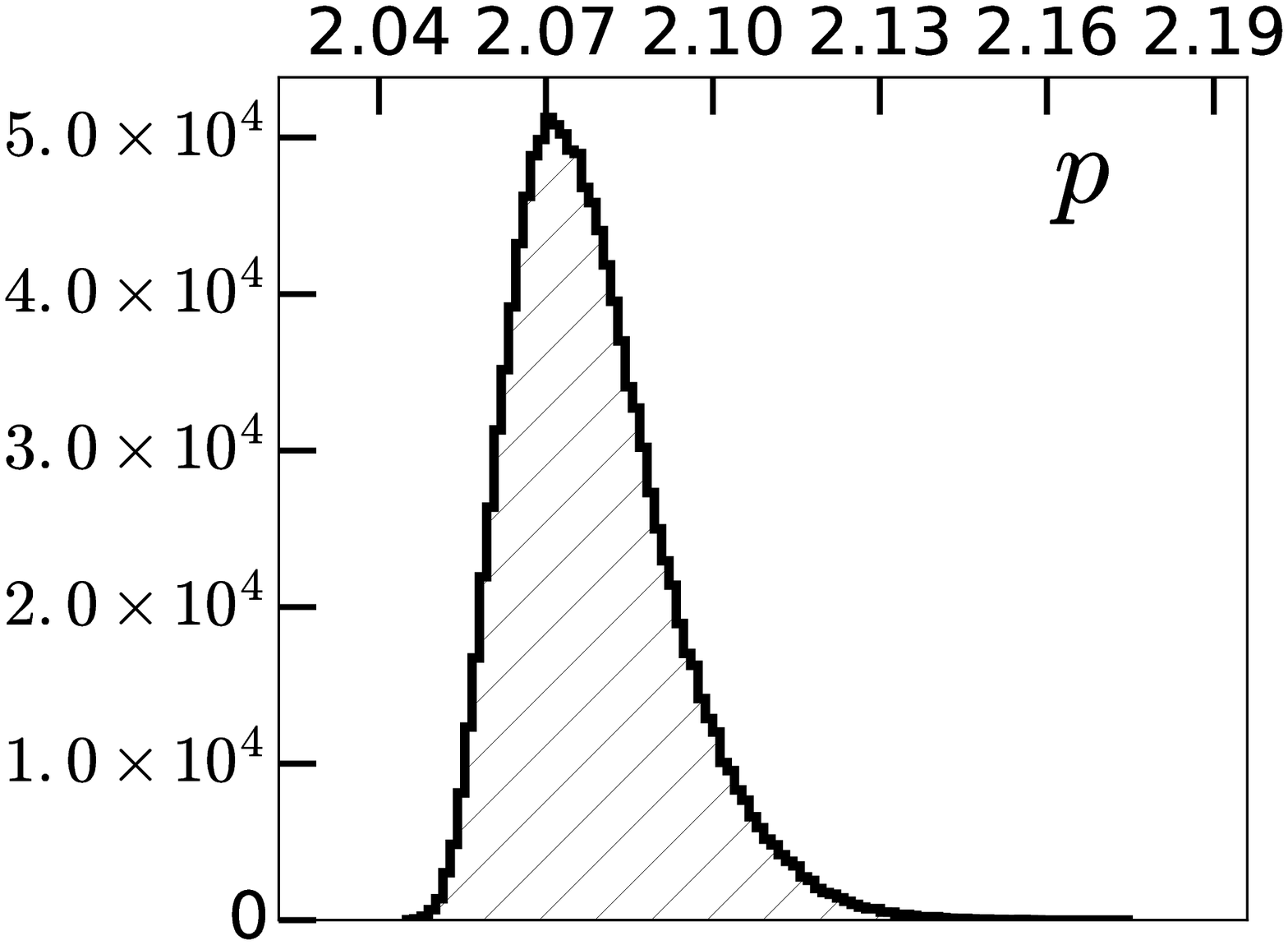} &
 \includegraphics[width=0.31\textwidth]{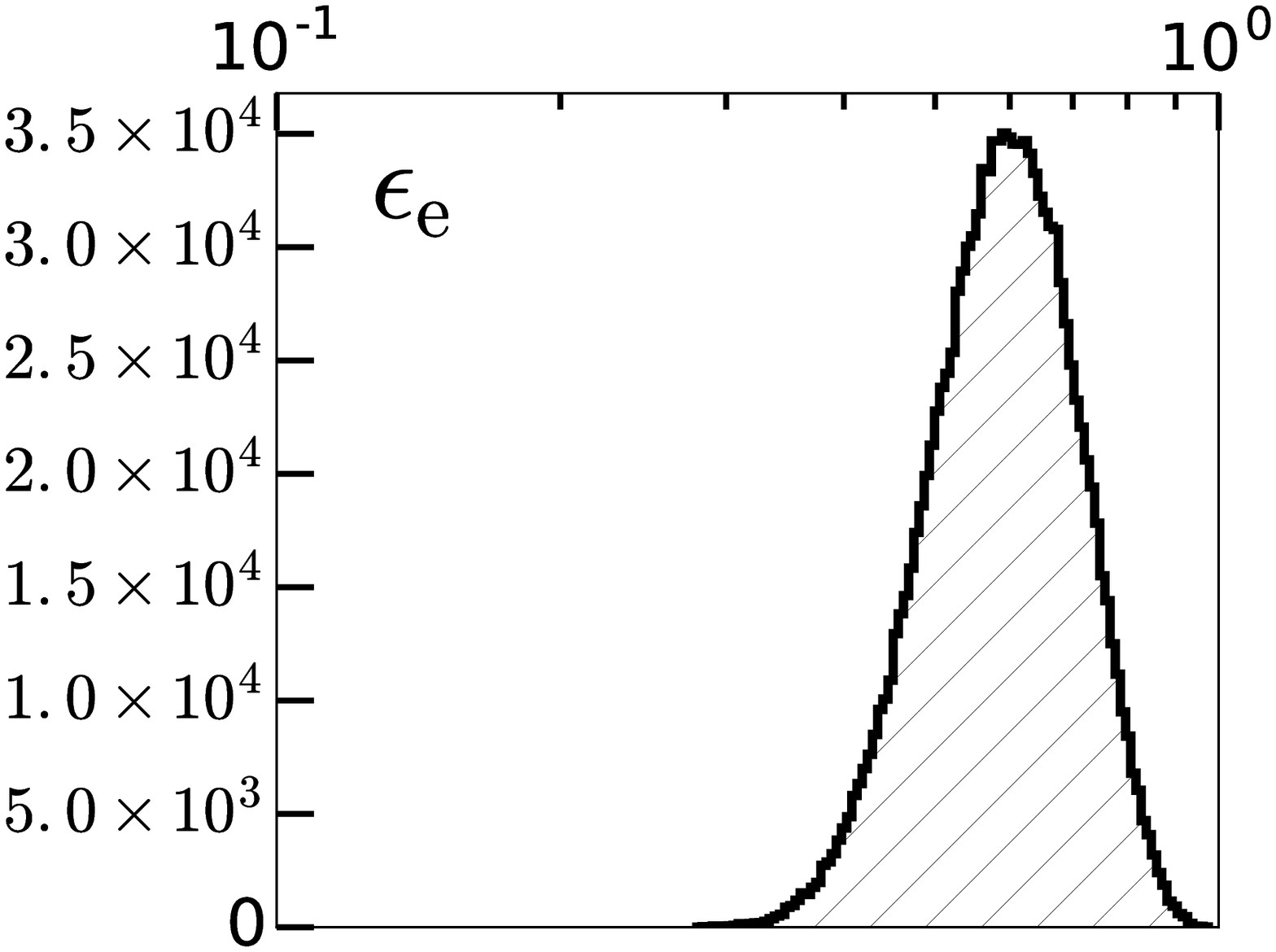} &
 \includegraphics[width=0.31\textwidth]{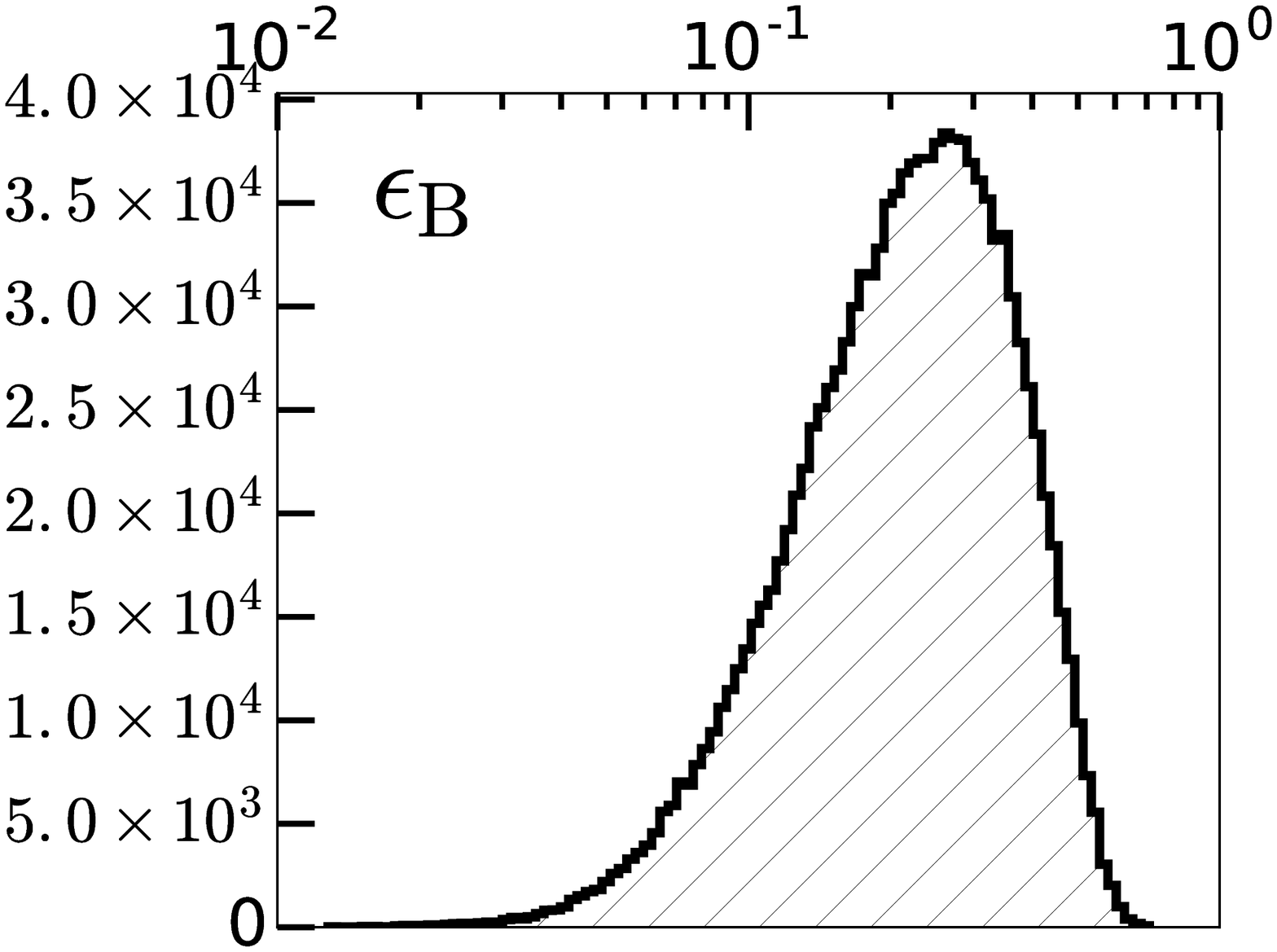} \\
 \includegraphics[width=0.31\textwidth]{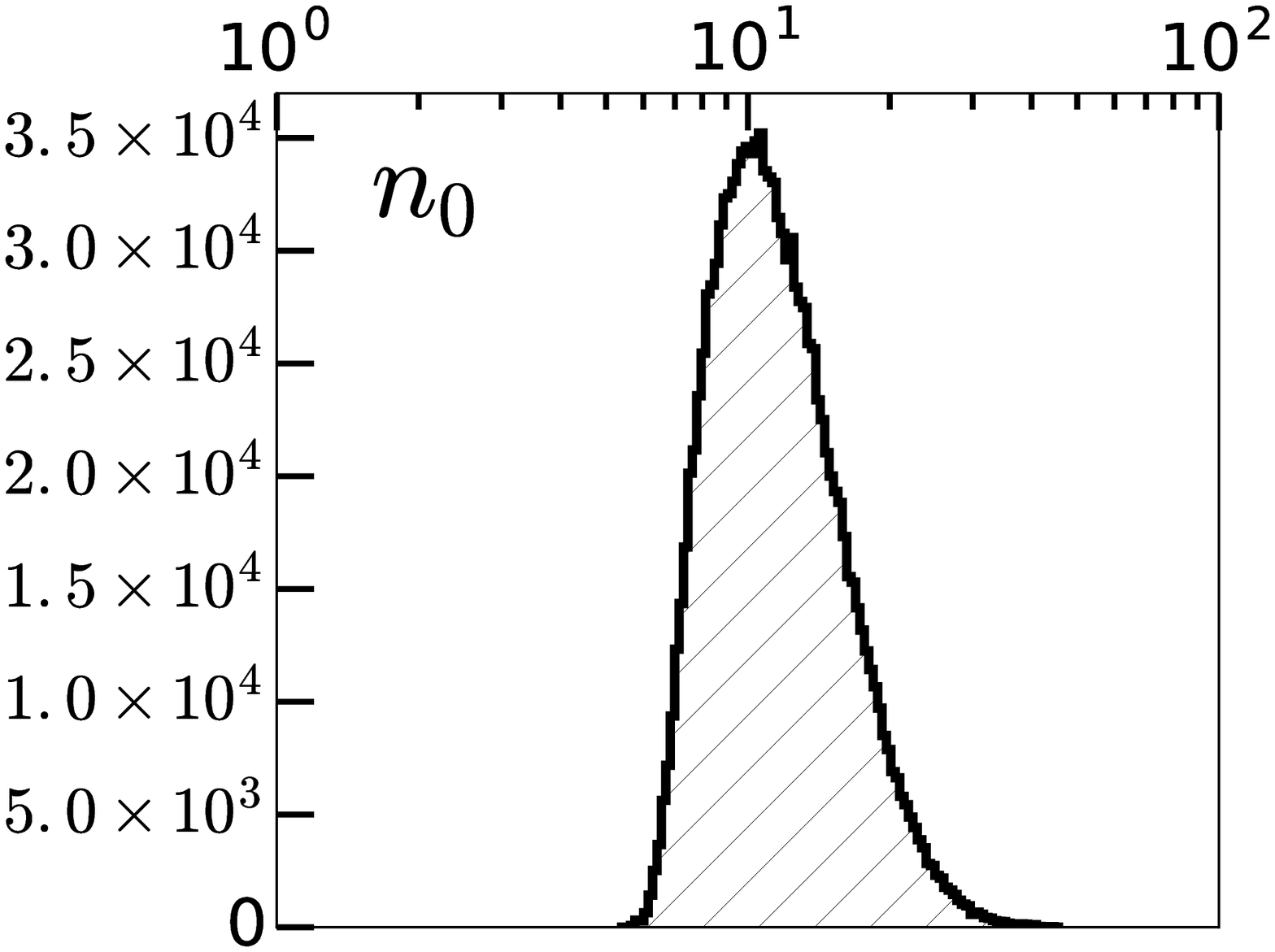} &
 \includegraphics[width=0.31\textwidth]{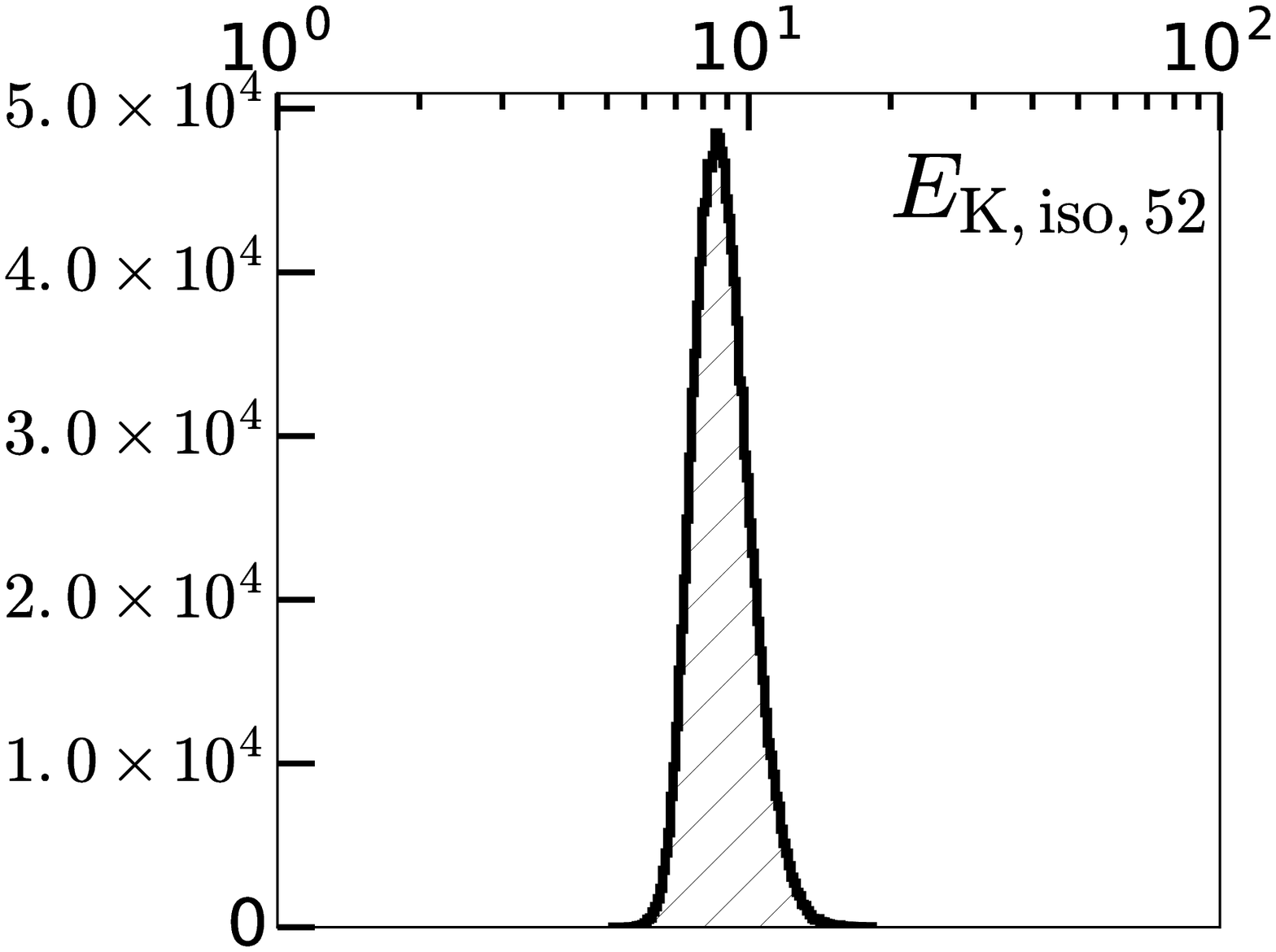} &
 \includegraphics[width=0.31\textwidth]{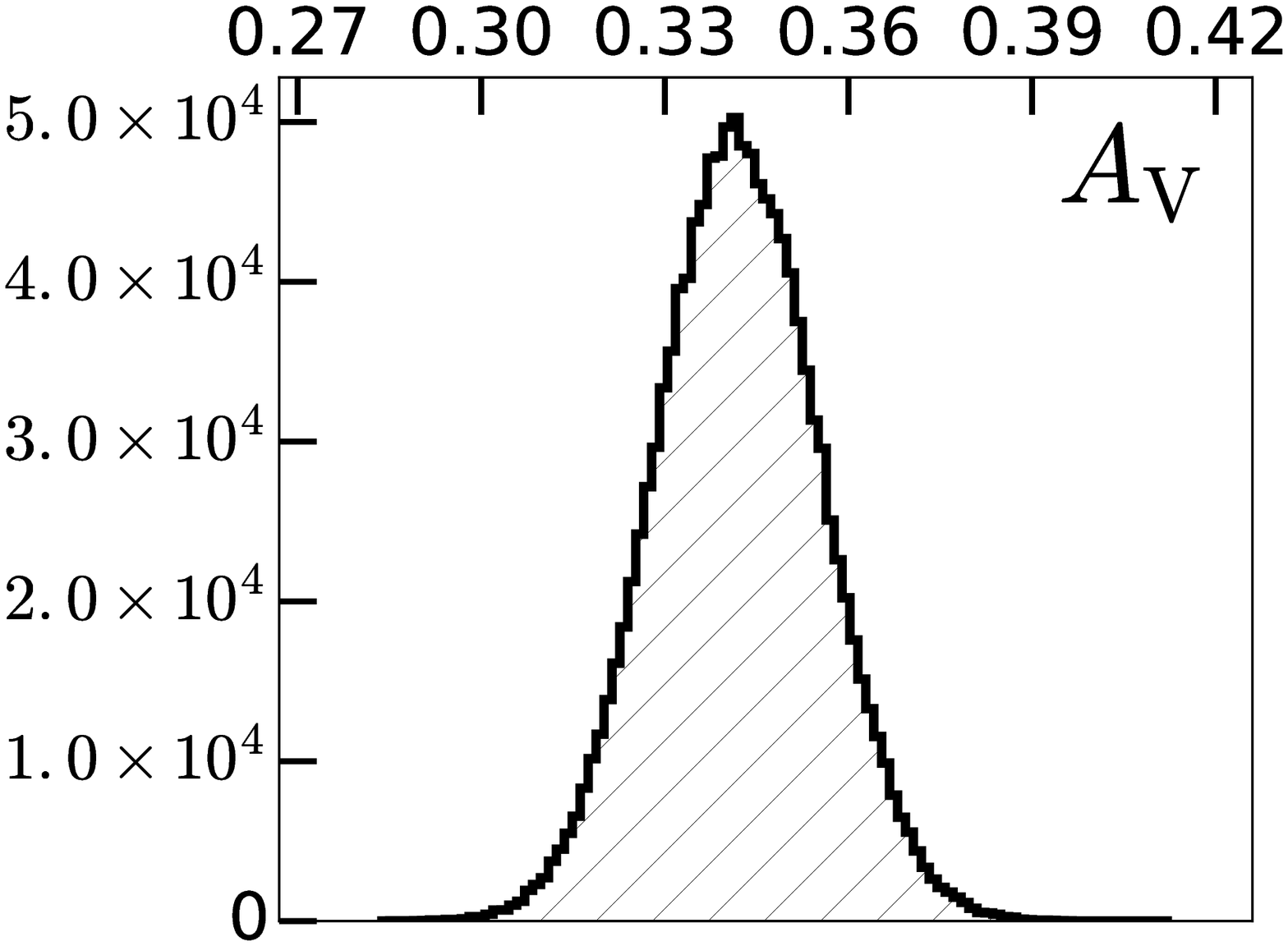} \\
 \includegraphics[width=0.31\textwidth]{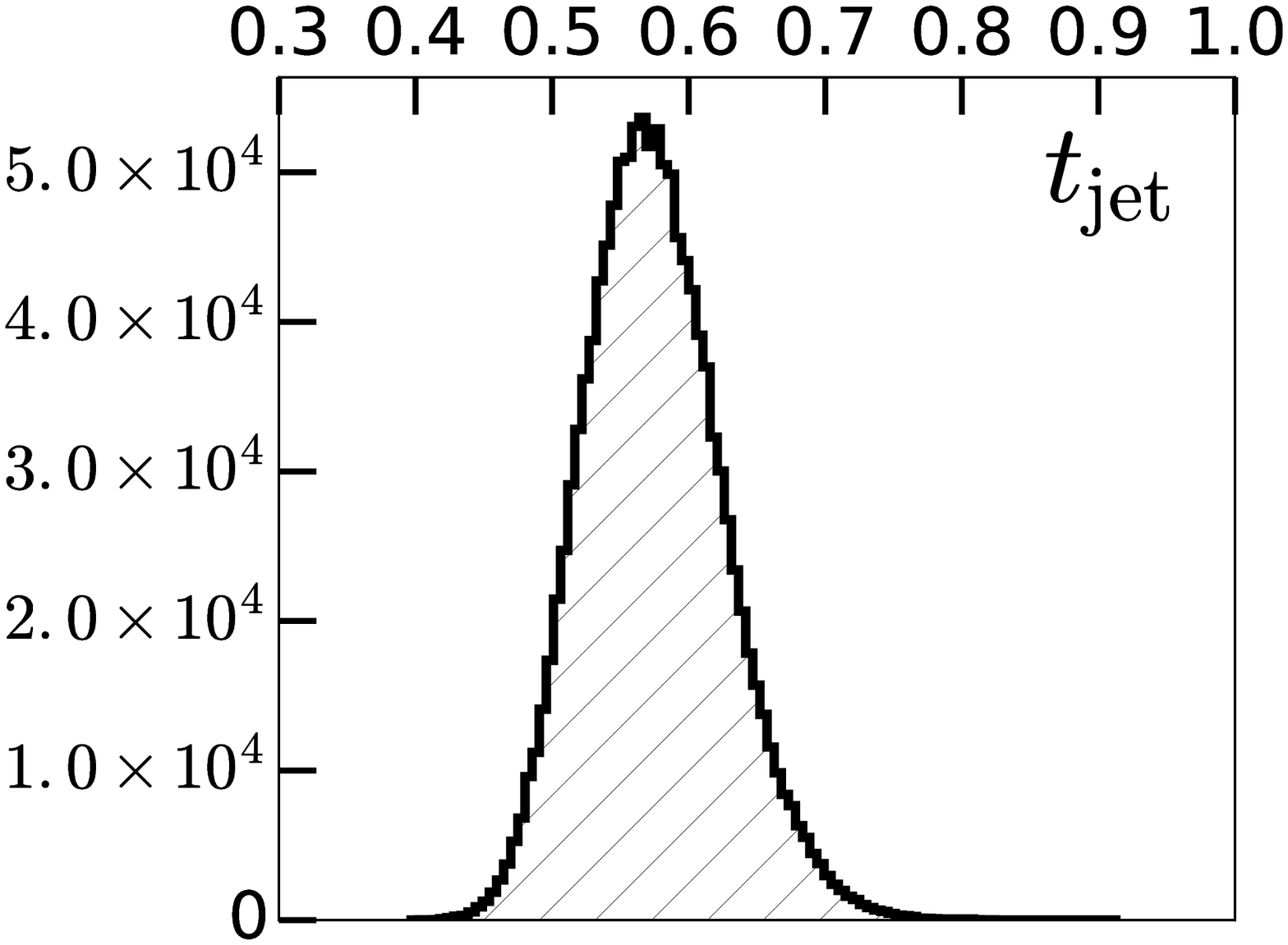} &
 \includegraphics[width=0.31\textwidth]{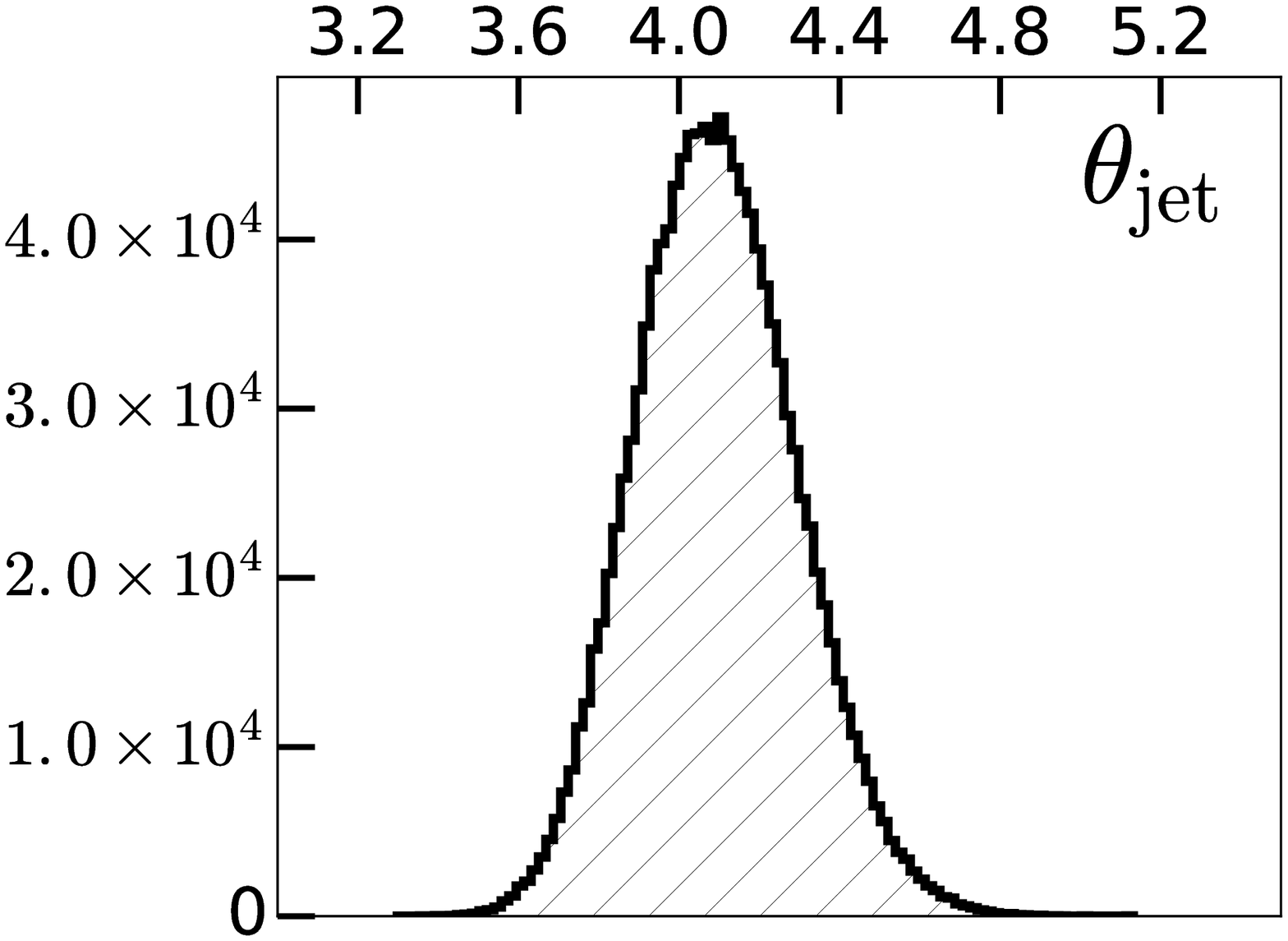} & 
 \includegraphics[width=0.31\textwidth]{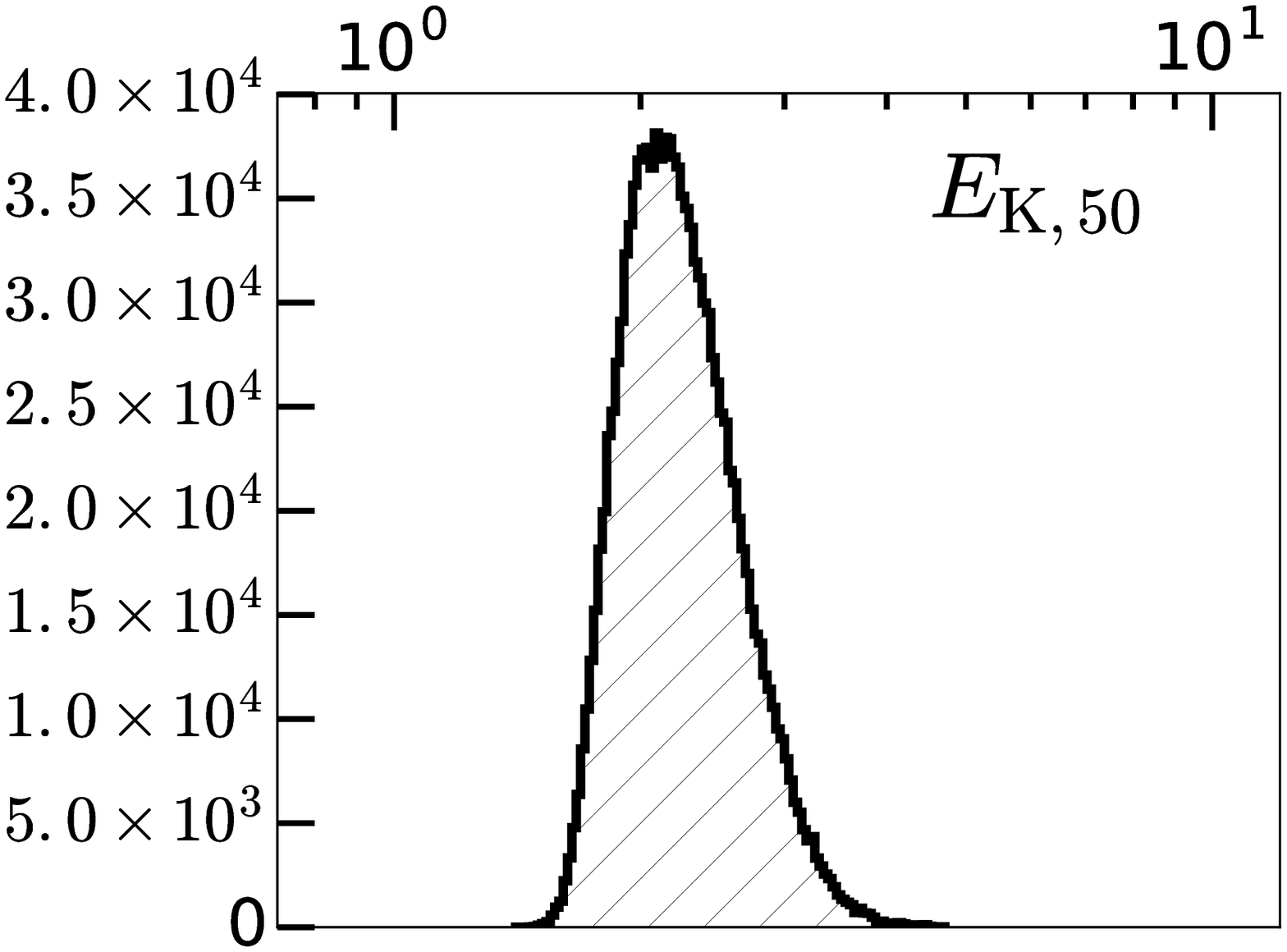}
\end{tabular}
\caption{Marginalized posterior probability density functions of the FS parameters from MCMC 
simulations. We have used the constraint $\epse+\epsb < 1$.}
\label{fig:hists}
\end{figure*}

\begin{figure*}
\begin{tabular}{ccc}
 \centering
 \includegraphics[width=0.31\textwidth]{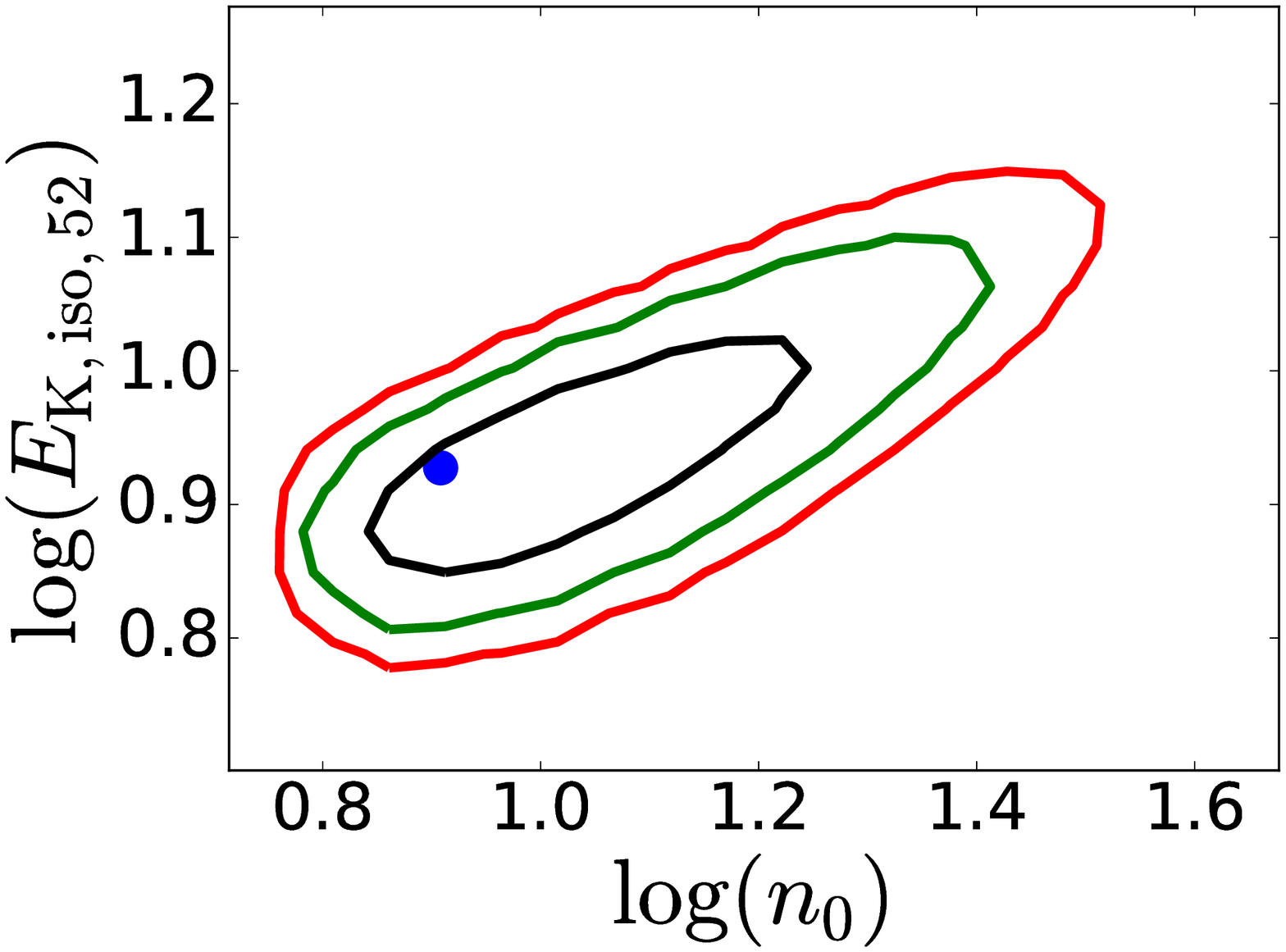} &
 \includegraphics[width=0.31\textwidth]{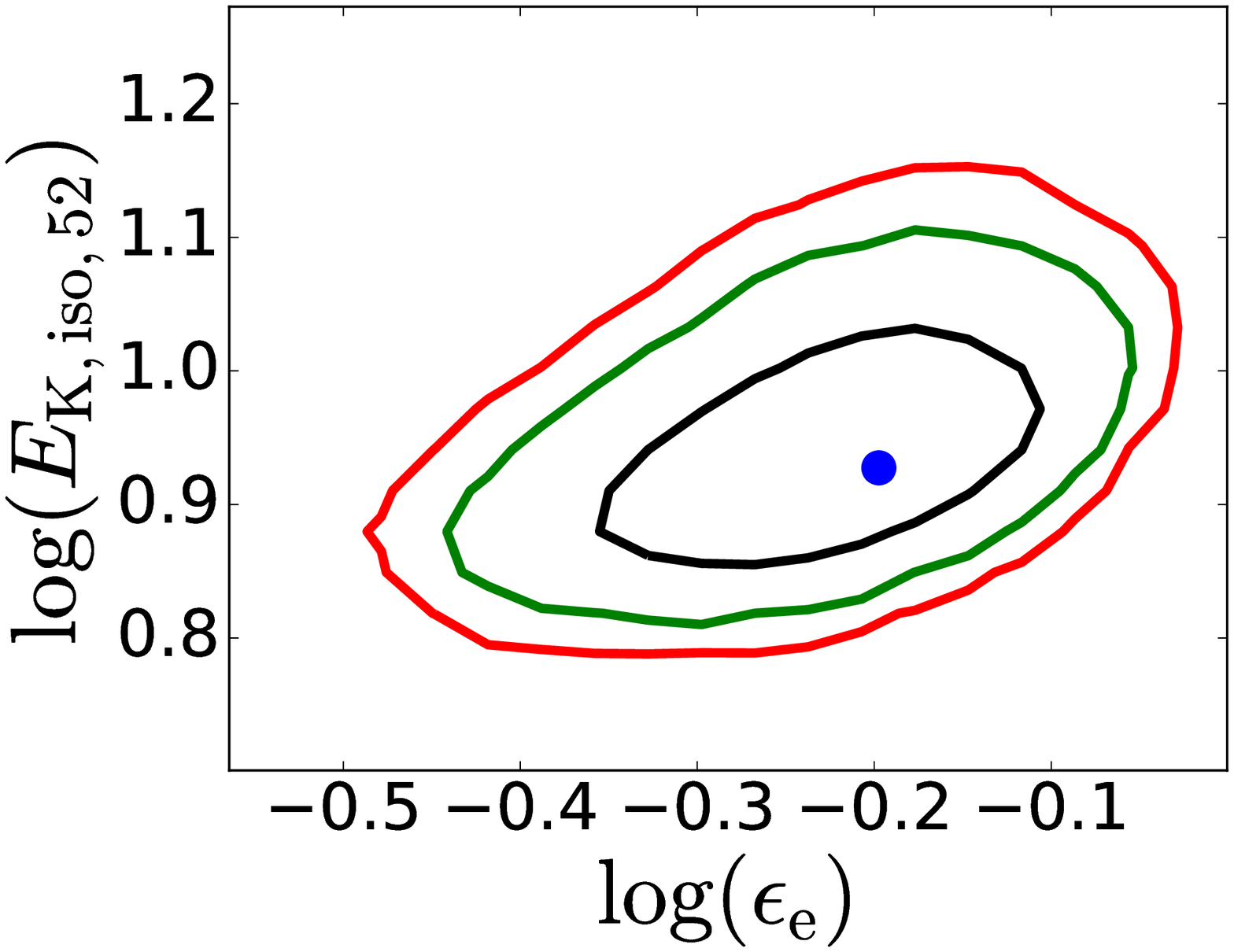} &
 \includegraphics[width=0.31\textwidth]{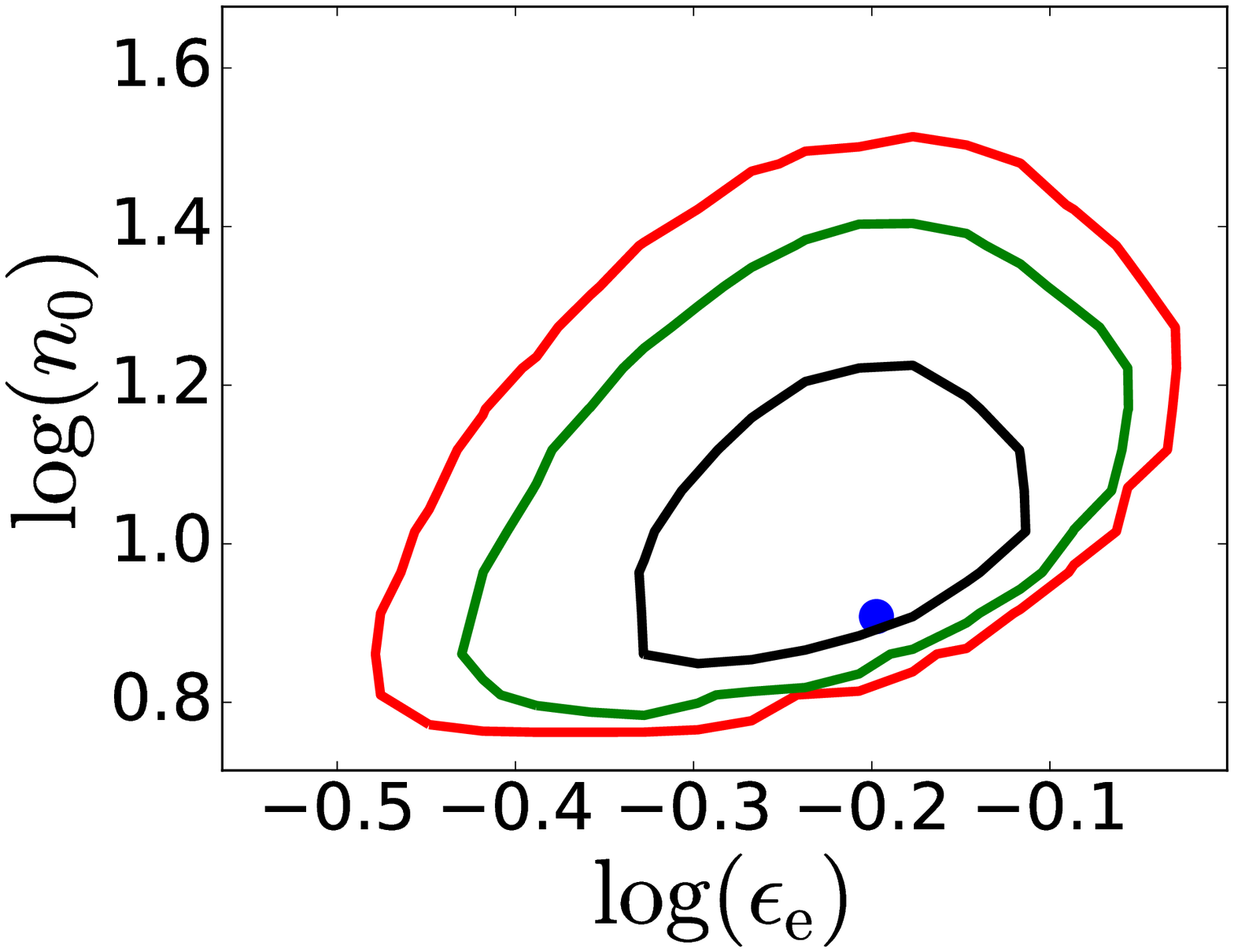} \\
 \includegraphics[width=0.31\textwidth]{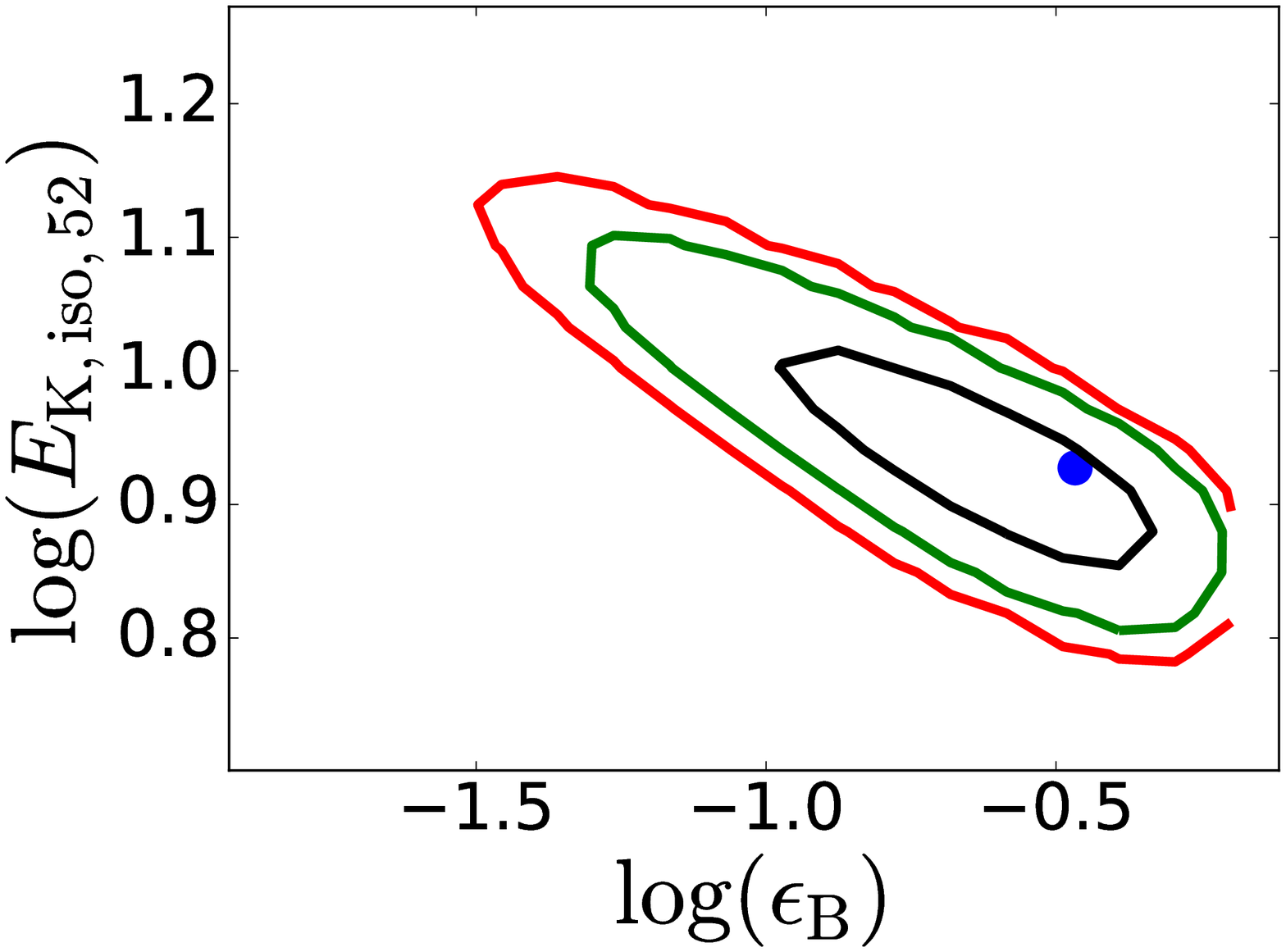} &
 \includegraphics[width=0.31\textwidth]{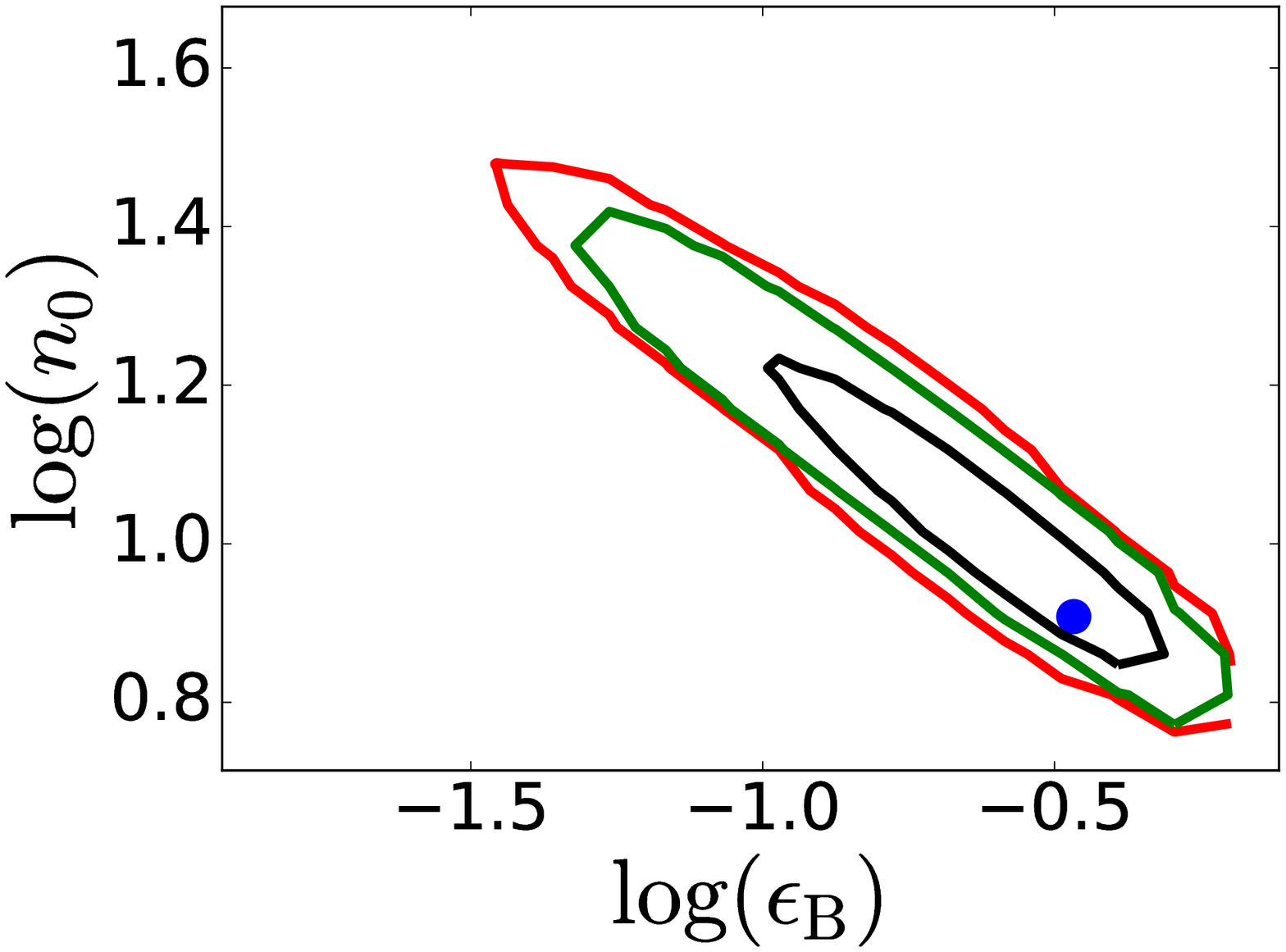} &
 \includegraphics[width=0.31\textwidth]{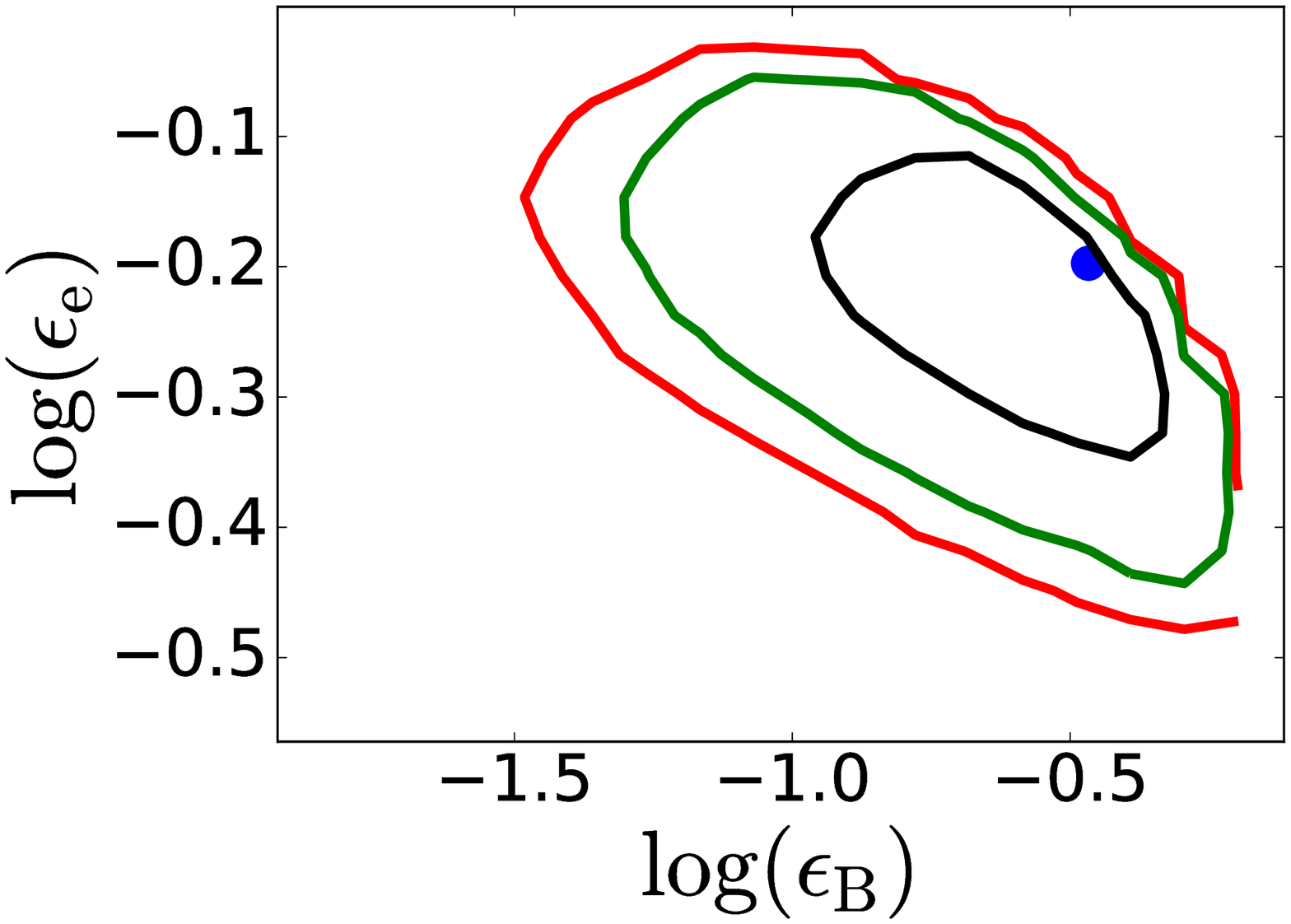} 
\end{tabular}
\caption{1$\sigma$ (red), 2$\sigma$ (green), and 3$\sigma$ (black) contours for correlations 
between the physical parameters \EKiso, \dens, \epse, and \epsb\ from Monte Carlo simulations,
together with the best-fit model (blue dot). 
We have used the constraint $\epse+\epsb<1$.}
\label{fig:corrplots}
\end{figure*}

We present the best fit model in Figure \ref{fig:modellc_ISM}, and list the derived parameters 
and their associated uncertainties in Table \ref{tab:params}. The SED remains in the fast cooling 
phase through the entire duration of the X-ray and optical light curves, and transitions to 
slow cooling at $\approx 9.8$\,d. The spectral ordering at 0.1\,d is $\nuc<\nuopt<\numax<\nux$.
The model requires a moderately high density of $\approx8$\,\pcc, as expected from the 
discussion in Section \ref{text:basic_considerations}.
The derived modest rest-frame host extinction of $\AV\approx0.3$~mag results in the 
correct spectral index both within the optical band and from the optical to the X-rays (Figure 
\ref{fig:betaoptx}). The jet break time of $\approx0.6$\,d yields an opening angle of $\approx 
3.9^{\circ}$. The kinetic energy corrected for beaming is $(2.2^{+0.4}_{-0.3})\times10^{50}$\,erg, 
and the beaming-corrected $\gamma$-ray energy is $\Egamma\approx7\times10^{50}$\,erg, yielding a 
$\gamma$-ray efficiency of $\eta\approx76\%$.
Histograms of the posterior density for each free parameter are provided in Figure \ref{fig:hists}, 
while correlation contours between pairs of the physical parameters (\epse, \epsb, \dens, \EKiso) 
are presented in Figure \ref{fig:corrplots}.

Since the afterglow remains in the fast cooling regime for a long period of time, significant 
radiative losses may be expected. A simple estimate under the assumption that the light curves can 
be modeled as produced by a blast wave with decreasing energy \citep{sar97}, yields a decrease in 
the kinetic energy by a factor of $\approx 24$ between the first $R$-band detection at 
$\approx3.8\times10^{-3}$\,d and the jet break at $\approx0.56$\,d. However, our model with 
constant energy fits the X-ray to radio observations well over this period, suggesting that this 
prescription for radiative losses may overestimate the effect \citep{nsg+13}. A detailed analysis of 
this effect requires better sampled optical light curves, as well as allowance for a variation in 
the Lorentz factor with radius for non-adiabatic blastwaves, and is beyond the scope of this work.

The ISM model under-predicts the flux density at $\lesssim10$\,GHz at 43.4\,d (Figures 
\ref{fig:modelsed_ISM1} and \ref{fig:modellc_ISM}), a feature that is 
true also of the wind model described in Appendix \ref{appendix:wind}. The light curves 
at $\lesssim 10$\,GHz rise between 18.5\,d and 43.4\,d, whereas we expect them to be either flat 
($F_{\nu < \nua} \propto t^{0}$) or slowly declining ($F_{\nua < \nu < \numax}\propto t^{-1/3}$). 
While a transition to non-relativistic expansion does allow for such a late-time rise in the radio 
light curve for $\nua<\nu<\numax$ \citep{fwk00}, the expected transition to non-relativistic 
expansion based on our best-fit model parameters is $\tnr \approx 500$\,d \citep{wkf98}.
One possible explanation for an early transition to non-relativistic expansion is a late encounter 
of the blast wave with a density enhancement, which decelerates the outflow rapidly and results 
in a re-brightening; however there are no other observations to test this hypothesis.

The model also marginally under-predicts the radio SED between 7\,GHz and 37\,GHz at 4.5\,d. While 
some of the excess flux may arise from additional processes such as emission from a reverse shock 
(RS), we note that there is no clear evidence for RS radiation at any other frequency at any time. 
On the other hand, the majority of the observed deviation appears consistent with the expected 
contribution of interstellar scintillation; thus there is no compelling evidence for a reverse shock 
component in the afterglow data for this event.

\section{Summary and Discussion}
\label{text:discussion}
Our multi-wavelength model explains the overall behavior of the afterglow over 8 orders 
of magnitude in frequency and 4 orders of magnitude in time, and indicates 
$\EK\approx2\times10^{50}$\,erg and $\dens\approx8$\,\pcc. These values are similar to those 
obtained for GRBs at $z\sim1$ \citep{pk02,yhsf03,ccf+08,cfh+10,cfh+11,lbt+14}, indicating no 
evolution in these properties with redshift to $z\sim5$. 

\begin{figure}
 \includegraphics[width=\columnwidth]{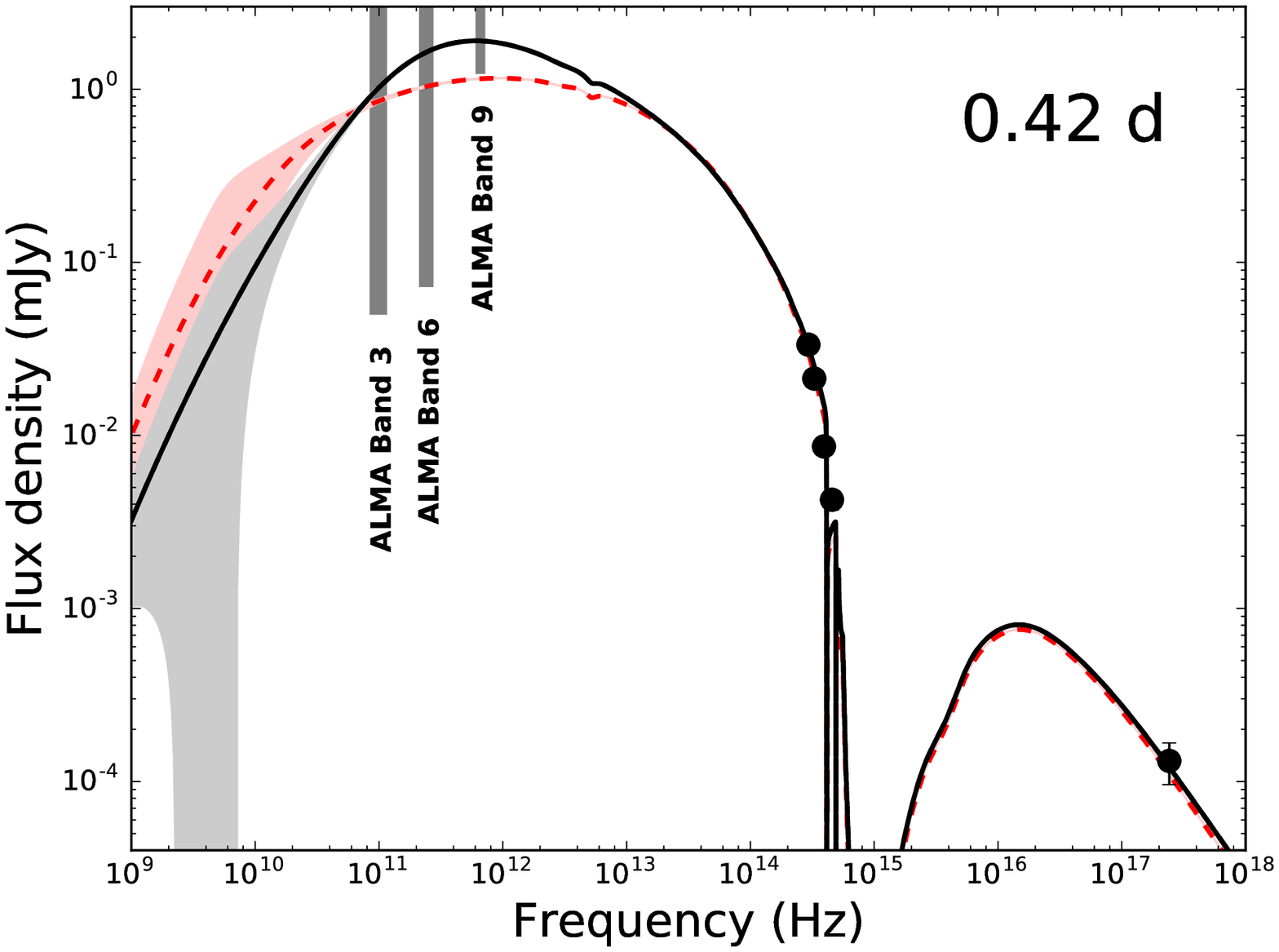}
\caption{Optical to X-ray spectral energy distribution of the afterglow of GRB~140311A with the 
best-fit ISM model (black; solid), together with an $\epsb\approx0.01$ model (red; dashed) for 
comparison. The shaded regions reflect the $1\sigma$ uncertainty due to scintillation along the 
line of sight (the grey band largely overlaps the red band), while the vertical bands indicate 
the $3\sigma$ sensitivity of ALMA in three different observing bands with 30 minutes on source.}
\label{fig:modelcomparison}
\end{figure}

Whereas our best fit values of the microphysical parameters are high ($\epse+\epsb\approx1$), we 
note that there is significant uncertainty in both of these parameters (in particular, the value of 
$\epsb$). We test this by fixing $\epsb\approx0.01$, and find that the consequent best-fit model 
results in a higher cooling frequency, lower peak flux (at $\nuc$) and lower self-absorption 
frequency, while the resulting mm-band flux density is unchanged within the error bars of the 
CARMA measurements, and the radio observations remain marginally consistent within the expected 
scatter from ISS. The parameter distributions resulting from these, and all other related families 
of models, are summarized in our plots of the correlation contours and the posterior density 
functions. We note that deeper mm-band observations at higher frequencies than were possible with 
CARMA (such as at $\approx200$\,GHz with ALMA in Band 6) would break some of these 
degeneracies, reducing the uncertainty in the physical parameters (Figure 
\ref{fig:modelcomparison}).

We note that the progenitor population of GRBs is also believed to produce type Ib/c 
supernovae, and whenever the latter are detected at radio wavelengths, the inferred density 
profile is consistent with a wind-like environment \citep[e.g. ][]{che98,bkc02,cf06,csc+15}. 
However, we find that an ISM model is a better fit than a wind environment for this burst, as
the latter over-predicts the radio emission before 2.5\,d. Whereas multi-wavelength studies of 
$z\sim1$ events have found no preference for either kind of density profile \citep[but see 
also][]{skb+11}, we previously inferred constant-density environments for all GRBs at $z\gtrsim6$ 
with radio detections (\lbt). Increasing the sample of $z\gtrsim5$ GRBs beyond the current 
set of four events is critical for exploring the statistical significance of this result. We add 
that mm-band observations have the strongest diagnostic power in distinguishing between constant 
density and wind-like environments, owing to scintillation effects at the cm-bands 
\citep[e.g.][]{yhsf03,lbm+15}. Thus mm-band observations at facilities such as ALMA in conjunction 
with cm-band observations at the VLA will play an important role in answering the question of the 
circumburst density profile and the mass loss rates of GRB progenitors in the the last moments 
before core collapse.

Our inferred jet opening angle of $\thetajet \approx 4^{\circ}$ for this burst is narrower than 
the median value of $\thetajet = 7.4^{+11}_{-6.6}$ (95\% confidence interval) for GRBs at  at 
$z\sim1$, but similar to the mean value for GRBs at $z\gtrsim6$ ($\thetajet = 
3.6^{\circ}\pm0.7^{\circ}$; \lbt), which may indicate that higher redshift GRBs are more strongly 
collimated. Our best fit wind model also yields a narrow opening angle and a high density, and 
therefore these results are robust to the choice of circumburst density profile. We note that a 
selection effect arising from the trigger criteria for $\gamma$-ray telescopes, which may select 
more tightly beamed events at higher redshift owing to the sensitivity threshold of the detectors, 
remains a possibility. The observed $\gamma$-ray fluence of this event and all the bursts studied 
in \lbt\ is within one standard deviation of the mean for lower-redshift events \citep{mzb+13}, 
with no systematic trend toward lower fluences; however, two out of four $z\gtrsim5$ events, 
GRB~140311A and 050904, were located through BAT image triggers, which traditionally find lower 
luminosity events \citep{lsb+16}. We therefore caution that detailed statistical studies
should account for possible selection biases due to the criteria used to discover the event.

Our observations afford no compelling evidence for emission from a reverse shock. We 
have previously found strong RS signatures only in low density environments, which we attribute to 
the slow cooling RS SEDs expected in such environments \citep{lbz+13,lab+16,pcc+14,alb+17}. Our 
best-fit model results in an afterglow SED in the fast cooling regime, increasing the likelihood 
that the RS SED is fast cooling as well. We speculate that the high density observed in the case of 
GRB~140311A may suppress RS emission, and that absence of RS signatures may not, therefore, be a 
signature of highly magnetized ejecta \citep{uzh+12}.

\section{Conclusions}
Our observations of GRB~140311A are the most detailed joint radio and millimeter observations 
of a GRB at $z\gtrsim5$ to date. They reveal an afterglow with parameters typical of events at 
$z\sim1$, with the exception of the opening angle, which is narrower, but similar to the values 
derived for events at $z\gtrsim6$ and consistent with the hypothesis that 
GRBs detected at high redshift may 
be more tightly beamed (LBT14). We find no evidence for a strong reverse shock, and note that 
synchrotron cooling may suppress the RS emission in this case. Upcoming papers in this series will 
address the connection between the circumburst density and the likelihood of observing RS emission, 
as well as the redshift evolution of the opening angles of GRB jets.

\acknowledgements
TL is a Jansky Fellow of the National Radio Astronomy Observatory. 
The Berger Time-Domain Group at Harvard is supported in part by the NSF under grant 
AST-1411763 and by NASA under grant NNX15AE50G.
BAZ acknowledges support from NSF AST-1302954. 
RC acknowledges support from NASA Swift grant NNX16AB04G.
VLA observations for this study were obtained via project 14A-344. 
The National Radio Astronomy Observatory is a facility of the National Science Foundation operated 
under cooperative agreement by Associated Universities, Inc.
This research has made use of data supplied by the UK Swift Science Data Centre at the
University of Leicester, and of data obtained through the High Energy Astrophysics Science Archive 
Research Center On-line Service, provided by the NASA/Goddard Space Flight Center.

\appendix

\section{A wind model}
\label{appendix:wind}
\begin{figure*}
\begin{tabular}{ccc}
 \centering
 \includegraphics[width=0.31\textwidth]{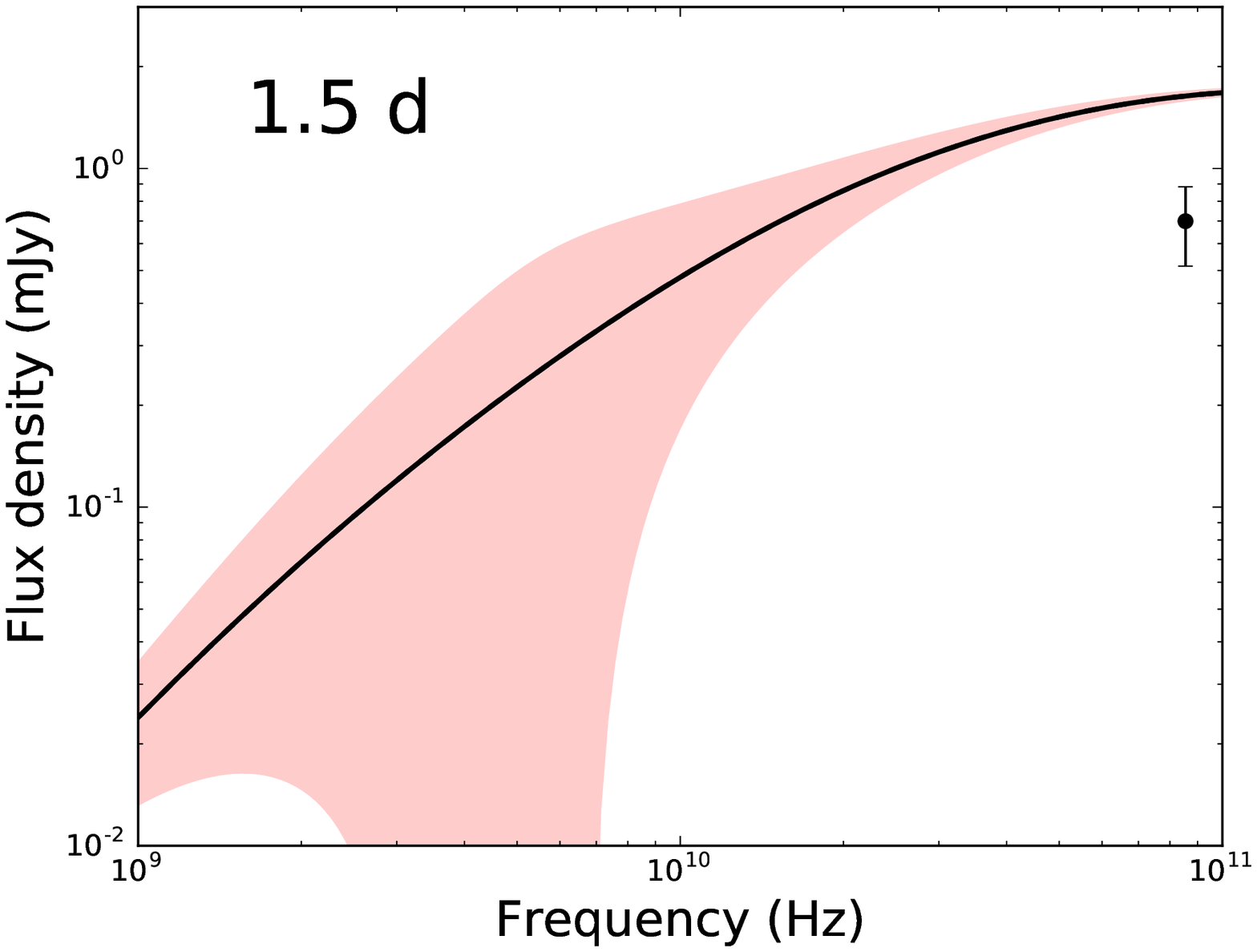} &
 \includegraphics[width=0.31\textwidth]{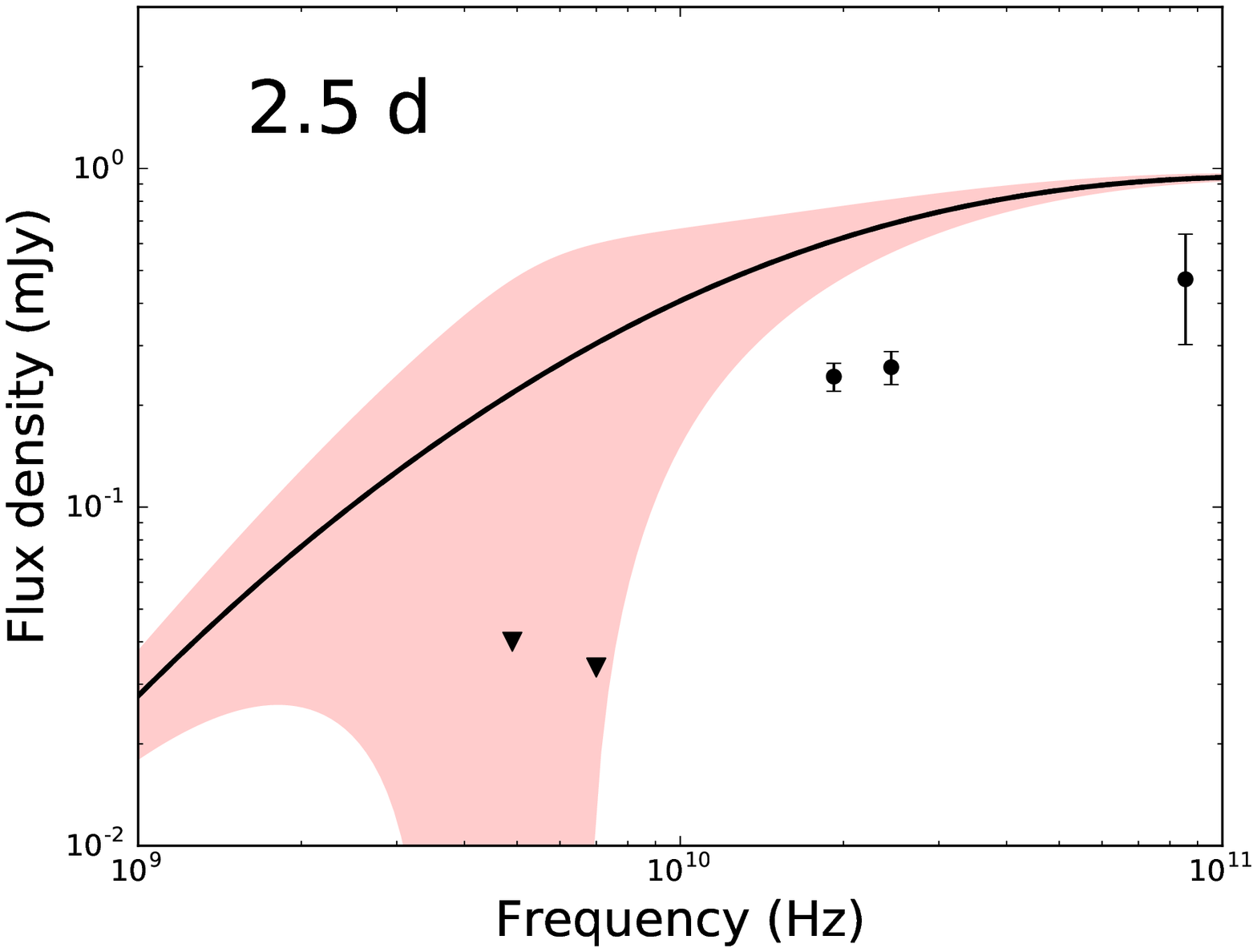} &
 \includegraphics[width=0.31\textwidth]{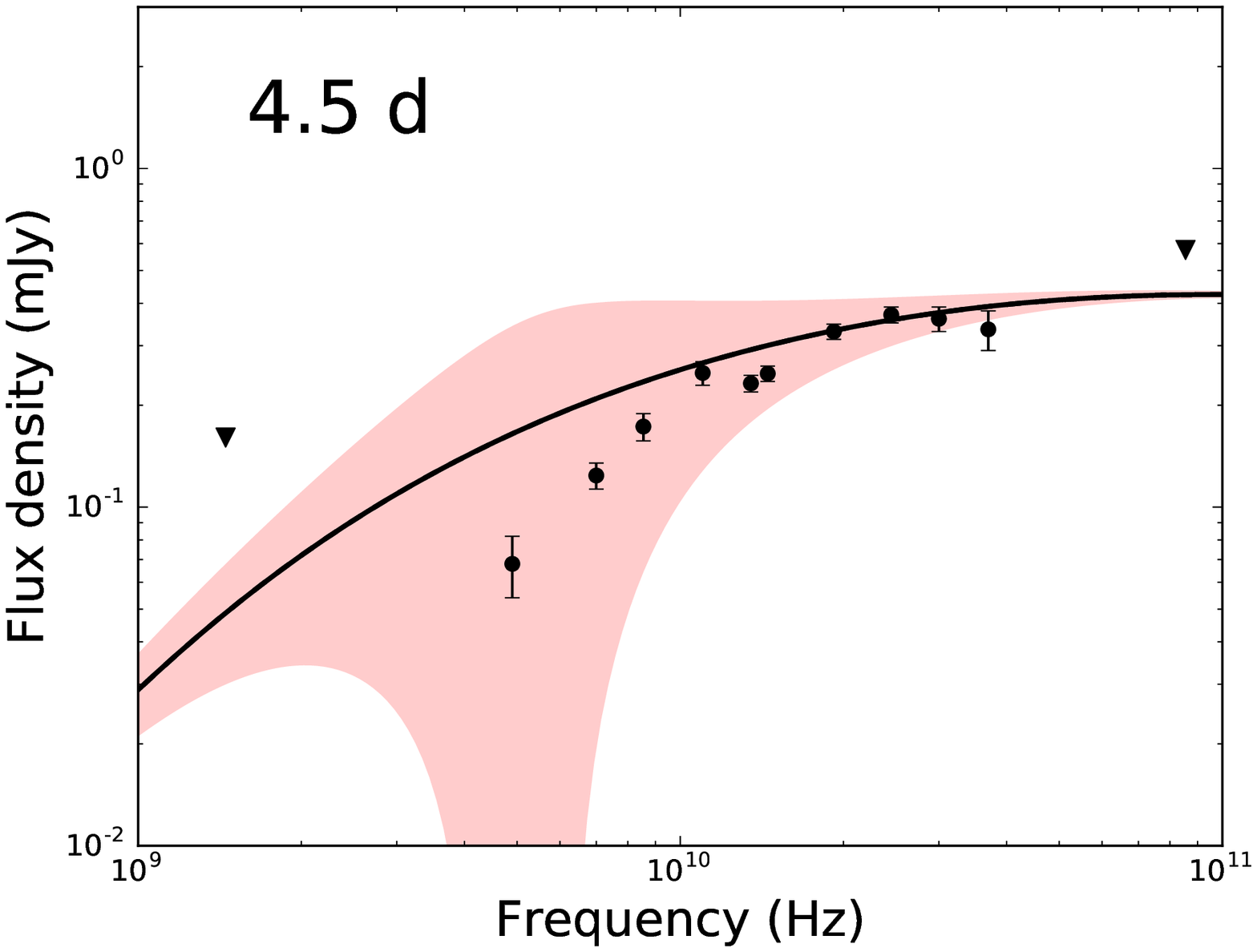} \\
 \includegraphics[width=0.31\textwidth]{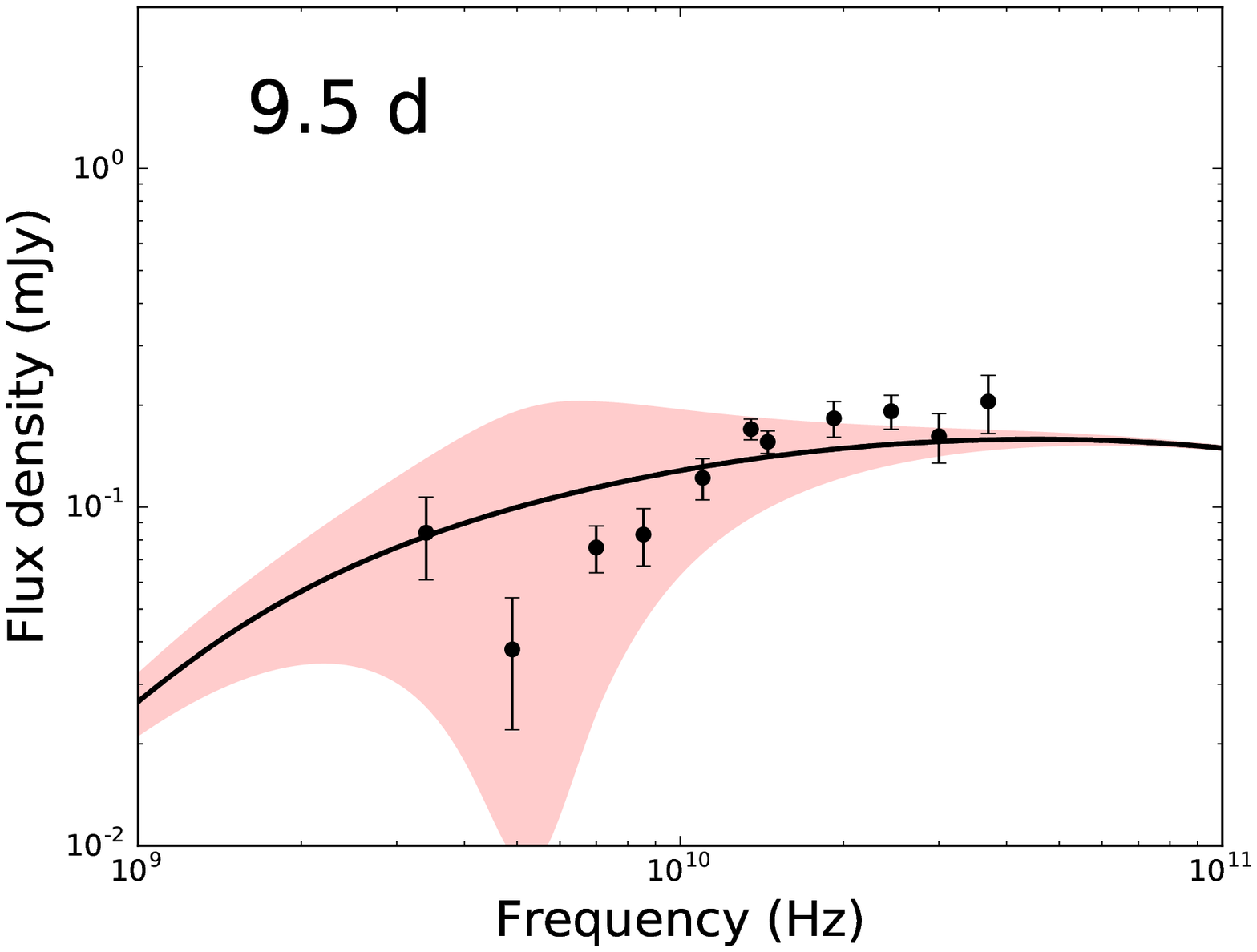} &
 \includegraphics[width=0.31\textwidth]{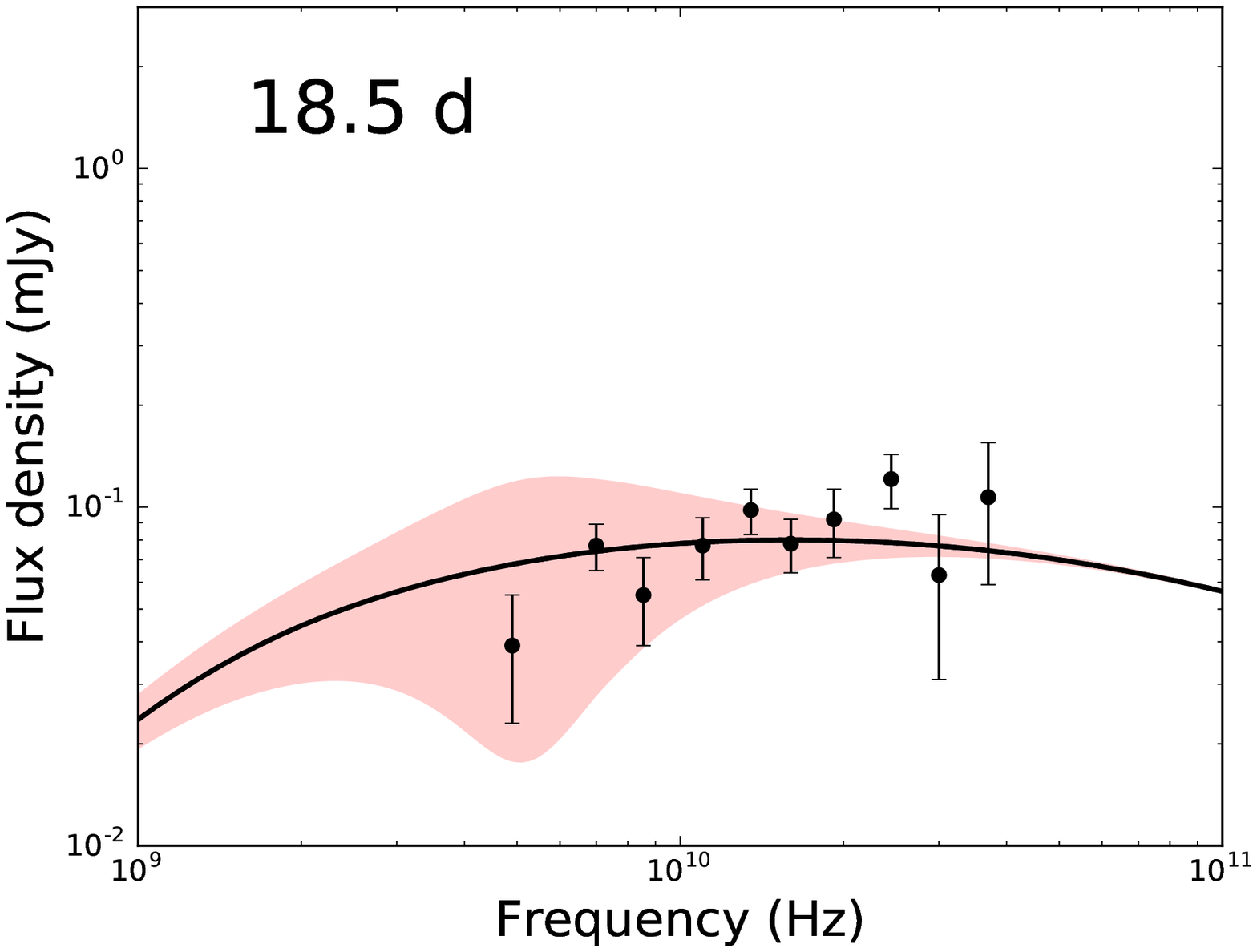} &
 \includegraphics[width=0.31\textwidth]{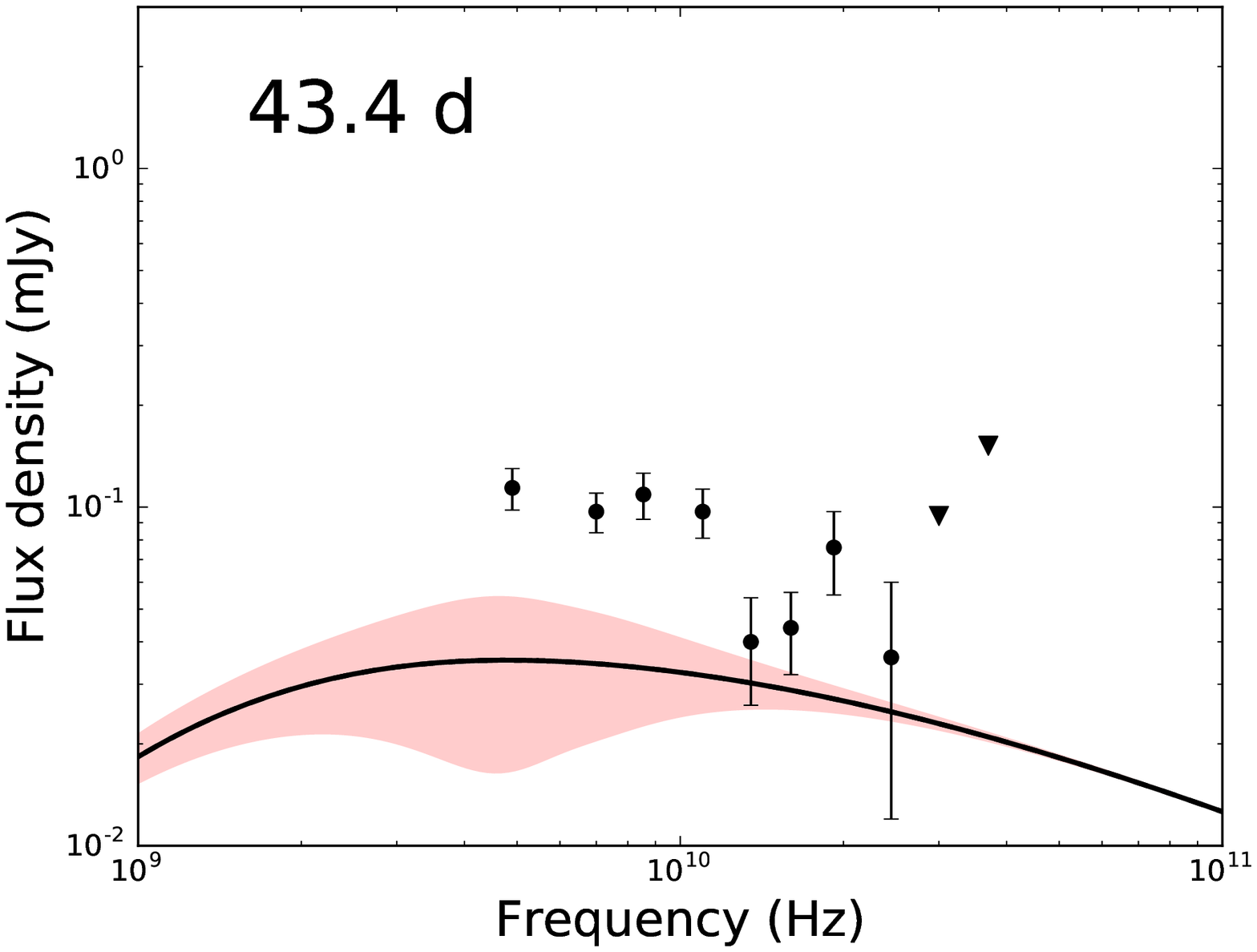} \\
 \includegraphics[width=0.31\textwidth]{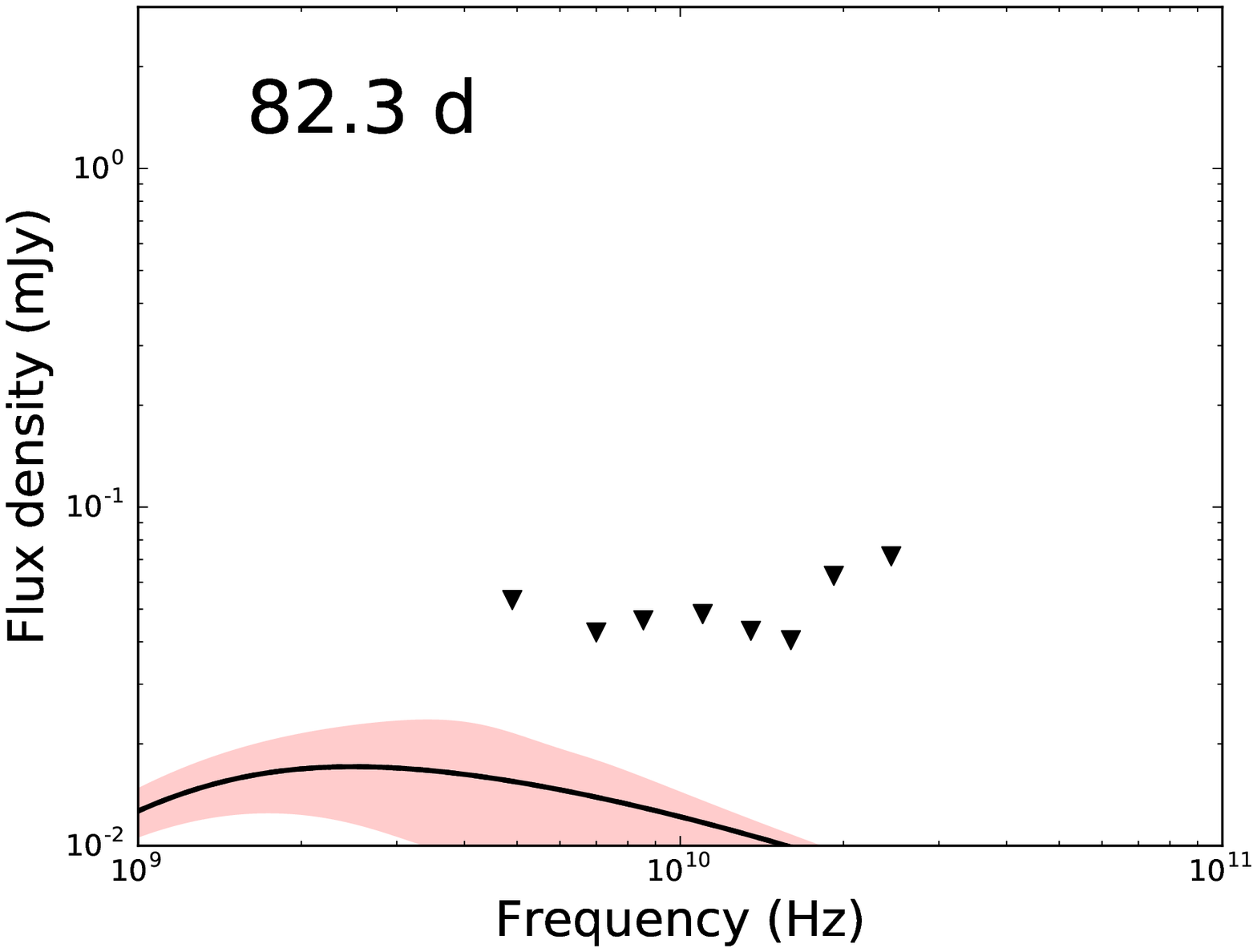} & 
\end{tabular}
\caption{Same as Figure \ref{fig:modelsed_ISM1}, but for a wind environment. The model 
over-predicts the millimeter and radio observations at 1.5\,d and 2.5\,d, requiring an 
additional source of opacity at low frequencies.}
\label{fig:modelsed_wind4}
\end{figure*}

\begin{figure*} 
 \begin{tabular}{cc}
  \includegraphics[width=0.47\textwidth]{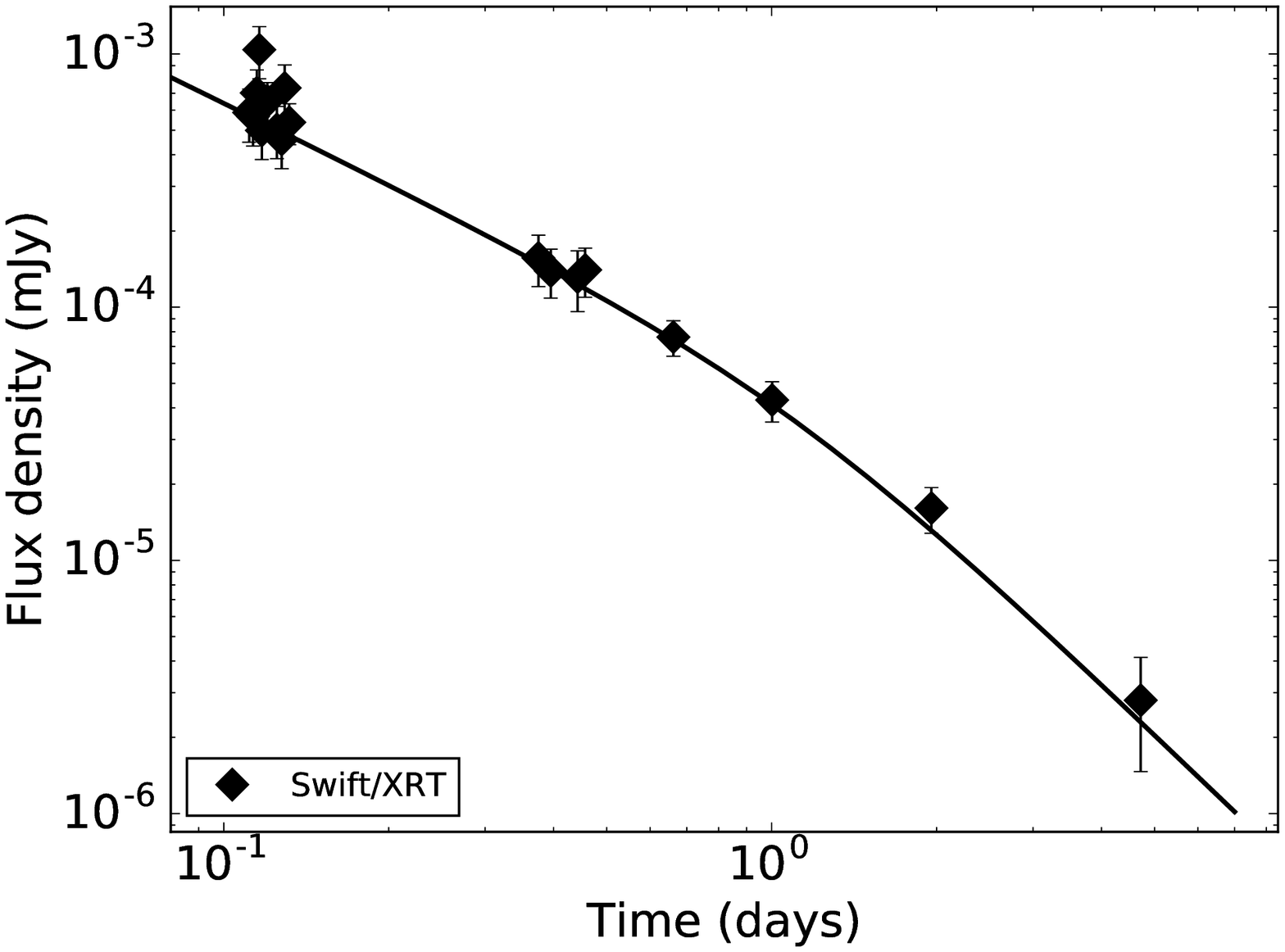} &
  \includegraphics[width=0.47\textwidth]{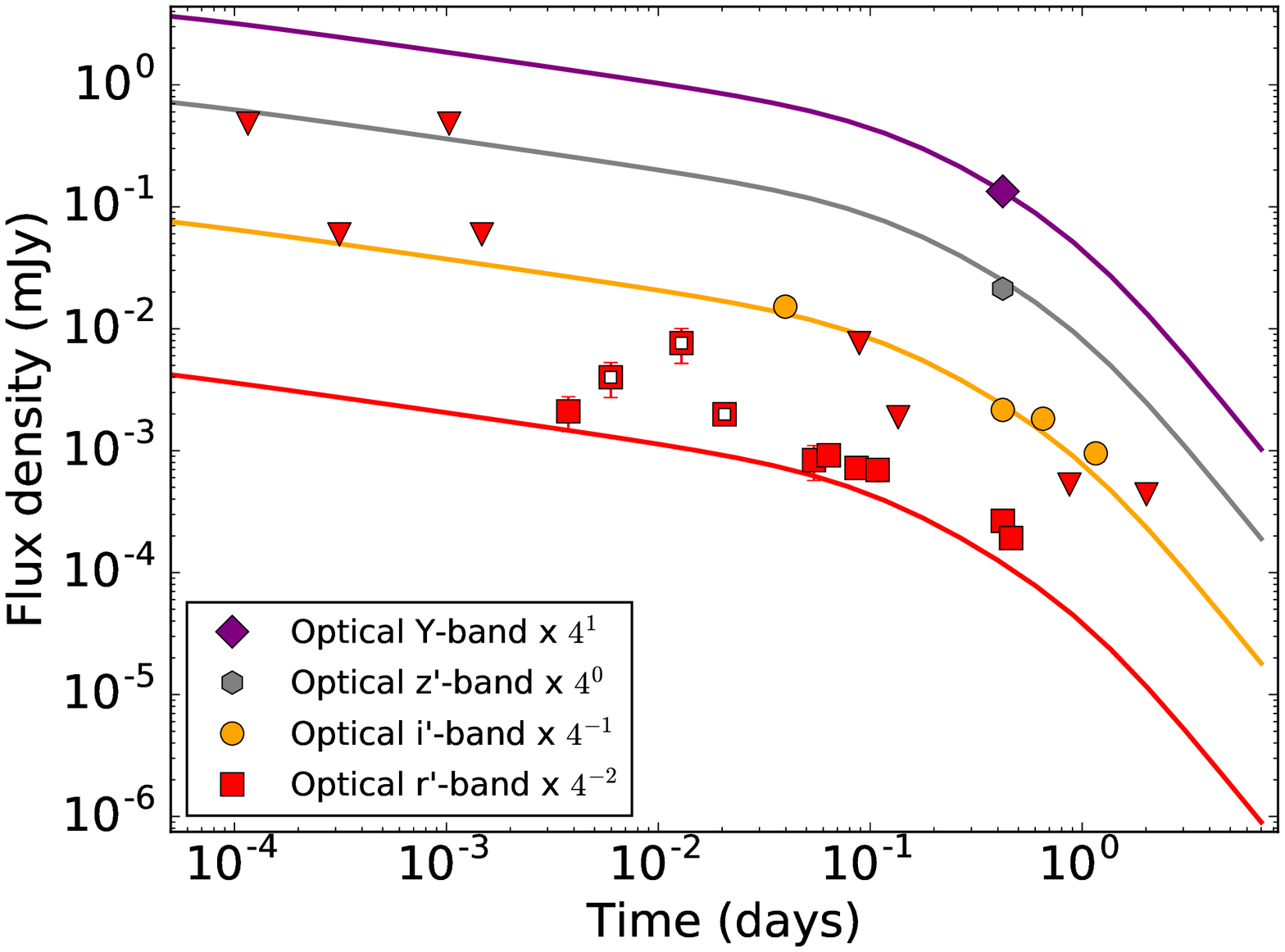} \\
  \includegraphics[width=0.47\textwidth]{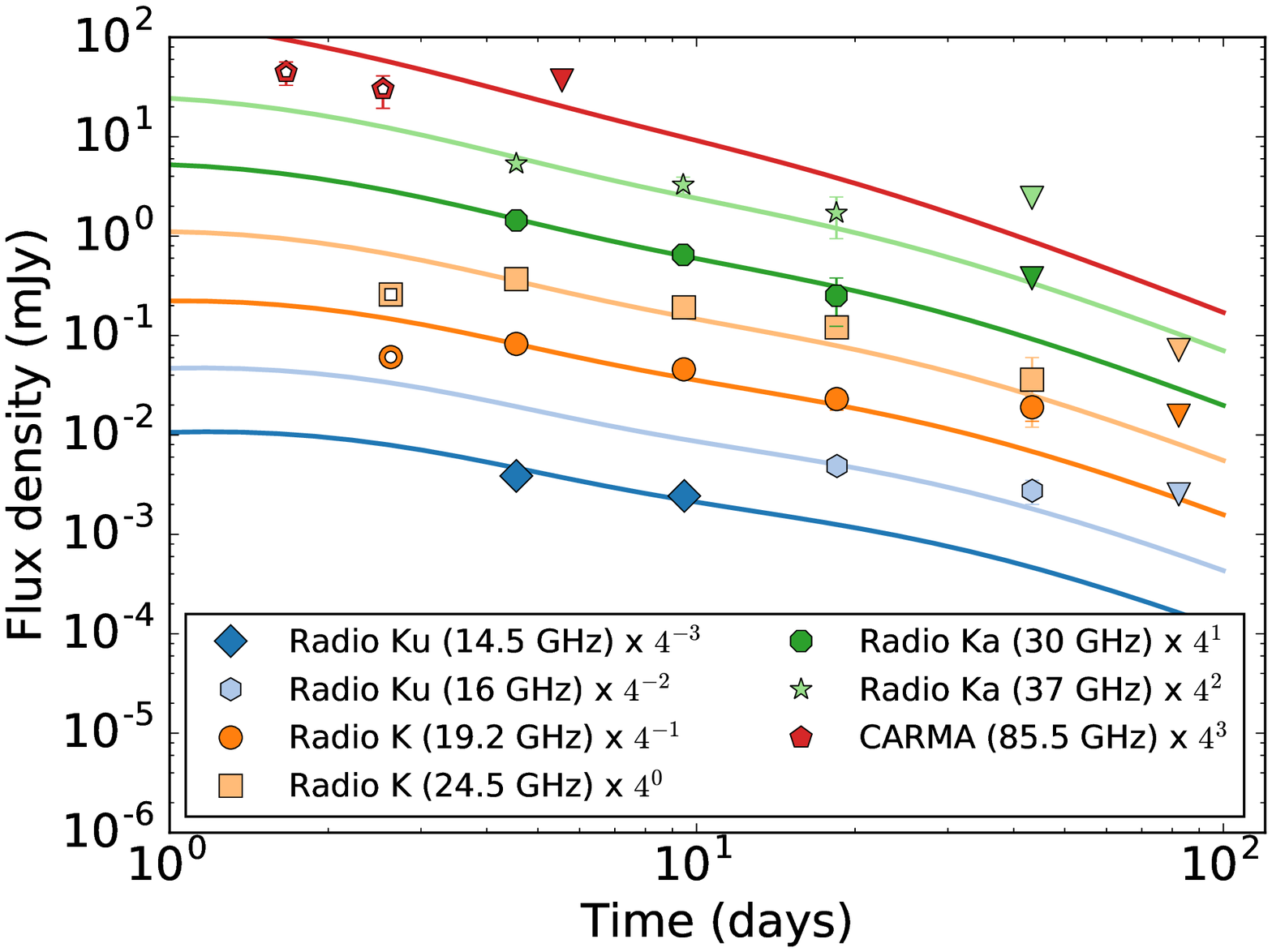} &
  \includegraphics[width=0.47\textwidth]{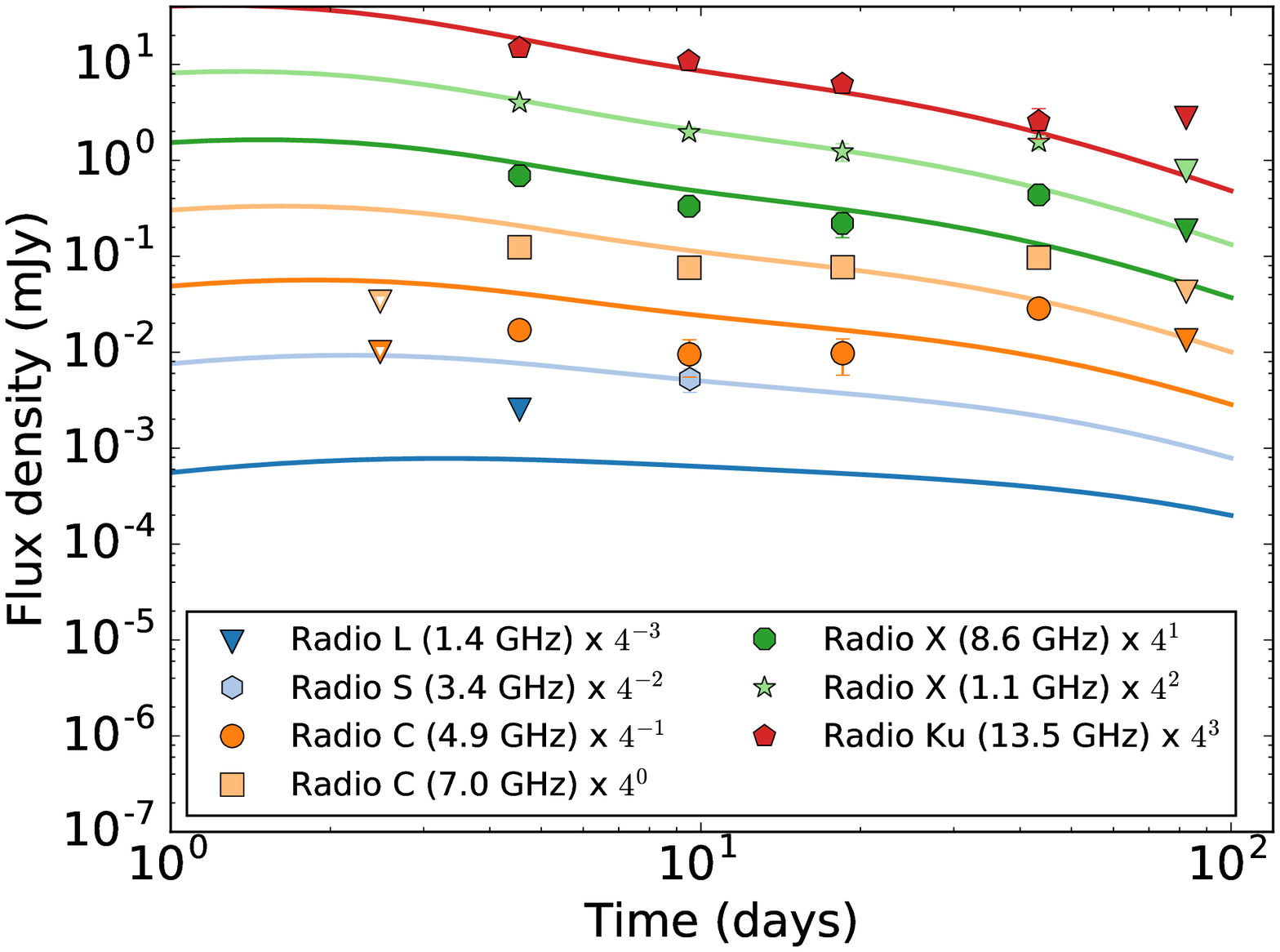} \\
 \end{tabular}
 \caption{Same as Figure \ref{fig:modellc_ISM}, but for a wind environment. The model fits the 
X-ray and optical light curves well, but over-predicts the radio and millimeter observations at 
1.5\,d and 2.5\,d (Figure \ref{fig:modelsed_wind4}). The data points with open symbols are not 
included in the multi-wavelength fit.}
\label{fig:modellc_wind}
\end{figure*}

\begin{figure*}
\begin{tabular}{ccc}
 \centering
 \includegraphics[width=0.31\textwidth]{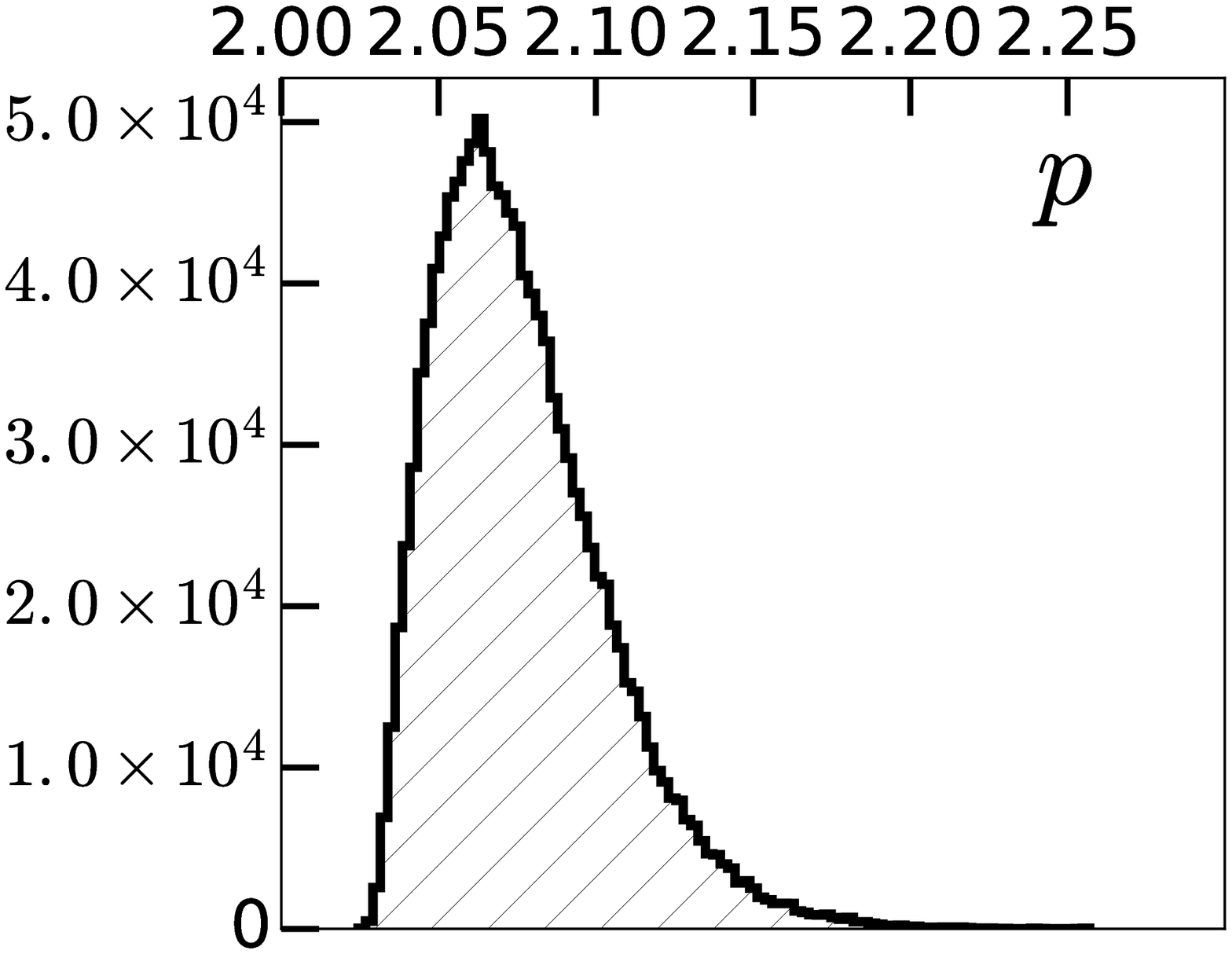} &
 \includegraphics[width=0.31\textwidth]{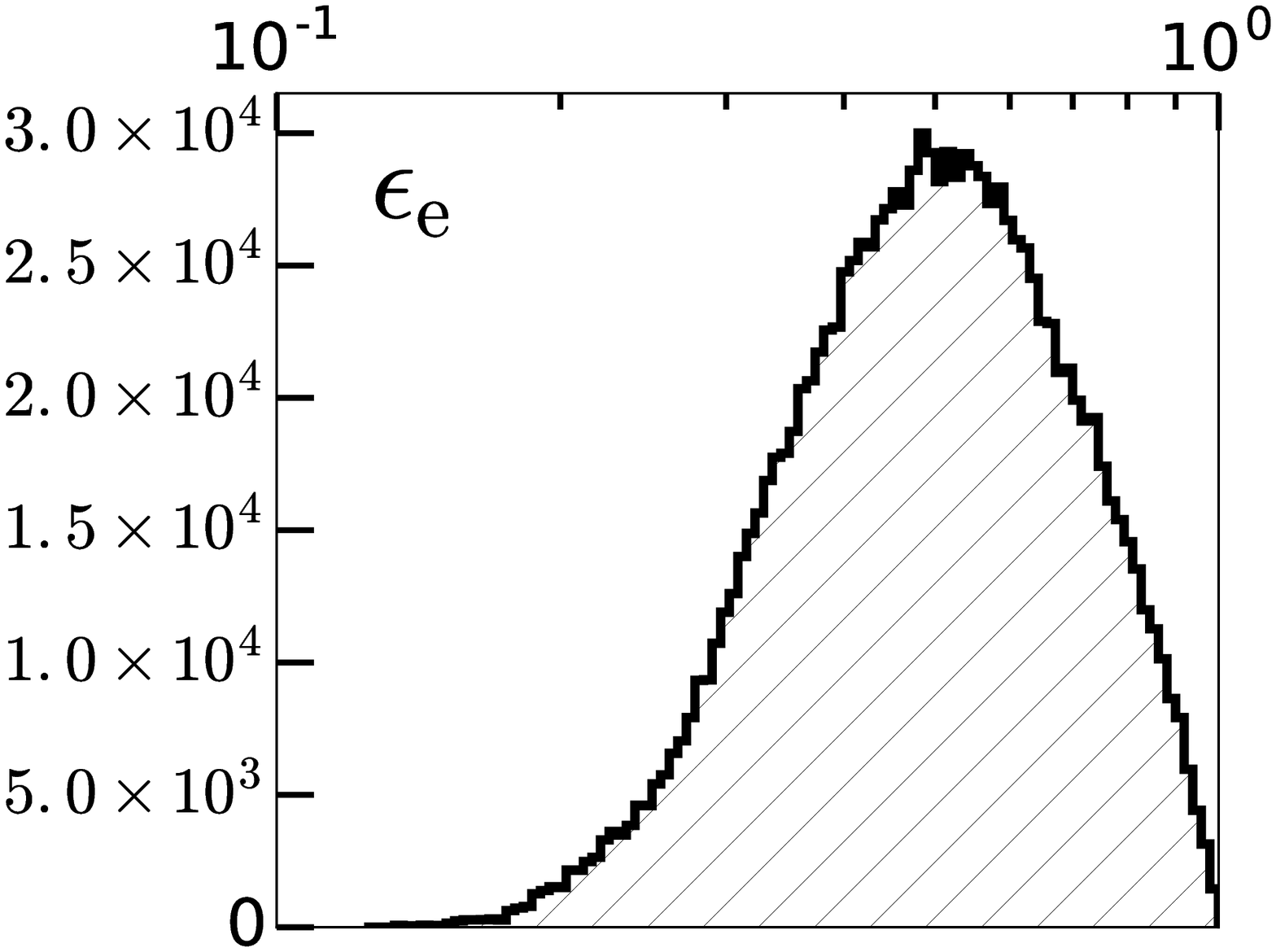} &
 \includegraphics[width=0.31\textwidth]{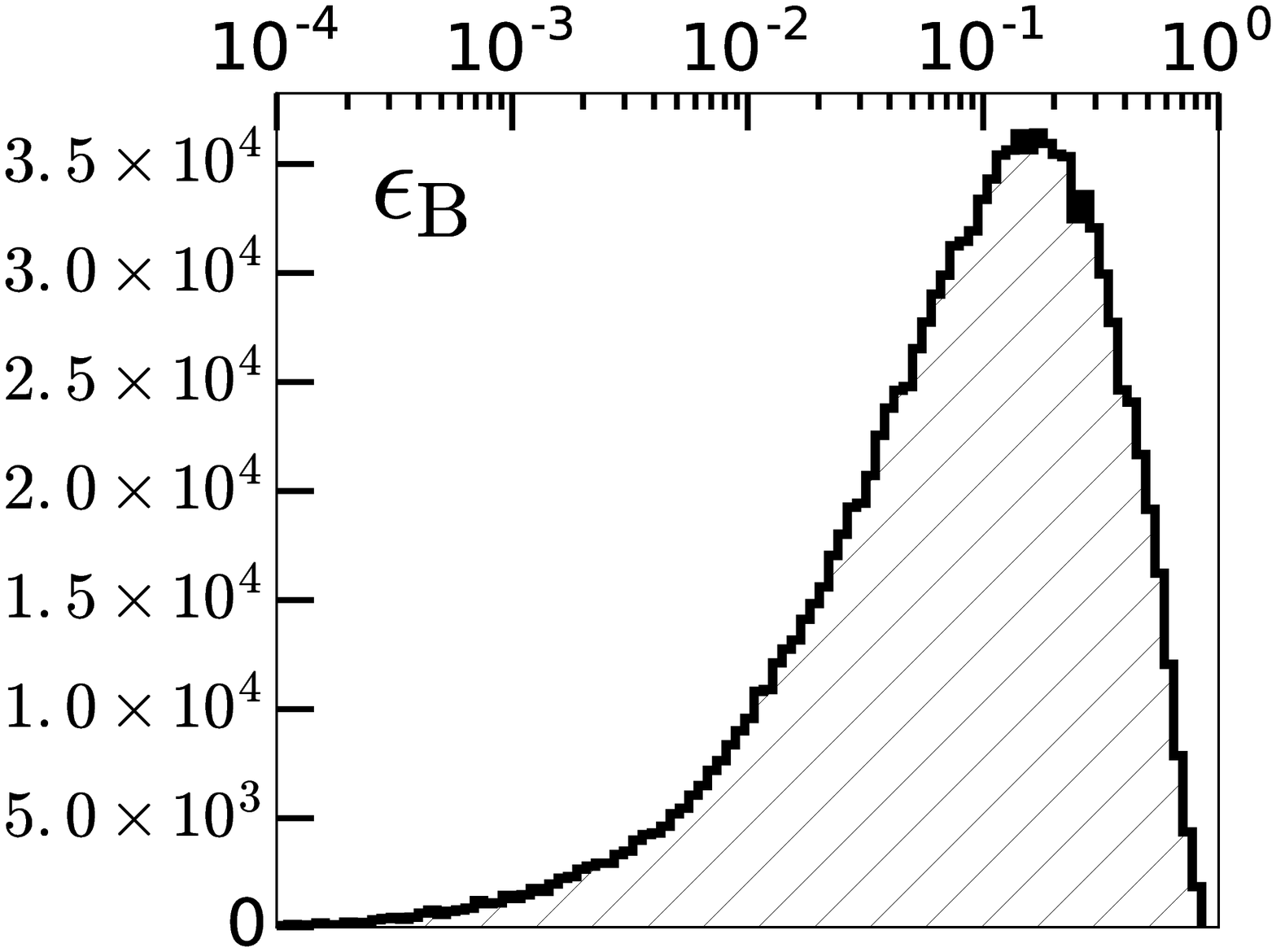} \\
 \includegraphics[width=0.31\textwidth]{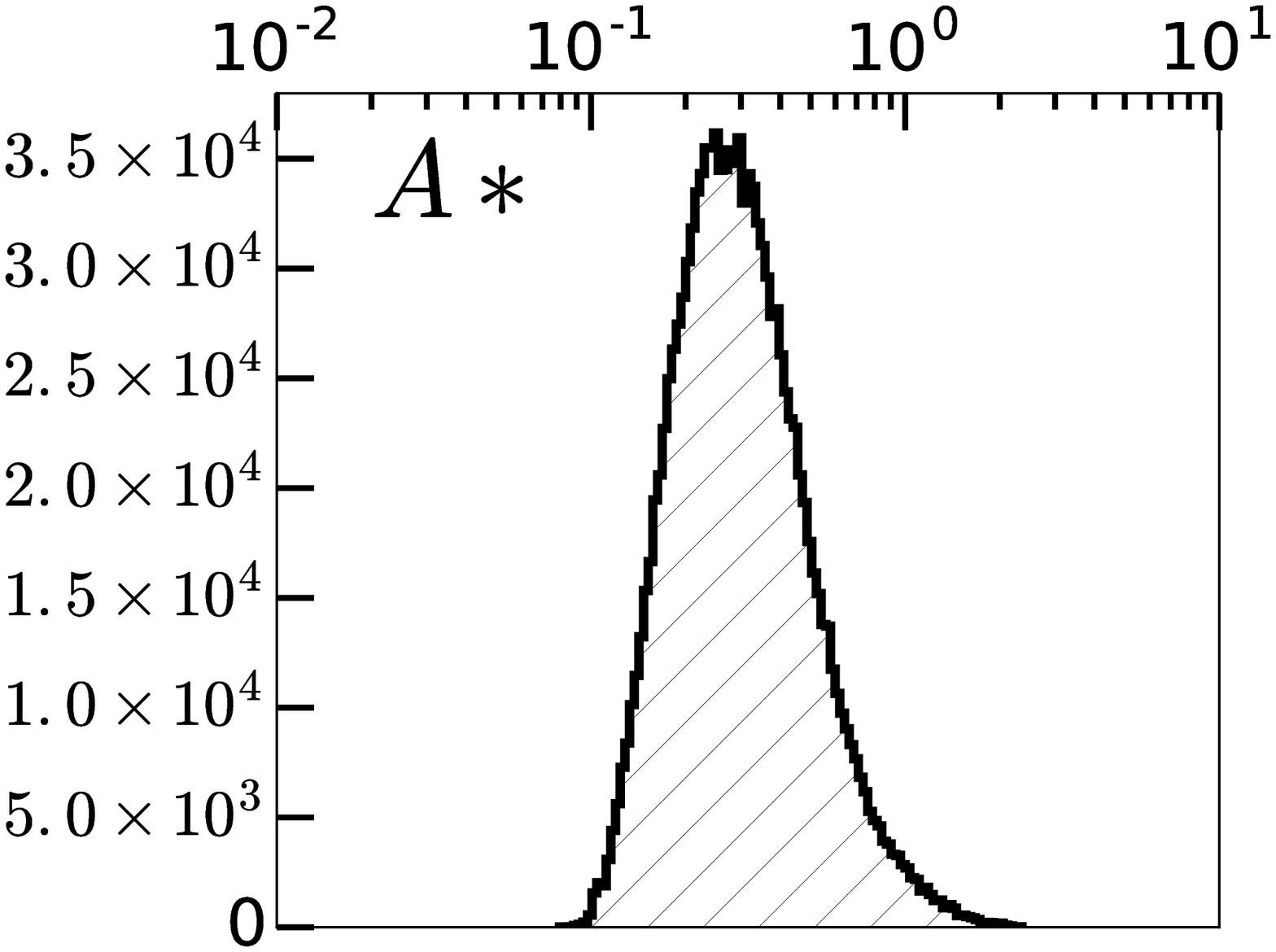} &
 \includegraphics[width=0.31\textwidth]{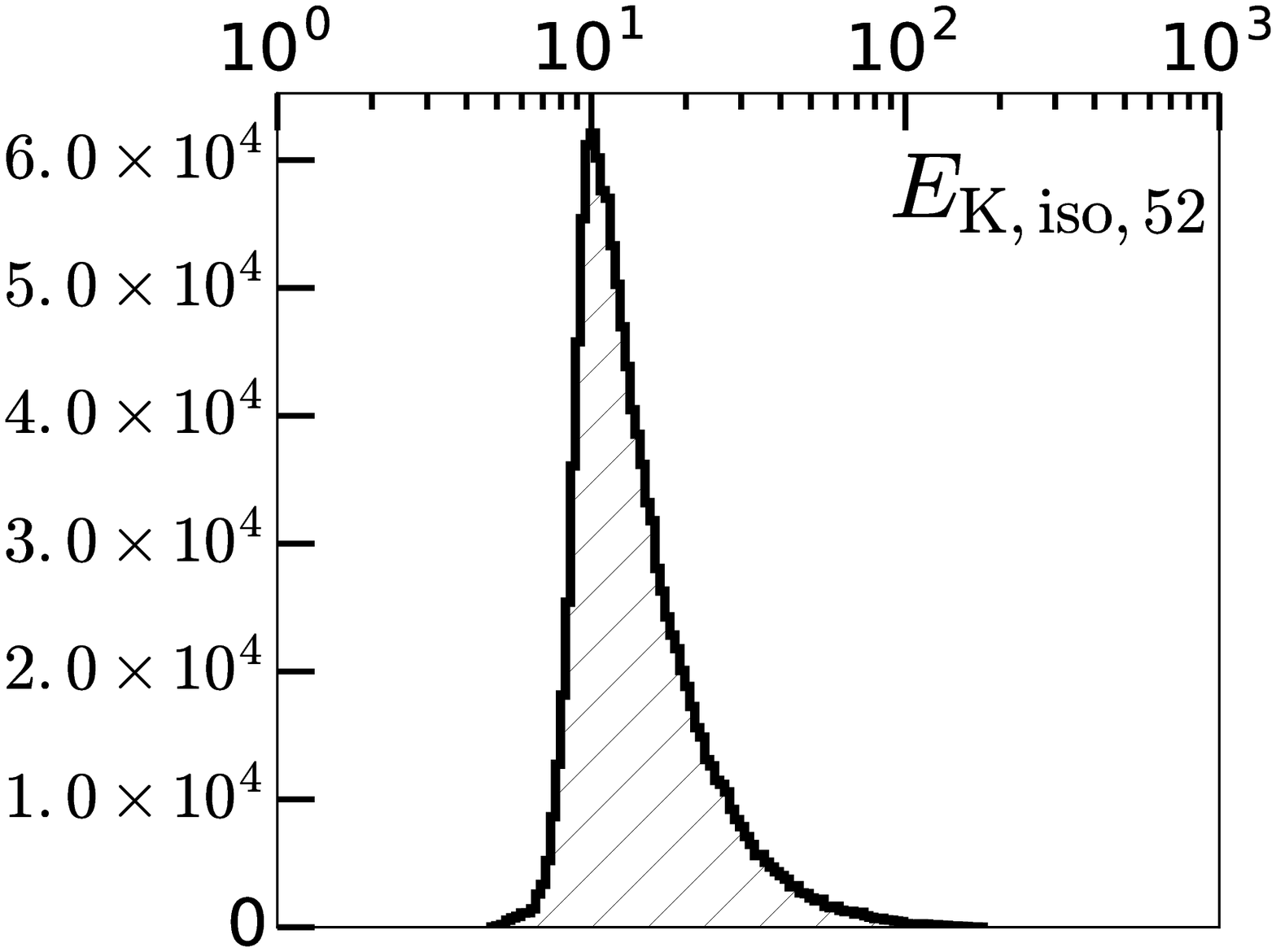} &
 \includegraphics[width=0.31\textwidth]{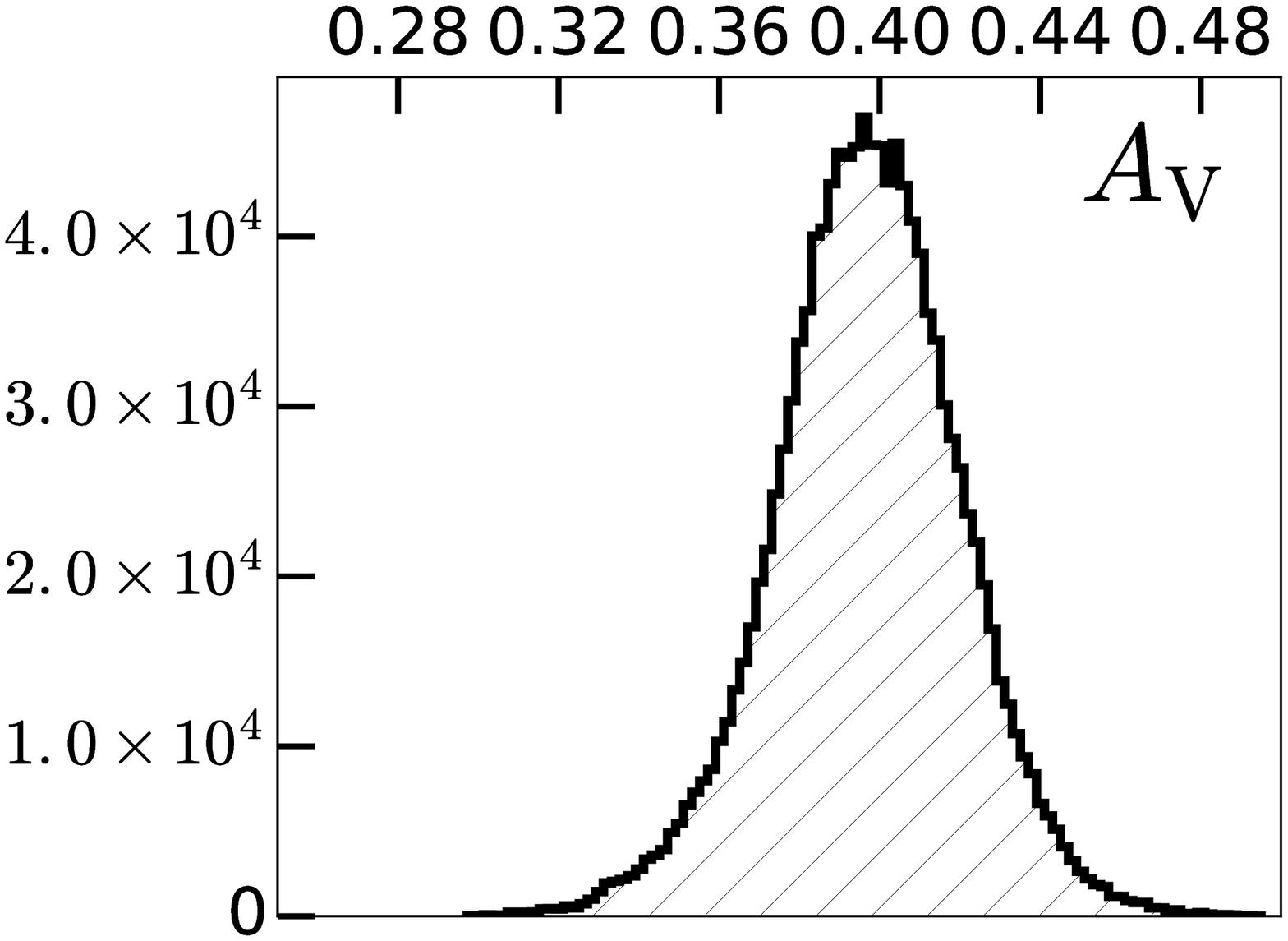} \\
 \includegraphics[width=0.31\textwidth]{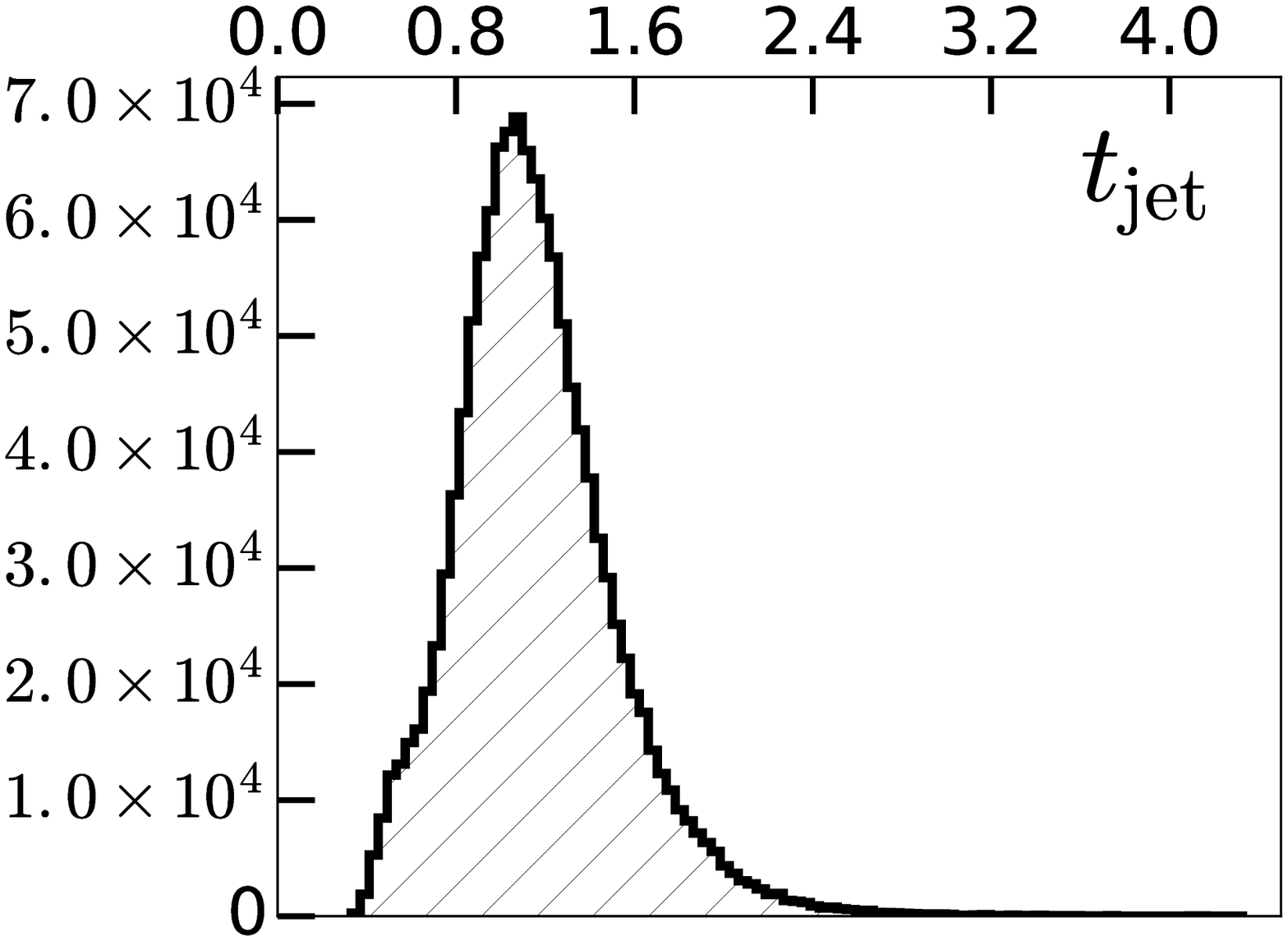} &
 \includegraphics[width=0.31\textwidth]{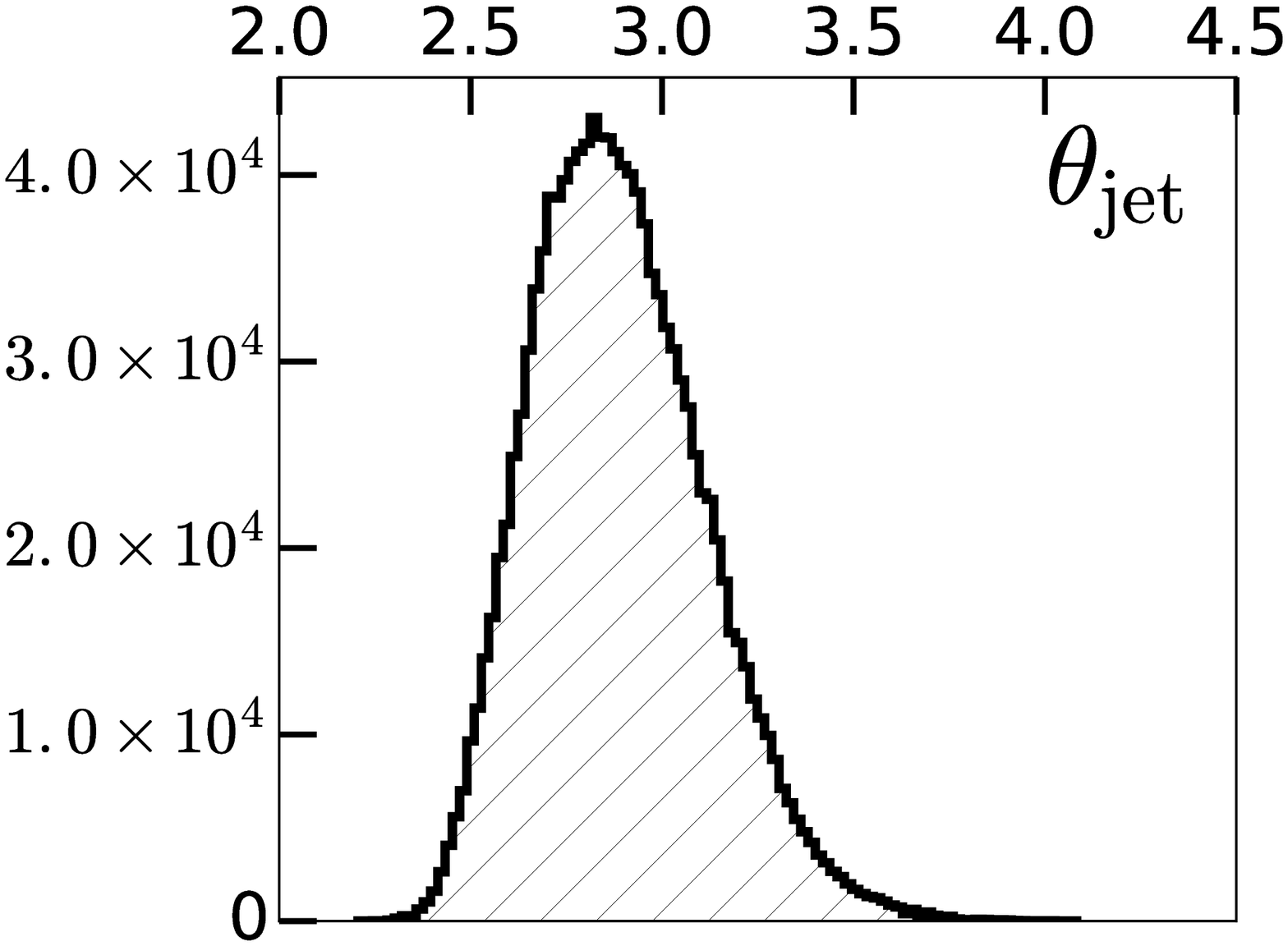} & 
 \includegraphics[width=0.31\textwidth]{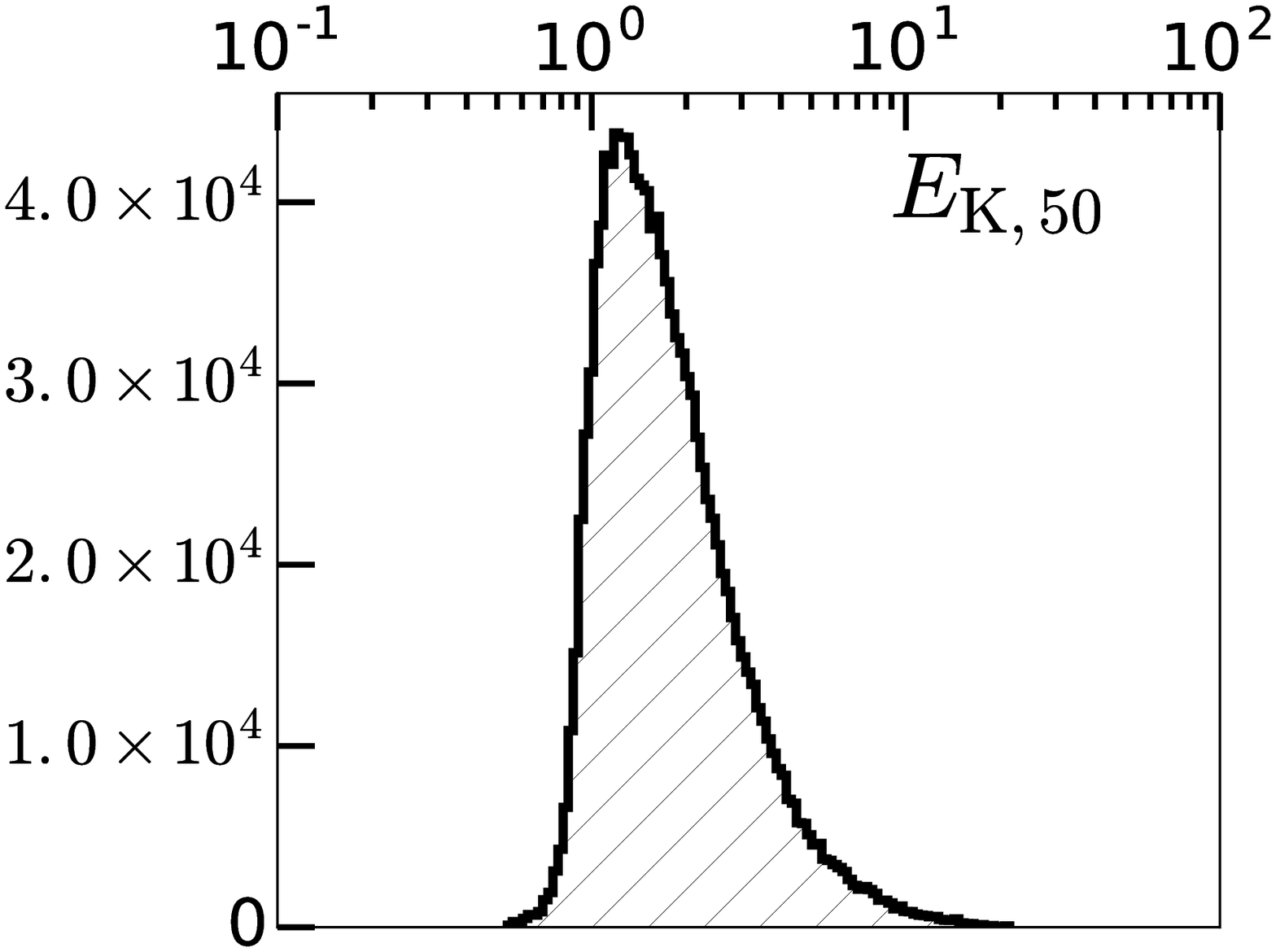}
\end{tabular}
\caption{Marginalized posterior probability density functions of the FS parameters from MCMC 
simulations for a wind environment. We have restricted $\epse+\epsb < 1$, and do not include the 
radio data before 2.5\,d in the analysis.}
\label{fig:hists_wind}
\end{figure*}

\begin{figure*}
\begin{tabular}{ccc}
 \centering
 \includegraphics[width=0.31\textwidth]{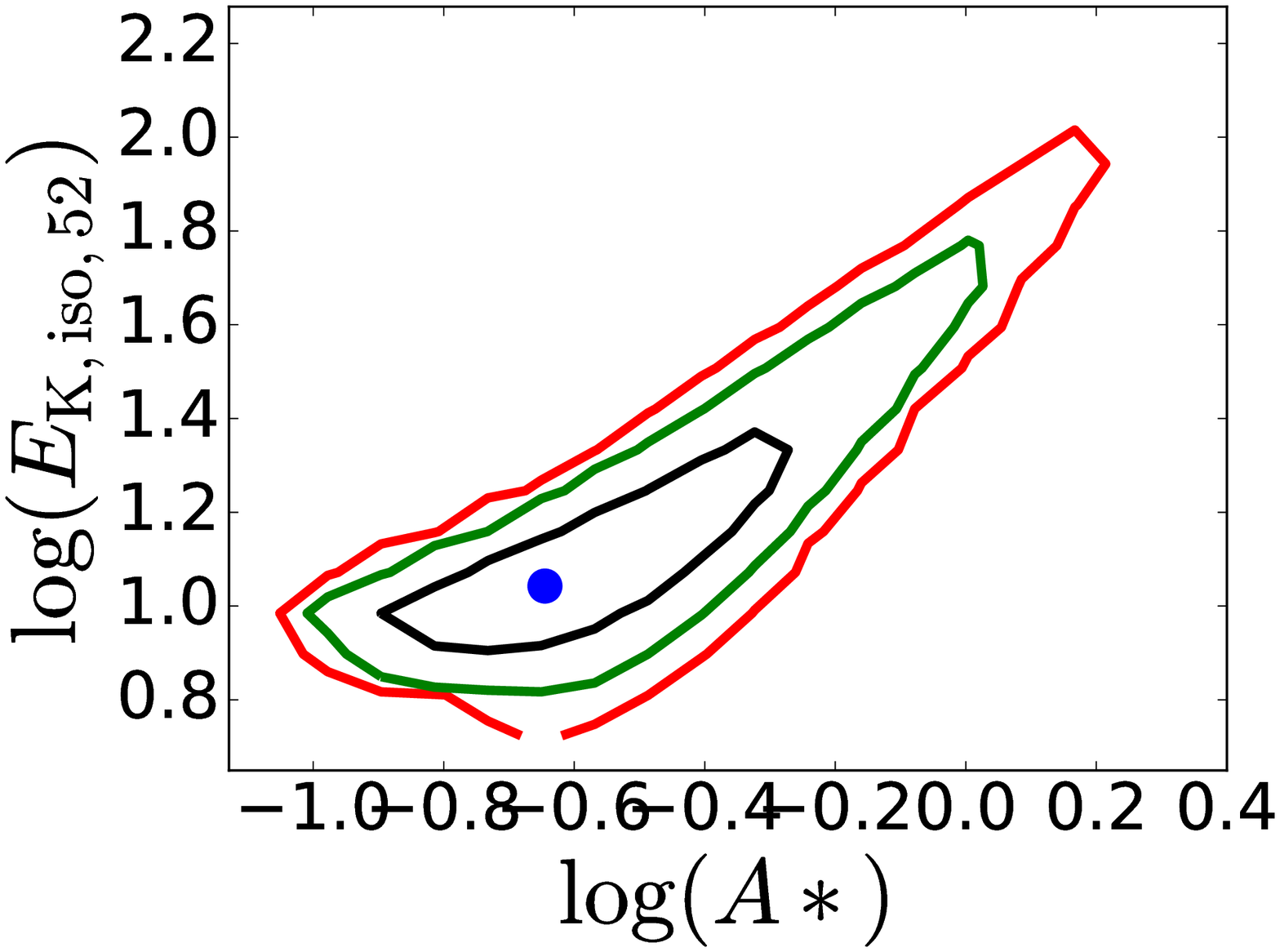} &
 \includegraphics[width=0.31\textwidth]{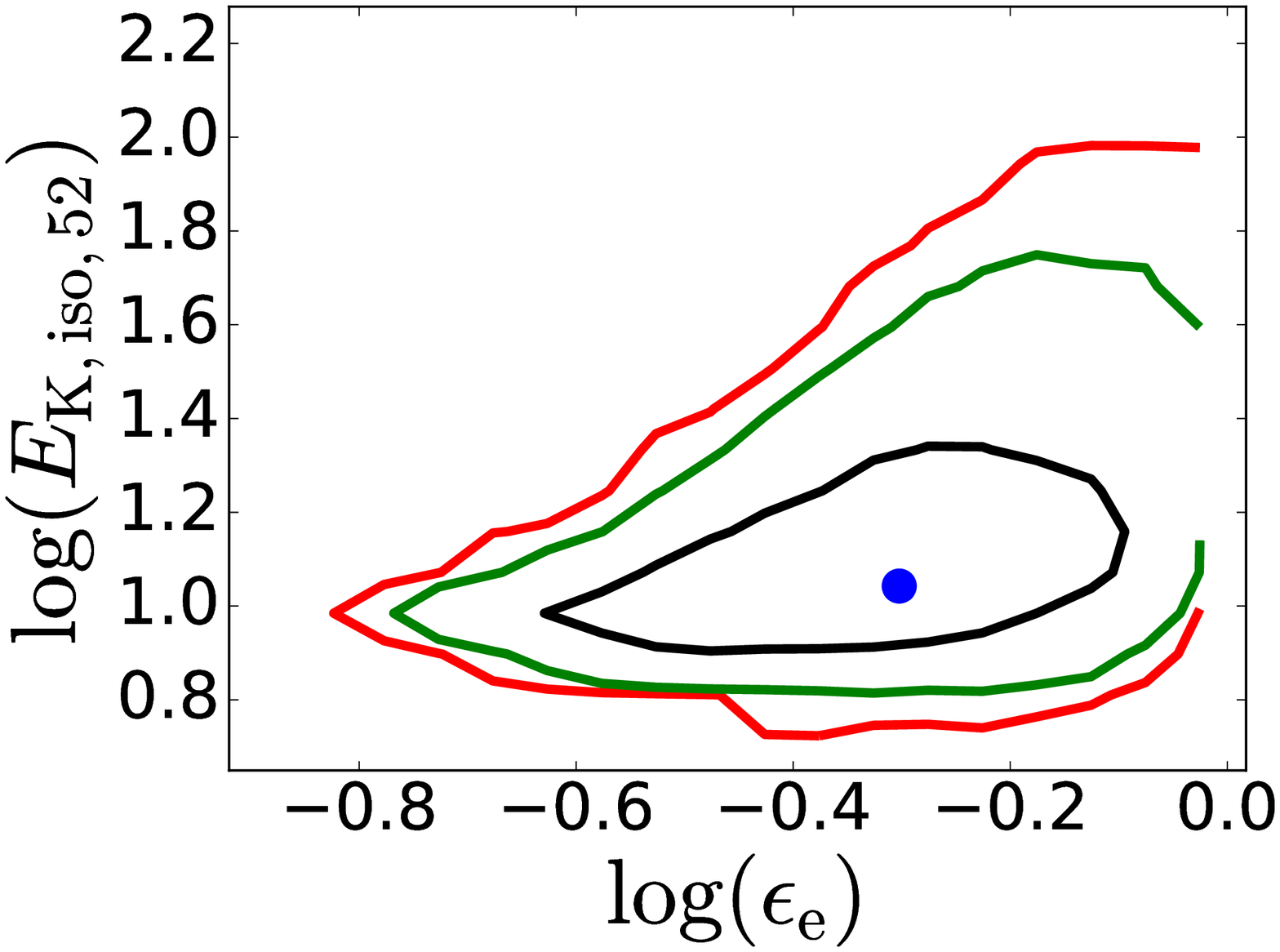} &
 \includegraphics[width=0.31\textwidth]{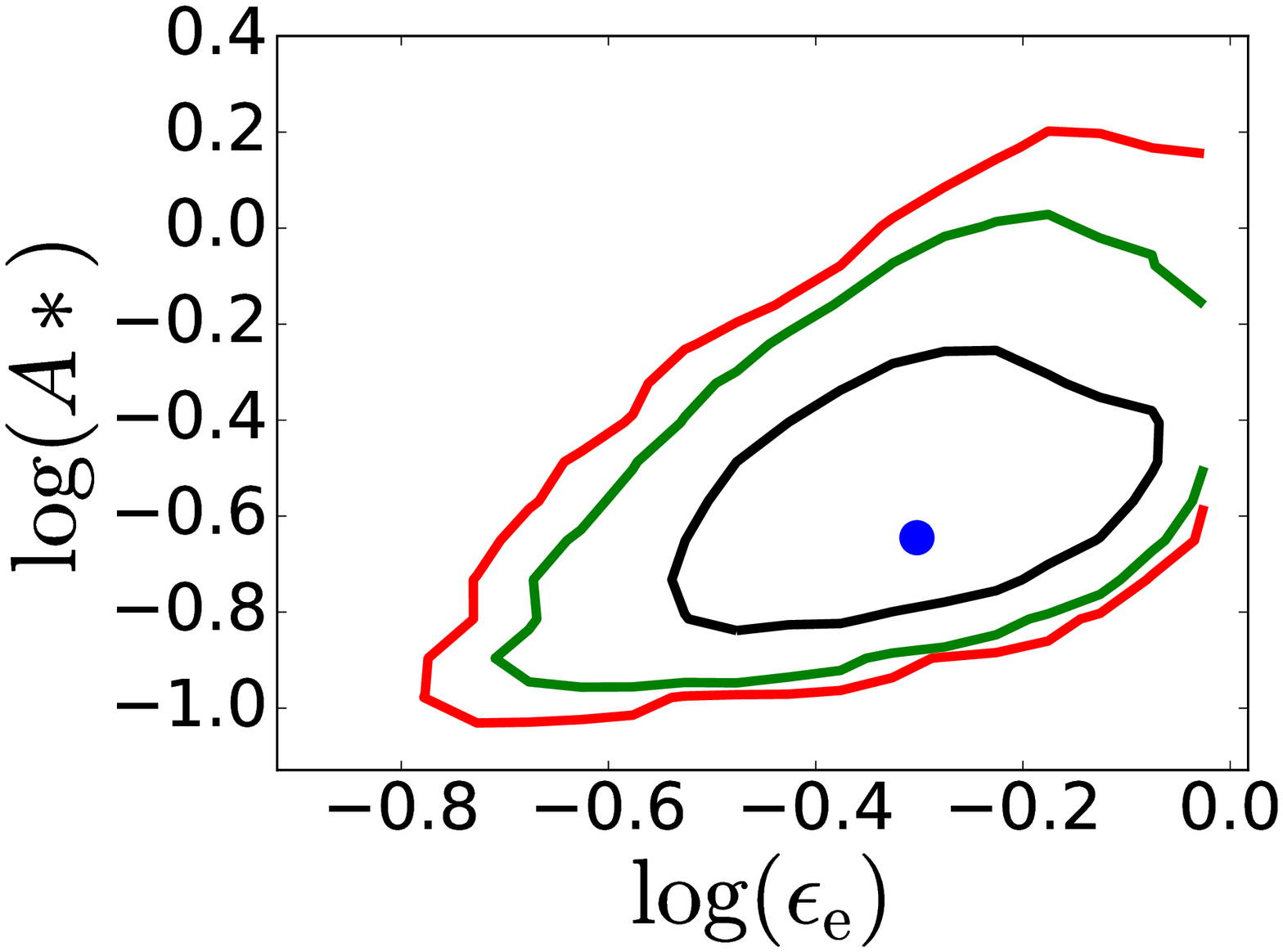} \\
 \includegraphics[width=0.31\textwidth]{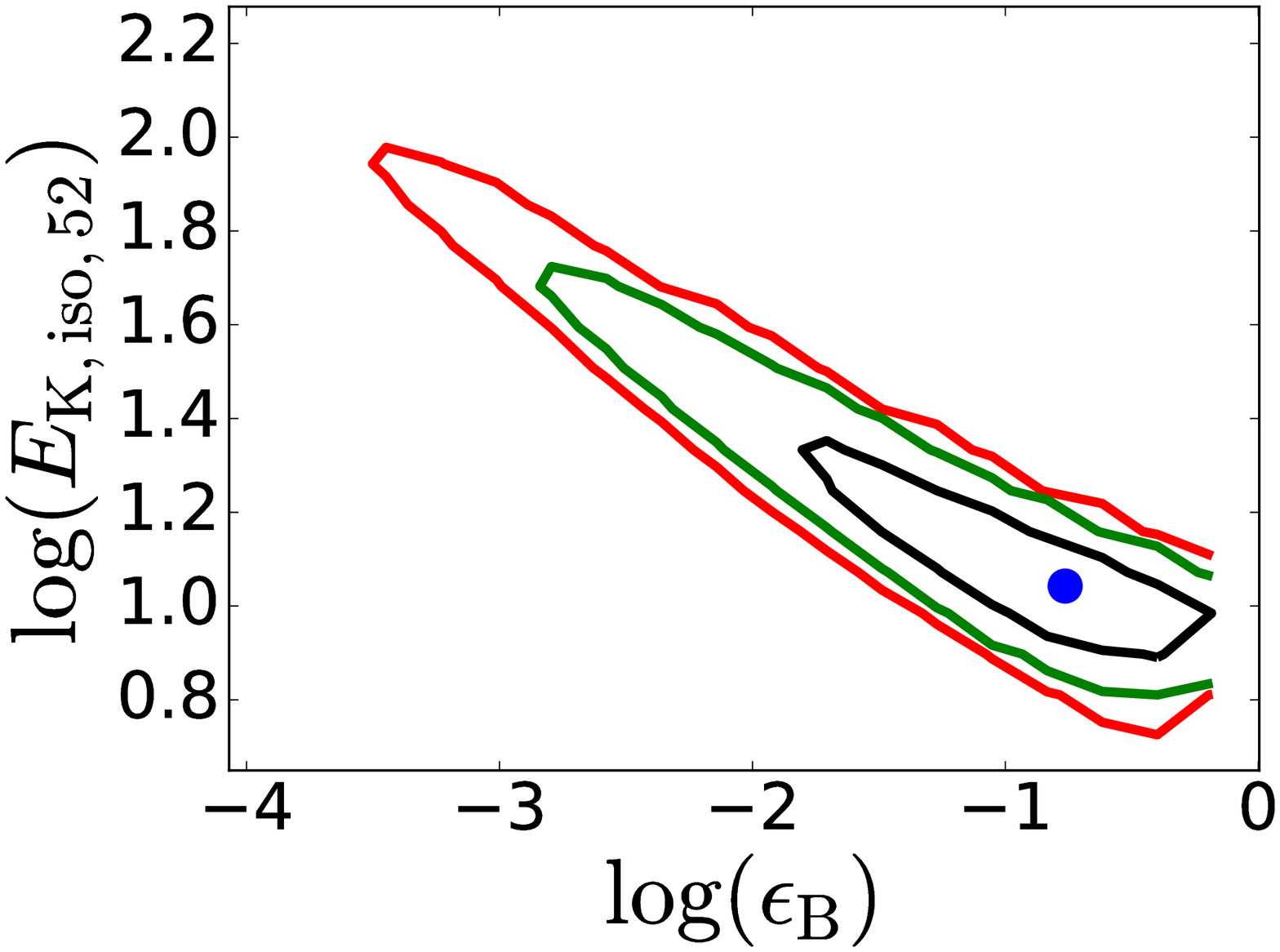} &
 \includegraphics[width=0.31\textwidth]{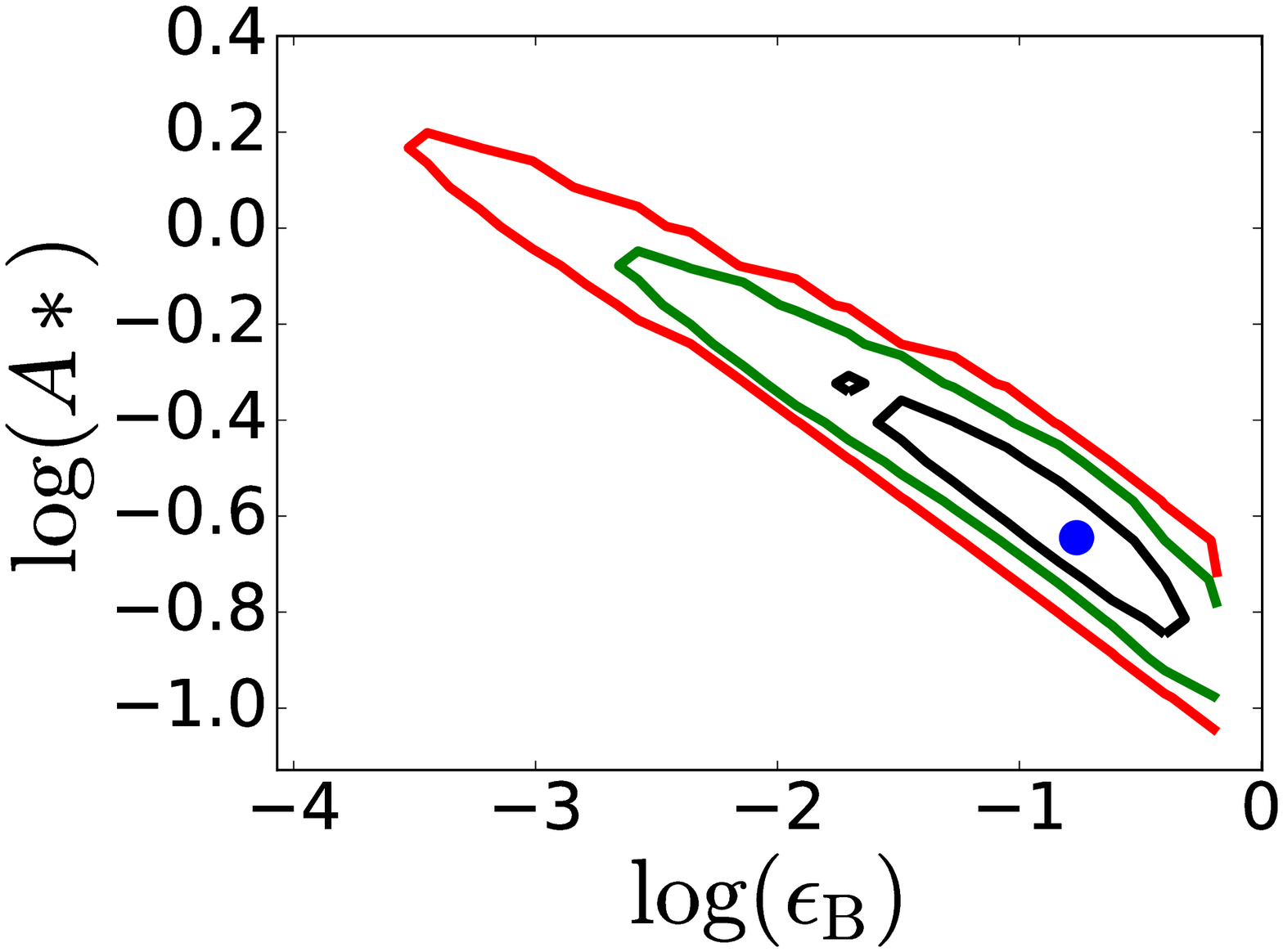} &
 \includegraphics[width=0.31\textwidth]{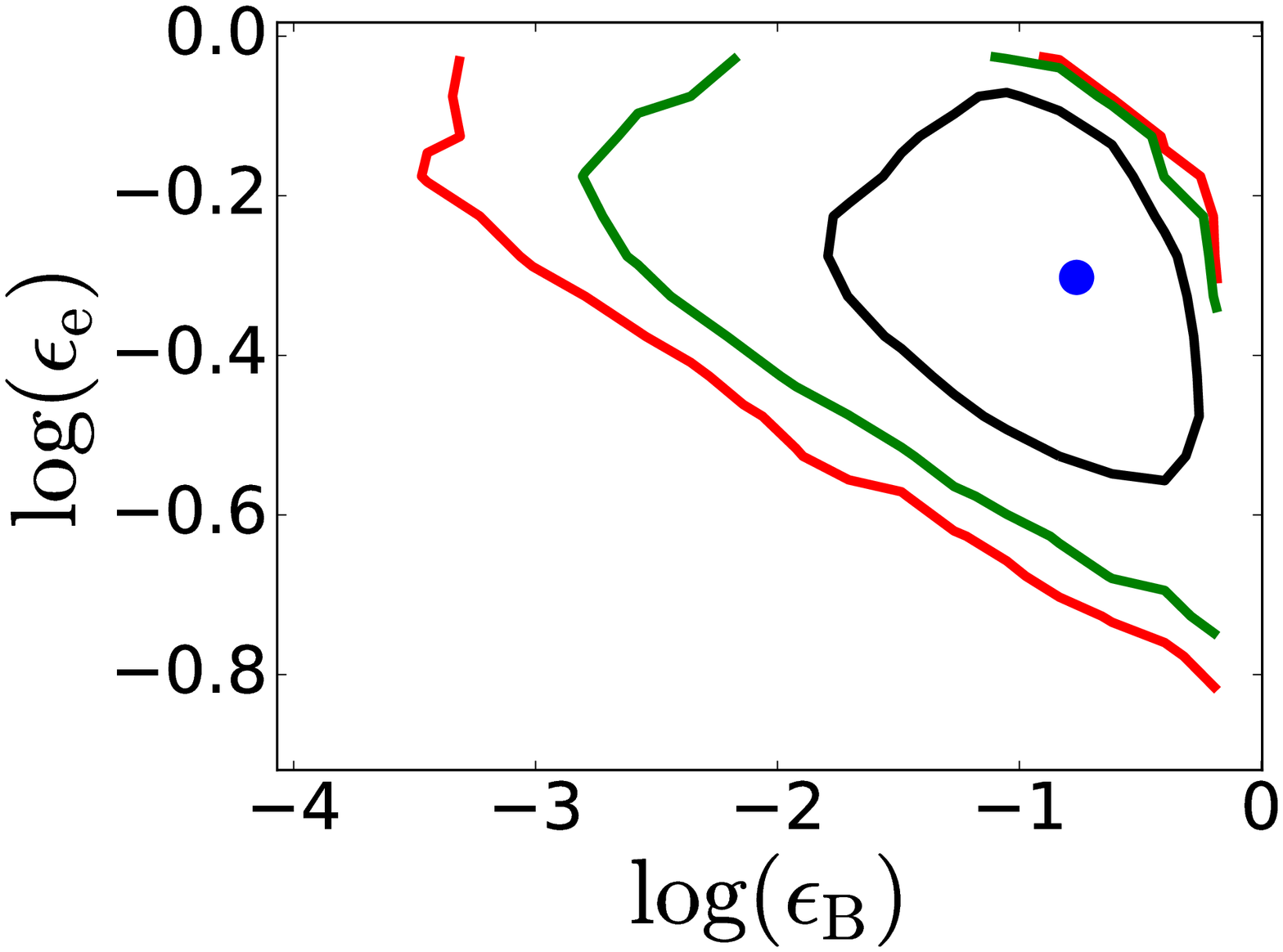} 
\end{tabular}
\caption{1$\sigma$ (red), 2$\sigma$ (green), and 3$\sigma$ (black) contours for correlations 
between the physical parameters \EKiso, \Astar, \epse, and \epsb\ from Monte Carlo simulations,
together with the best-fit model (blue dot). We have restricted $\epse+\epsb<1$, and do not include 
the radio data before 2.5\,d in the analysis.}
\label{fig:corrplots_wind}
\end{figure*}

We carried out an MCMC analysis for the wind environment similar to the analysis for the ISM case 
described in Section \ref{text:model}. The parameters of our best-fit model and the results of the 
MCMC analysis are listed in Table \ref{tab:params_wind}. Light curves and radio SEDs are presented 
in Figures \ref{fig:modelsed_wind4} and \ref{fig:modellc_wind}, and histograms of the posterior 
density and correlation contours between the physical parameters are presented in Figures 
\ref{fig:hists_wind} and \ref{fig:corrplots_wind}. All figures and tables pertaining to the wind 
model are available in the on-line edition of this article.

Our best fit model reproduces the X-ray and optical light curves well, but over-predicts the radio 
SED in the first two epochs. The spectral break frequencies are in the order $\nuc<\nua<\numax$ at 
0.1\,d, a scenario that occurs more frequently in wind environments due to the higher density at 
small radii \citep{kmz04}. We note that in this scenario, synchrotron self-absorption prevents the 
electrons from cooling efficiently and re-distributes the electron energy, thus changing the 
underlying distribution function, an effect that has not been modeled in detail. Accurately fitting 
the radio SEDs then requires an additional source of opacity in the radio and mm-bands, which must 
disappear by 9.5\,d. We note that increased opacity at radio wavelengths is expected when non-shock 
accelerated electrons are present \citep{ew05,webn17}, and the increased self-absorption from these 
`thermal electrons' is indeed expected to decline with time \citep{rl17}. A detailed analysis of 
this effect requires a treatment of the observed radiation spectrum including thermal electrons in 
jetted GRB afterglows, and is beyond the scope of this work.

\bibliographystyle{apj}
\bibliography{grb_alpha,gcn}

\clearpage
\
\vfil
\hfil \textsc{On-line only material} \hfil
\vfil
\clearpage

\begin{deluxetable}{lcc}
 \tabletypesize{\footnotesize}
 \tablecolumns{2}
 \tablecaption{Parameters for best-fit wind model}
 \tablehead{   
           \colhead{Parameter} &
           \colhead{Best-fit}  &
           \colhead{MCMC}
   }
 \startdata    
   $p$                  & 2.06 & $2.07^{+0.03}_{-0.02}$  \\[2pt]
   \epse                & 0.50 & $0.49^{+0.20}_{-0.15}$  \\[2pt]
   \epsb                & 0.17 & $(9.7^{+20.2}_{-7.8})\times10^{-2}$  \\[2pt]
   \Astar               & 0.23 & $0.29^{+0.20}_{-0.10}$  \\[2pt]
   $E_{\rm K, iso, 52}$ & 11.0 & $12.5^{+8.6}_{-3.0}$    \\[2pt]
   \tjet\ (d)           & 1.05 & $1.1^{+0.4}_{-0.3}5$    \\[2pt]
   \thetajet\ (deg)     & 2.84 & $2.9\pm0.2$             \\[2pt]
   \AV\ (mag)           & 0.39 & $0.40\pm0.02$           \\[2pt]
   \nuac\  (Hz)         & $8.6\times10^{8}{}^\dag$ & \ldots \\[2pt]
   \nuc\   (Hz)         & $8.9\times10^{10}$    & \ldots \\[2pt]
   \nusa\  (Hz)         & $3.7\times10^{11}$    & \ldots \\[2pt]
   \numax\ (Hz)         & $4.8\times10^{14}$    & \ldots \\[2pt]
   $F_{\nu, \rm peak}$ (mJy) & 14.5             & \ldots \\[2pt]
   \hline\\[-4pt]
   $E_{\gamma}$ (erg)   & \multicolumn{2}{c}{$(3.4^{+1.3}_{-1.2})\times10^{50}$} \\[2pt]
   $E_{\rm K}$ (erg)    & \multicolumn{2}{c}{$(1.6^{+1.3}_{-0.5})\times10^{50}$}  \\[2pt]
   $E_{\rm tot}$ (erg)  & \multicolumn{2}{c}{$\approx5\times10^{50}$}  \\[2pt]
   $\eta_{\rm rad}$     & \multicolumn{2}{c}{$\approx68\%$}
 \enddata
 \tablecomments{All break frequencies are listed at 0.1\,d. 
 ${}^\dag$This break frequency is not directly constrained by the data.}
\label{tab:params_wind}
\end{deluxetable}

\end{document}